\documentclass[preprint2,numberedappendix]{emulateapj-rtx4}
\usepackage{graphicx}
\usepackage{times}
\usepackage{url}
\usepackage{bm}
\usepackage{color}
\usepackage{natbib}
\graphicspath{{./figs/}{./png/}}

\newcommand{\EQ}{\begin{equation}}
\newcommand{\EE}{\end{equation}}
\newcommand{\EQA}{\begin{eqnarray}}
\newcommand{\EEA}{\end{eqnarray}}
\newcommand{\brac}[1]{\langle #1 \rangle}
\newcommand{\pd}{\partial}

\newcommand{\ve}[1]{\boldsymbol{#1}}
\newcommand{\mean}[1]{\overline{#1}}

\newcommand{\meanF}{\overline{F}}
\newcommand{\cs}{c_{\rm s}}

\newcommand{\etat}{\eta_{\rm t}}

\newcommand{\urms}{u_{\rm rms}}
\newcommand{\Urms}{u_{\rm rms}^{\rm tot}}
\newcommand{\brms}{B_{\rm rms}}
\newcommand{\Brms}{B_{\rm rms}^{\rm tot}}
\newcommand{\Beq}{B_{\rm eq}}

\newcommand{\kef}{k_{\rm f}}

\newcommand{\chit}{\chi_{\rm SGS}}
\newcommand{\chitm}{\overline{\chi}_{\rm SGS}}

\newcommand{\Pm}{{\rm Pm}}
\newcommand{\Rm}{{\rm Rm}}
\newcommand{\Rey}{{\rm Re}}

\newcommand{\Pra}{{\rm Pr}}
\newcommand{\Ta}{{\rm Ta}}

\newcommand{\Rat}{{\rm Ra}_{\rm t}}
\newcommand{\Ma}{{\rm Ma}}

\newcommand{\Co}{{\rm Co}}

\def\onethird{{\textstyle{1\over3}}}
\def\onehalf{{\textstyle{1\over2}}}

\begin{document}

\title{Spoke-like differential rotation in a convective dynamo with a coronal envelope}

\author{J\"orn Warnecke$^{1,2}$, Petri J.\ K\"apyl\"a$^{1,3}$, Maarit J.\ Mantere$^{3,4}$ and Axel Brandenburg$^{1,2}$}
\affil{$^1$NORDITA, KTH Royal Institute of Technology and Stockholm University,
Roslagstullsbacken 23, SE-10691 Stockholm, Sweden; joern@nordita.org\\
$^2$Department of Astronomy, AlbaNova University Center,
Stockholm University, SE-10691 Stockholm, Sweden\\
$^3$Physics Department, Gustaf H\"allstr\"omin katu 2a, P.O. Box 64,
FI-00014 University of Helsinki, Finland\\
$^4$Aalto University, Department of Information and Computer Science, P.O. Box 15400, FI-00076 Aalto, Finland\\
}
\email{joern@nordita.org
($ $Revision: 1.329 $ $)
}

\begin{abstract}
We report on the results of four convective dynamo simulations with an
outer coronal layer.
The magnetic field is self-consistently generated by the convective
motions beneath the surface.
Above the convection zone, we include a polytropic layer
that extends to 1.6 solar radii.
The temperature increases in this region
to $\approx8$ times the value at the surface, corresponding to
$\approx1.2$ times the value at the bottom of the spherical shell.
We associate this region with the solar corona.
We find solar-like differential rotation with radial contours of
constant rotation rate, together with a near-surface shear layer.
This non-cylindrical rotation profile is caused by a non-zero latitudinal
entropy gradient that offsets the Taylor--Proudman balance through the
baroclinic term.
The meridional circulation is multi-cellular with a solar-like
poleward flow near the surface at low latitudes. 
In most of the cases, the mean magnetic field is 
oscillatory with equatorward migration in two cases.
In other cases, the equatorward migration is overlaid by stationary or
even poleward migrating mean fields.
\end{abstract}


%________________________________________________________________

\section{Introduction}

The Sun has an activity cycle of 11\,yr, which is manifested
by sunspots occurring at the solar surface.
The sunspot number changes from a few during
minimum to over 200 during maximum.
The sunspot locations display a latitudinal dependence during the cycle.
At solar minimum (maximum),
sunspots emerge preferably at higher (lower) latitudes.
By plotting the sunspot latitudes for several cycles, one obtains the
``butterfly diagram''.  Every 11\,yr the polarity of sunspot pairs
changes sign, which is characteristic of the 22\,yr magnetic cycle.
To understand this cyclical behavior, one has to connect the fluid
motions in the Sun with magnetic field generation to construct dynamo
models.  These dynamo models should be able to reproduce the 22\,yr
magnetic activity cycle as well as the large-scale magnetic field
evolution at the surface of the Sun.  It is widely believed that
sunspots are correlated with the large-scale magnetic field
distribution.  Therefore, a successful solar dynamo model should
reproduce the equatorward migration of the large-scale field as we
observe it indirectly from sunspots and more directly from synoptic
magnetograms.

Until recently, only kinematic mean-field models, where turbulent effects are
parameterized through transport coefficients \citep[e.g.,][]{KR80}, have
been able to show equatorward migration \citep[e.g.,][]{DC99,KKT06,KO11}.
Such models have been used to reproduce certain features of the solar cycle,
such as the Maunder minimum \citep[e.g.,,][]{K10}.
However, those models are only valid in the kinematic regime in which the
fluid motions are assumed to be given, so they are not self-consistently
generated. The backreaction from the magnetic field is either ignored
or taken into account in a rudimentary way involving ad hoc quenching
of the turbulent transport coefficients.
Until recently, direct numerical simulations (DNS)
of the solar dynamo have been unsuccessful
in producing migration toward the equator using convective motions to drive a
dynamo \citep[e.g.,][]{G83,BMT04,KKBMT10,GCS10,BMBBT11,NBBMT13}.
This was presumably due to the low fluid and magnetic
Reynolds numbers of those simulations.
Equatorward migration was, for the
first time, found in DNS by \cite{KMB12}. The exact cause is
not yet fully understood, but the amount of density stratification
seems to play an important role \citep{KMCWB13}.

An important ingredient of the solar dynamo is differential rotation.
It is believed that strong shear at the bottom of the convection zone
\citep{SW80} or near the surface \citep{B05}, plays an important
role in amplifying the magnetic field.
However, even today it is not straightforward to reproduce a solar-like
differential rotation profile.
Mean-field simulations \citep{BMT92,KR95} have been able
to reproduce a solar-like rotation profile by modeling small-scale
effects through mean-field coefficients such as the $\Lambda$ effect
and anisotropic heat transport \citep[see, e.g.,][]{R80,R89}.
These models reproduce the positive (negative) latitudinal gradient
of angular velocity in
the northern (southern) hemisphere---i.e., the equator rotates faster than
the poles---together with ``spoke-like'' contours in the meridional plane.
DNS of convective dynamos are able to reproduce a rapidly
rotating equator at sufficiently large Coriolis numbers
\citep{BMT04,KMGBC11}.
Spoke-like differential rotation has only been found
in purely hydrodynamical large-eddy simulations (LES)
by imposing a latitudinal entropy
gradient \citep{MBT06} or, recently, by adding a stably stratified layer \citep{BMT11} at
the bottom of the convection zone.
A self-consistently generated spoke-like profile in DNS of
magnetohydrodynamics has not yet been
found.

An important issue with solar dynamo models is the effect of
catastrophic quenching of the dynamo at high magnetic
Reynolds numbers; see \cite{BS05}.
This is caused by the accumulation of magnetic helicity in
the dynamo region.
DNS provide evidence that magnetic helicity fluxes both within and through the
boundaries of the dynamo domain can prevent the dynamo from being
catastrophically quenched \citep[e.g.,][]{BS04,HB12}.
In the case of the Sun, magnetic helicity flux can emerge through the
solar surface and can be transported away from the Sun by
coronal mass ejections or by the solar wind \citep{BB03}.
In earlier work this was modeled by using an upper layer
at the top of a dynamo region to allow for magnetic helicity fluxes leaving the
domain \citep{WB10,WBM11,WBM12,WKMB12}.
This two-layer model was successful in showing that
the dynamo is not only enhanced, but that it can actually trigger the emergence of
coronal ejections.
These ejections have a similar shape as coronal mass ejections and
carry a significant amount of magnetic helicity out of the dynamo region.
In these models, the temperature in the coronal layer was the same
as at the surface of the convection zone, which did not allow for a large
density jump to develop.
Furthermore, in the polytropic convection zone of \cite{WKMB12},
the convective flux was smaller than the radiative flux.
Besides dynamo models, this two-layer approach was successful in
combination with stratified turbulence in
producing a bipolar magnetic region \citep{WLBKR13} as a possible
mechanism of sunspot formation.

In this work we use the two-layer approach to investigate the
influence of the coronal layer as an upper boundary condition for a
convective dynamo.
We focus on the physical properties and dynamics in the
convection zone.
The effects of varying the strength of stratification on a convective
dynamo without a corona is studied in a companion paper \citep{KMCWB13}.

\section{Model and Setup}
\label{model}

We use a two-layer model in spherical polar coordinates
($r,\theta,\phi$), where the lower layer ($r\leq R$) represents the 
convection zone and the upper layer represents the corona.
The simulations are performed in a spherical wedge with radial extent
$r_0 \leq r\leq R_{\rm c}=1.6\,R$, where $r_0=0.7\,R$ corresponds to the
bottom of the convection zone and $R$ to the solar radius,
for colatitudes $15^{\circ} \leq\theta\leq 165^{\circ}$ and an
azimuthal extent $0\leq\phi\leq 45^{\circ}$.
We solve the following
equations of compressible magnetohydrodynamics,
\begin{equation}
{\partial\ve{A}\over\partial
  t}=\ve{u}\times\ve{B}-\mu_0\eta\ve{J},
\end{equation}
\begin{equation}
{{\rm D}\ln\rho\over{\rm D} t} =-\ve{\nabla}\cdot\ve{u},
\end{equation}
\begin{equation}
{{\rm D}\ve{u}\over{\rm D} t}=  \ve{g} - 2\ve{\Omega}_0 \times \ve{u} + {1\over\rho}
\left(\ve{J}\times\ve{B} - \ve{\nabla} p+\ve{\nabla}\cdot
  2\nu\rho\mbox{\boldmath ${\sf S}$}\right),
\end{equation}
\begin{equation}
T{{\rm D} s\over{\rm D} t}=-{1\over\rho}\ve{\nabla}\cdot
\left({\bm F^{\rm rad}}+ {\bm F^{\rm SGS}}\right) +
2\nu\mbox{\boldmath ${\sf S}$}^2+{\mu_0\eta\over\rho}\ve{J}^2 -
\Gamma_{\rm cool},
\label{entro}
\end{equation}
where the magnetic field is given by $\ve{B}=\ve{\nabla}\times\ve{A}$
and thus obeys ${\bm \nabla}\cdot{\bm B}=0$ at all times,
${\bm J}=\mu_0^{-1}\bm\nabla\times{\bm B}$ is the current density,
$\mu_0$ is the vacuum permeability,
$\eta$ and $\nu$ are the magnetic diffusivity and kinematic viscosity,
respectively,
${\rm D}/{\rm D} t =\partial/\partial t+\ve{u}\cdot\ve{\nabla}$ is the
advective time derivative, $\rho$ is the density, and $\ve{u}$ is the
velocity.
The traceless rate-of-strain tensor is given by
\begin{equation}
{\mathsf
  S}_{ij}=\onehalf(u_{i;j}+u_{j;i})-\onethird\delta_{ij}\ve{\nabla}\cdot\ve{u},
\end{equation}
where semicolons denote covariant differentiation; see \cite{MTBM09}
for details.
Furthermore,
$\ve{\Omega}_0 =\Omega_0(\cos\theta,-\sin\theta,0)$ is the rotation vector
and $p$ is the pressure.
The gravitational acceleration is given by
\begin{equation}
\ve{g}=-GM\ve{r}/r^3,
\end{equation}
where $G$ is Newton's gravitational constant and $M$ is the
mass of the star.
The radiative and sub-grid scale (SGS) heat fluxes are defined as
\begin{equation}
{\bm F^{\rm rad}}=-K\ve{\nabla} T,\quad {\bm F^{\rm SGS}} =-\chit \rho  T\ve{\nabla} s,
\end{equation}
where $K$ is the radiative heat conductivity and $\chit$ is the turbulent heat
diffusivity, which represents the unresolved convective transport of heat.
The fluid obeys the ideal gas law, $p=(\gamma-1)\rho e$,
where $\gamma=c_{\rm P}/c_{\rm V}=5/3$ is the ratio of specific heats at constant
pressure and constant volume, respectively, and $e=c_{\rm V} T$ is the
internal
energy density, defining the temperature $T$.
Finally, $\Gamma_{\rm cool}$ is the cooling profile
that is specified in Equation~(\ref{cool}).

\begin{figure}[t!]
\begin{center}
\includegraphics[width=0.9\columnwidth]{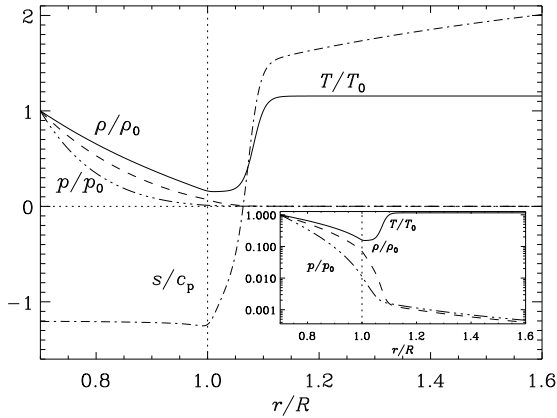}
%\includegraphics[width=0.6\columnwidth]{strat_A.eps}
%ApJ
\end{center}\caption[]{
%\small
%ApJ
Averaged radial profiles of stratification for Run~A.
The normalized density $\rho/\rho_0$ (dashed lines), pressure $p/p_0$
(triple-dot-dashed), and temperature $T/T_0$ (solid lines) are plotted together
with the specific entropy $s/c_{\rm P}$ (dash-dotted lines) over radius.
The inset shows various profiles in logarithmic representation to
emphasize the steep decrease of 
the pressure and density in the coronal layer.
The subscript $0$ refers to the value at the bottom of the convection zone.
}
\label{strat}
\end{figure}

The two-layer model is similar to that used in previous work
\citep{WB10,WBM11,WBM12,WKMB12}, except that here
we improve the model of \cite{WKMB12} in two important ways.
First, we use a more realistic model for the convection zone 
than in \cite{KMB11,KMB12}.
Instead of using a polytropic setup with $m=1$, we lower the radiative
flux by using a profile for $m$ (defined in Equation~(\ref{eq-m})) and
introducing a turbulent heat conductivity $\chit$ \citep[referred to
as $\chi_{\rm t}$ in][]{KMB12}.  We apply a piecewise constant profile
for $\chit$ such that in the interval of $0.75\,R\leq r\leq 0.97\,R$
it is equal to a quantity $\chitm$
(whose value is related to $\nu$ via the Prandtl number specified below),
and it goes smoothly to zero above and below the boundaries of the interval.
Additionally, we change the temperature profile compared with
our earlier isothermal cold
corona to a temperature-stratified corona, which is $\approx 8$ times hotter
than the surface and $\approx 1.2$ times hotter than the bottom of the
convection zone.
The profiles of averaged temperature, density, pressure, and entropy
for a typical run are shown in Figure~\ref{strat}.

We initialize the simulations with precalculated radial profiles
of temperature, density, and pressure.
In the convection zone ($r \leq R$) we have an isentropic and hydrostatic
initial state for the temperature, whose gradient is given by
\begin{equation}
\frac{\pd T}{\pd r}=\frac{-|\ve{g}(r)|}{c_{\rm V}(\gamma-1)(m_{\rm ad}+1)},
\end{equation}
where $m_{\rm ad}=1.5$ is the polytropic index for an adiabatic stratification.
This leads to a temperature minimum $T_{\rm min}$ above the
surface of the convective layer at $r=R$.
In the corona ($R \leq r \leq R_{\rm c}$), we prescribe the temperature as
\begin{equation}
T_{\rm ref}(r)=T_{\rm min}+\onehalf\left(T_{\rm cor}-T_{\rm
    min}\right)\left[1+\tanh{\left({r-r_{\rm tra}\over w}\right)}\right],
\label{eq:tem}
\end{equation}
where $T_{\rm cor}$ is temperature in the corona, and $r_{\rm tra}$
and the width $w=0.02\, R$ are chosen to produce a smooth temperature
profile as shown in Figure~\ref{strat}.
The cooling profile $\Gamma_{\rm cool}$ in Equation~(\ref{entro})
maintains the temperature profile,
\begin{equation}
\Gamma_{\rm cool}=\Gamma_0 f(r){T-T_{\rm ref}(r)\over T_{\rm ref}(r)},
\label{cool}
\end{equation}
where $f(r)$ is a profile function equal to unity in $r>R$ and
going smoothly to zero in $r\leq R$, and $\Gamma_0$ is a
cooling luminosity chosen such that the temperature in the corona
relaxes towards the reference temperature profile $T_{\rm ref}(r)$ given in
Equation~(\ref{eq:tem}).
As stated in Equation (\ref{cool}), the cooling function is sensitive
to the total temperature consisting of a mean part and a fluctuating part.
Nevertheless, temperature fluctuations can still develop.
The stratification of density follows from hydrostatic equilibrium.
The density contrast within the convection zone is $\rho_0/\rho_{\rm
s}\approx14$ (see the sixth column in Table~\ref{tab:runs}),
while in the whole domain $\rho_0/\rho_{\rm t}\approx2000$, where
$\rho_0$ is the density at the bottom ($r=r_0$),
$\rho_{\rm s}$ is the density at the surface ($r=R$) and $\rho_{\rm
  t}$ is the density at the top of the corona ($r=1.6\,R$).
The location of the surface, $r=R$, is close to the position where
the radial entropy gradient changes sign, which is slightly below the
surface; see Figure~\ref{strat}.
This implies that, similar to the Sun, convection ceases just below
the surface.
The radial heat conductivity profile is chosen such that the energy in
the convection zone is transported mostly by convective motions.
We apply a profile for the viscosity $\nu$ that is
constant in the convection zone ($r\leq R$) and increases smoothly
above the surface to a value that is 20 times higher in the corona.
This helps to suppress high velocities and sharp flow structures
aligned with the rotation vector in the corona---especially in the
beginning of the simulation, when the magnetic field is weak.
Compared with the use of velocity damping in \cite{WKMB12}, this approach
is Galilean invariant and allows the flow to develop more freely.
The magnetic diffusivity $\eta$ is constant throughout the convection
zone, but decreases by 20\% in the corona.
In the convection zone, the radiative heat conductivity $K$ is defined
via a polytropic index $m$ given by
\begin{equation}
m=2.5\,(r/r_0)^{-15}-1,
\label{eq-m}
\end{equation}
which has a value of 1.5 at the bottom of the convection zone.
The conductivity is proportional to $m+1$ and
decreases toward the surface as $r^{-15}$.
In the corona, $K$ is chosen such that $\chi=K/c_{\rm P}\rho=\rm const$.
The radiative diffusivity $\chi$ varies from $0.5\chitm$ at the
bottom of the convection zone to $0.04\chitm$ near the surface and
$0.3\chitm$ in the corona.
We initialize the magnetic field with weak Gaussian-distributed perturbations
inside the convection zone.

We use periodic boundary conditions in the azimuthal direction.
For the velocity field we apply stress-free boundary conditions at the
radial and latitudinal boundaries.
The magnetic field follows a perfect conductor condition at the lower
radial and the two latitudinal boundaries.
On the outer radial boundary, we force the field to be radial;
for a discussion on the applicability of this boundary condition for
the Sun, see \cite{WKMB12}.
We fix the gradient of the temperature at the lower radial boundary such that
it corresponds to a given radiative flux and we set the temperature to
a constant value at the radial outer boundary.
At the latitudinal boundaries, we impose a vanishing $\theta$-derivative
of entropy to have zero heat flux through the boundary.
 
Our runs are characterized by the values of the fluid and magnetic Reynolds
numbers, $\Rey=\urms/\nu\kef$ and $\Rm=\urms/\eta\kef$, respectively,
where $\urms$ is the volume averaged rms velocity in the convection zone,
and $\kef=2\pi/(R-r_0)\approx21/R$ is used as a reference wavenumber.
To represent the turbulent velocities in a proper way, we define
$\urms=\sqrt{3/2\brac{u_r^2+u_{\theta}^2}_{\theta\phi r\leq R}}$,
which corrects for the removal of the differential rotation-dominated
$\phi$-component of velocity.
In our case, $\chitm\approx0.02\chi_{t0}$, where $\chi_{t0}=\urms/3\kef$
is an estimate for the macro-physical turbulent diffusivity.
We also define the fluid and magnetic Prandtl numbers
$\Pra=\nu/\chitm$ and $\Pm=\nu/\eta=\Rm/\Rey$, the Coriolis number
$\Co=2\Omega_0/\urms\kef$, and the Taylor number $\Ta=(2\Omega_0 R^2/\nu)^2$.
Time is given in turnover times, $\tau=(\urms\kef)^{-1}$.
We measure the magnetic field strength as the rms value over
the convection zone, $\brms$, and we normalize this value with
the equipartition value of the magnetic field defined by
$\Beq^2=\mu_0\brac{\rho\bm{u}^2}_{r\leq R}$.
The typical diffusion time of the system is characterized by the fluid and
magnetic Reynolds numbers times the turnover time.
We use the (semi-) turbulent Rayleigh number $\Rat$ from the thermally
relaxed state of the run,
\begin{eqnarray}
\Rat\!=\!\frac{GM(R-r_0)^4}{\nu \chitm R^2} \bigg(-\frac{1}{c_{\rm
    P}}\frac{{\rm d}\brac{s}_{\theta\phi t}}{{\rm d}r} \bigg)_{r=0.85\, R}.
\label{equ:Ra}
\end{eqnarray}
To monitor the solutions in the convection zone, we use two different
heights, one near the surface at $r_1=0.97\, R$, and one in the middle of
the convection zone at $r_2=0.84\, R$.
We use the 
{\sc Pencil Code}\footnote{\texttt{http://pencil-code.googlecode.com}}
with sixth-order centered finite differences in space and 
a third-order Runge-Kutta scheme in time;
see \cite{MTBM09} for the extension of the {\sc Pencil Code} to
spherical coordinates.

\section{Results}

\begin{deluxetable*}{llccccccccccrrc}
%ApJ
%\begin{deluxetable}{llccccccccccrrc}
\tabletypesize{\scriptsize}
\tablecaption{Summary of the Runs}
\tablecomments{The second to sixth columns show 
quantities that are input parameters to the models, whereas the quantities 
in the last eight columns are results of the simulations computed from the
saturated state.
All quantities are volume averaged over the convection zone $r\leq R$,
unless explicitly stated otherwise.
The Mach number is defined as
$\Ma=\urms/\cs|_{r=0.97\, R}$ and the latitudinal differential rotation is quantified through
$(\Delta_\Omega=\partial\mean{\Omega}/\partial\!\cos^2\!\theta) /\Omega_0$
evaluated at ${r\leq R}$.
$\Delta T=(T_{\rm pol}-T_{\rm eq})/T_{\rm eq}$ is the normalized
temperature difference between pole $T_{\rm
  pol}=\left(T(\theta=15^\circ)+T(\theta=165^\circ)\right)/2$
and equator $T_{\rm eq}=T(\theta=90^\circ)$, measured at the
surface ($r=R$).
}
\tablewidth{0pt}
\tablehead{
\colhead{Run} & \colhead{grid} & \colhead{$\Pra$} & \colhead{$\Pm$} & 
\colhead{$\mbox{Ta}$} & \colhead{$\rho_0/\rho_{\rm s}$} & \colhead{$\Ma$} &
\colhead{${\Rat}$} & \colhead{$\Rey$} & \colhead{$\Rm$} &
\colhead{$\Co$} & \colhead{$\brms^2/\Beq^2$} &
\colhead{$\Delta_\Omega$} & \colhead{$\Delta T$} &
}
\startdata
A  & $400\times256\times192$ & $5$ & $1$ & $1.4\cdot10^{10}$ & $14$ &
$0.08$ & $1.8\cdot10^{6}$ & $25$ & $25$ & $11$&$0.25$&$-0.011$&$0.08$\\
Ab& $400\times256\times192$ & $5$ & $0.71$ & $1.4\cdot10^{10}$ & $14$
&$0.08$&$1.8\cdot10^{6}$&$25$&$18$&$11$&$0.22$&$-0.014$&$0.08$\\ 
Ac& $400\times256\times192$ & $5$ & $1.67$ & $1.4\cdot10^{10}$ & $14$
&$0.08$&$2.1\cdot10^{6}$&$25$&$41$&$11$&$0.27$&$0.009$&$0.08$\\ 
B&$400\times256\times192$&$4$&$1$&$\!7.2\cdot10^{9}$&$14$ 
&$0.09$&$1.2\cdot10^{6}$&$37$&$37$&$5.2$&$0.36$&$-0.06~~$&$0.12$
\enddata
\label{tab:runs}
\end{deluxetable*}
%ApJ
%\end{deluxetable}

\begin{figure}[t!]
\begin{center}
\includegraphics[width=0.9\columnwidth]{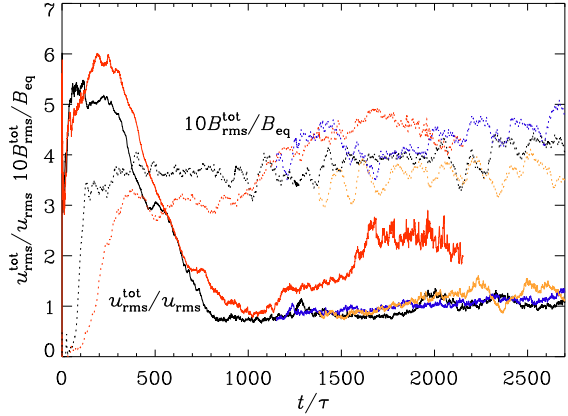}
\end{center}\caption[]{
%ApJ
%\small
%
Time evolution of the total rms velocity and magnetic field.
The rms velocity of the whole domain $\Urms$ is normalized by
$\urms$ (solid lines) and is plotted together with the rms magnetic field of
the whole domain $\Brms$ normalized by the equipartition field in the
convection zone, $\Beq$,
(dotted lines) and multiplied by 10 for visualization purposes, for Runs~A
(black line), Run~Ab (yellow), Run~Ac (blue), and Run~B (red).
}
\label{purms}
\end{figure}

\begin{figure}[t!]
\begin{center}
\includegraphics[width=0.9\columnwidth]{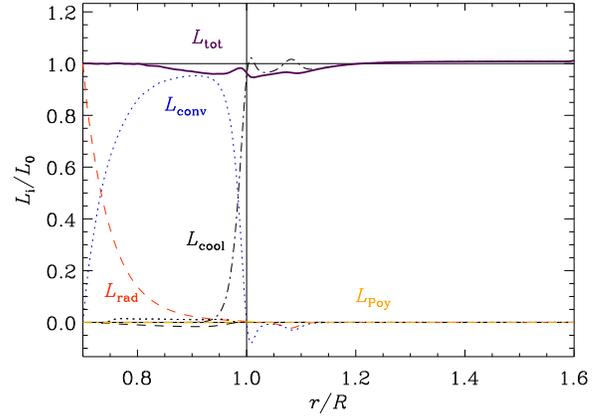}
\includegraphics[width=0.9\columnwidth]{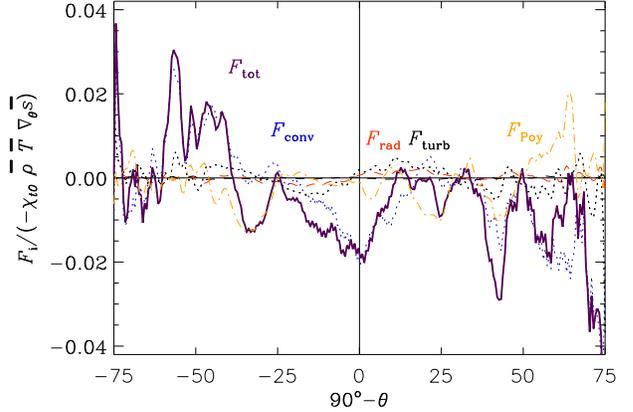}
%\includegraphics[width=0.6\columnwidth]{pflux_A.eps}
%\includegraphics[width=0.6\columnwidth]{pflux_theta_A.eps}
%ApJ
\end{center}\caption[]{
%\small
%ApJ
Top panel: the different contributions to the total radial luminosity (thick solid line)
are due to radiative diffusion (dashed red line), resolved convection
(blue dotted line), unresolved turbulent convection (black dotted line), viscosity
(yellow dashed line), cooling flux (dash-dotted line), and the Poynting flux 
(orange dash-dotted line) for Run~A.
The thin solid black lines denote the zero level and the total
luminosity through the lower boundary, respectively.
Bottom panel: latitudinal heat fluxes.
The various contributions to the latitudinal energy flux, $F_i$,
are normalized by the rms value of
$-\chi_{t0} \mean{\rho}\mean{T} {\bm \nabla}_\theta\mean{s}$.
The thin solid black lines indicate the zero line as well as the
equator at $\theta=90^\circ$.
}
\label{pflux}
\end{figure}

\begin{figure*}[t!]
\begin{center}
\includegraphics[width=1.04\columnwidth]{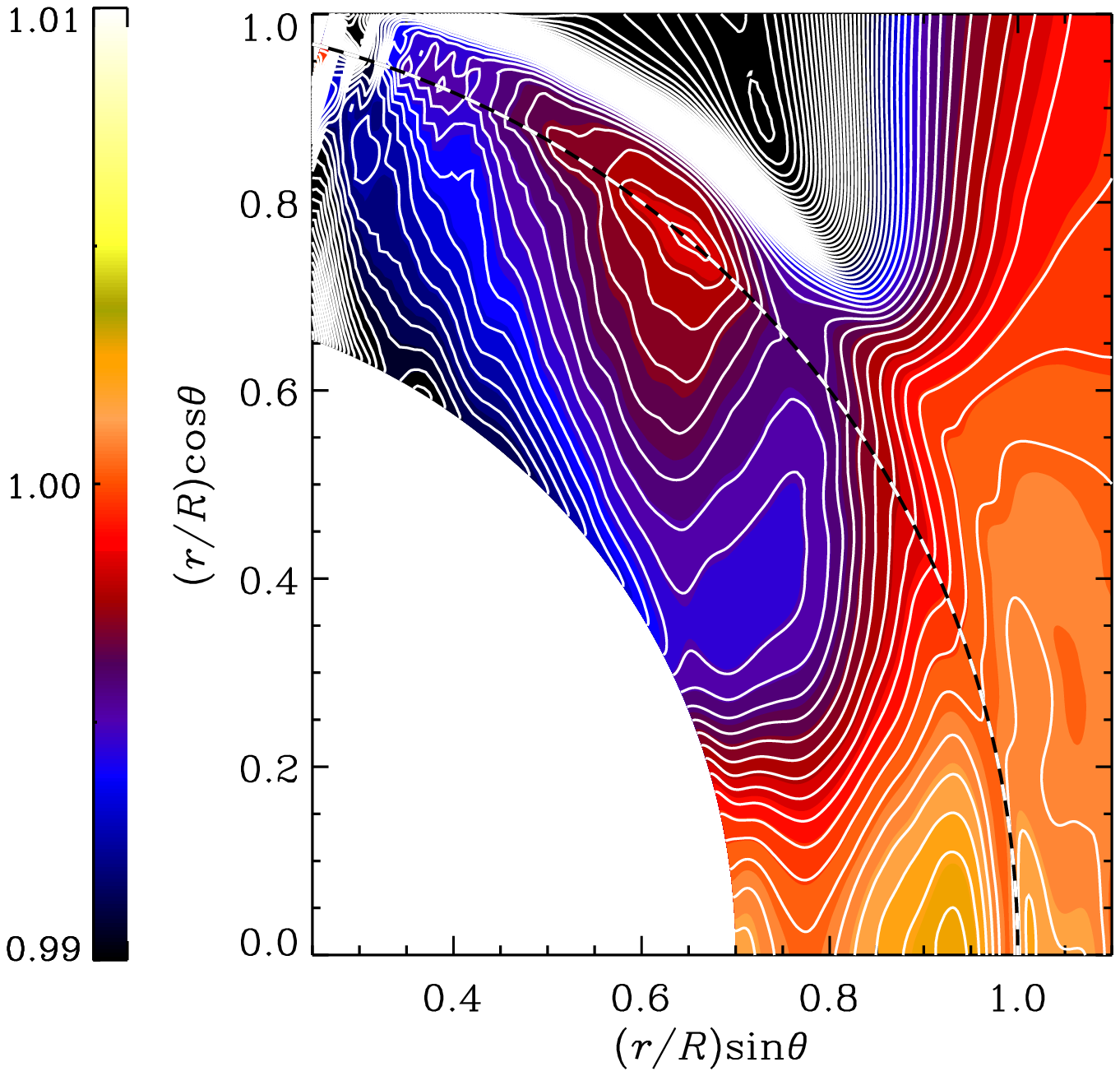}
\includegraphics[width=1.04\columnwidth]{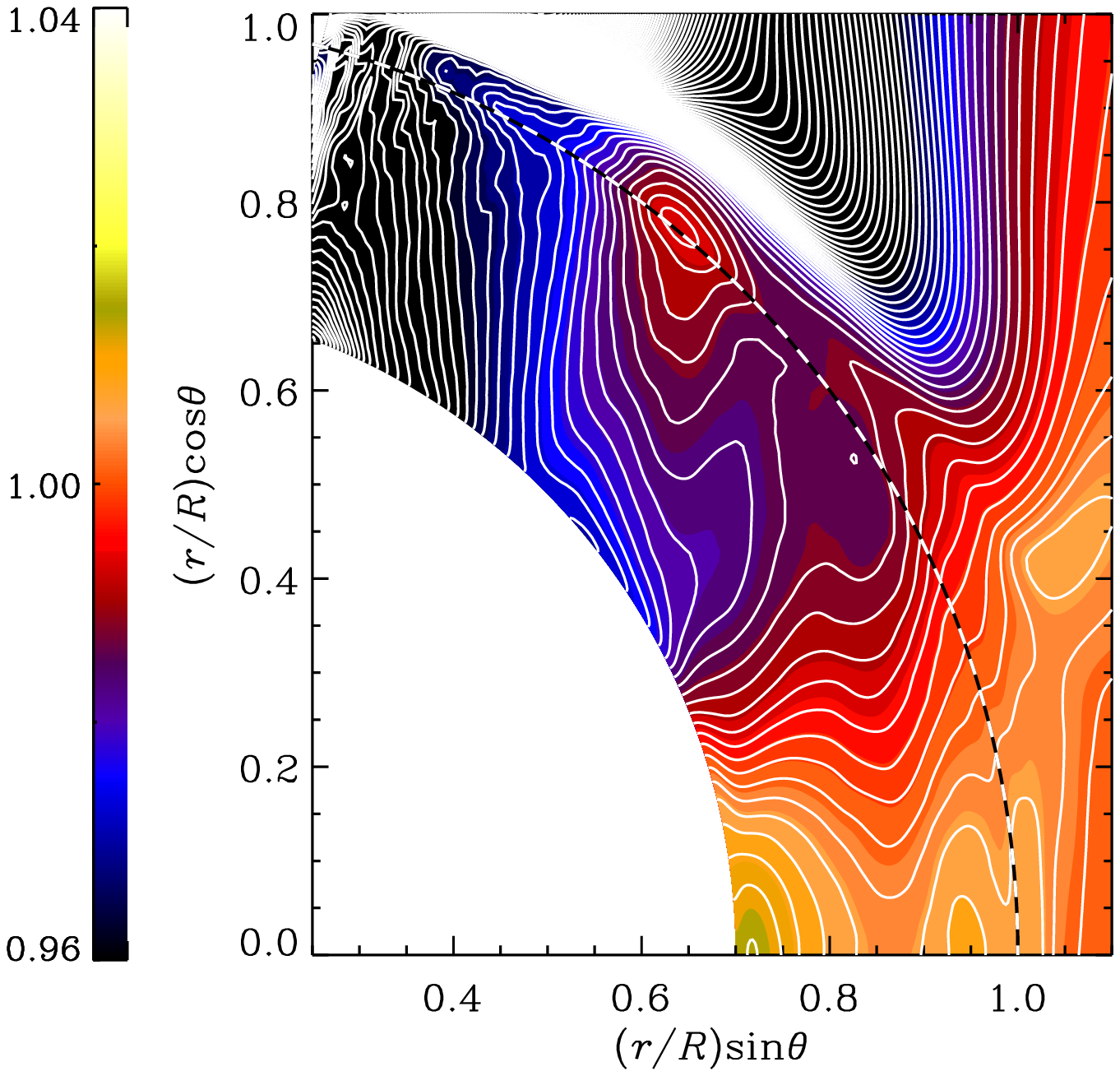}
\includegraphics[width=0.5\columnwidth]{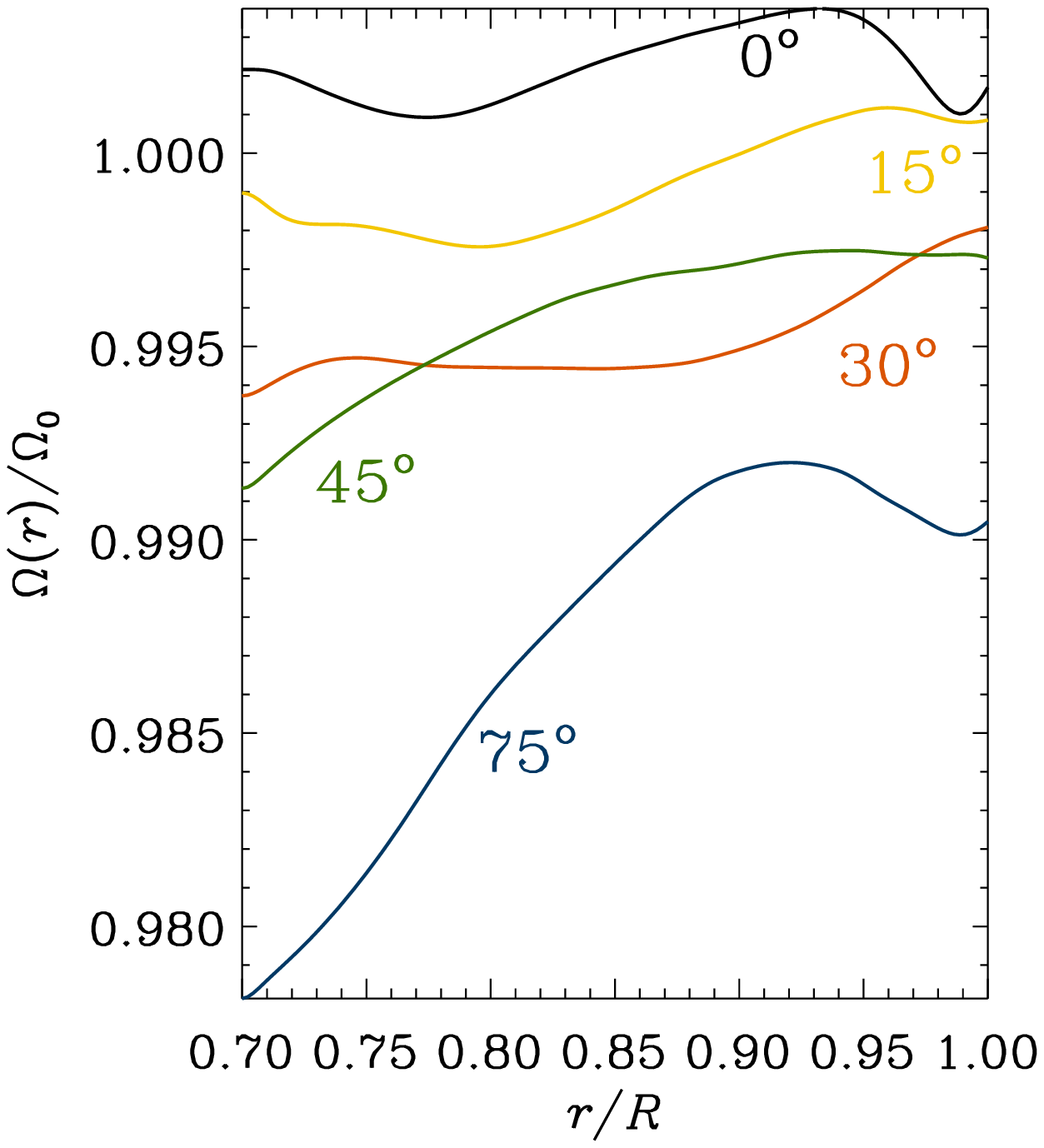}
\includegraphics[width=0.5\columnwidth]{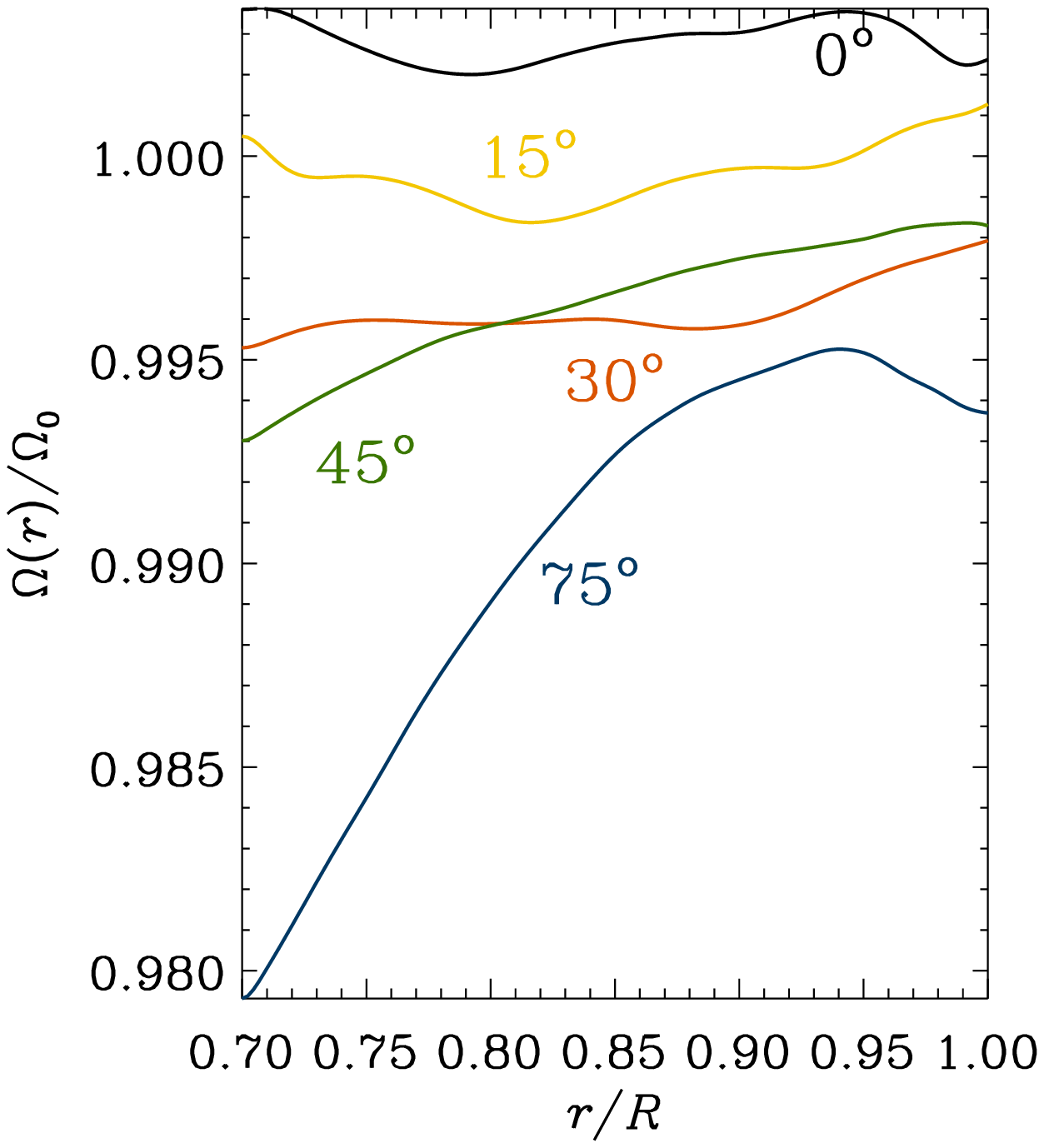}
\includegraphics[width=0.5\columnwidth]{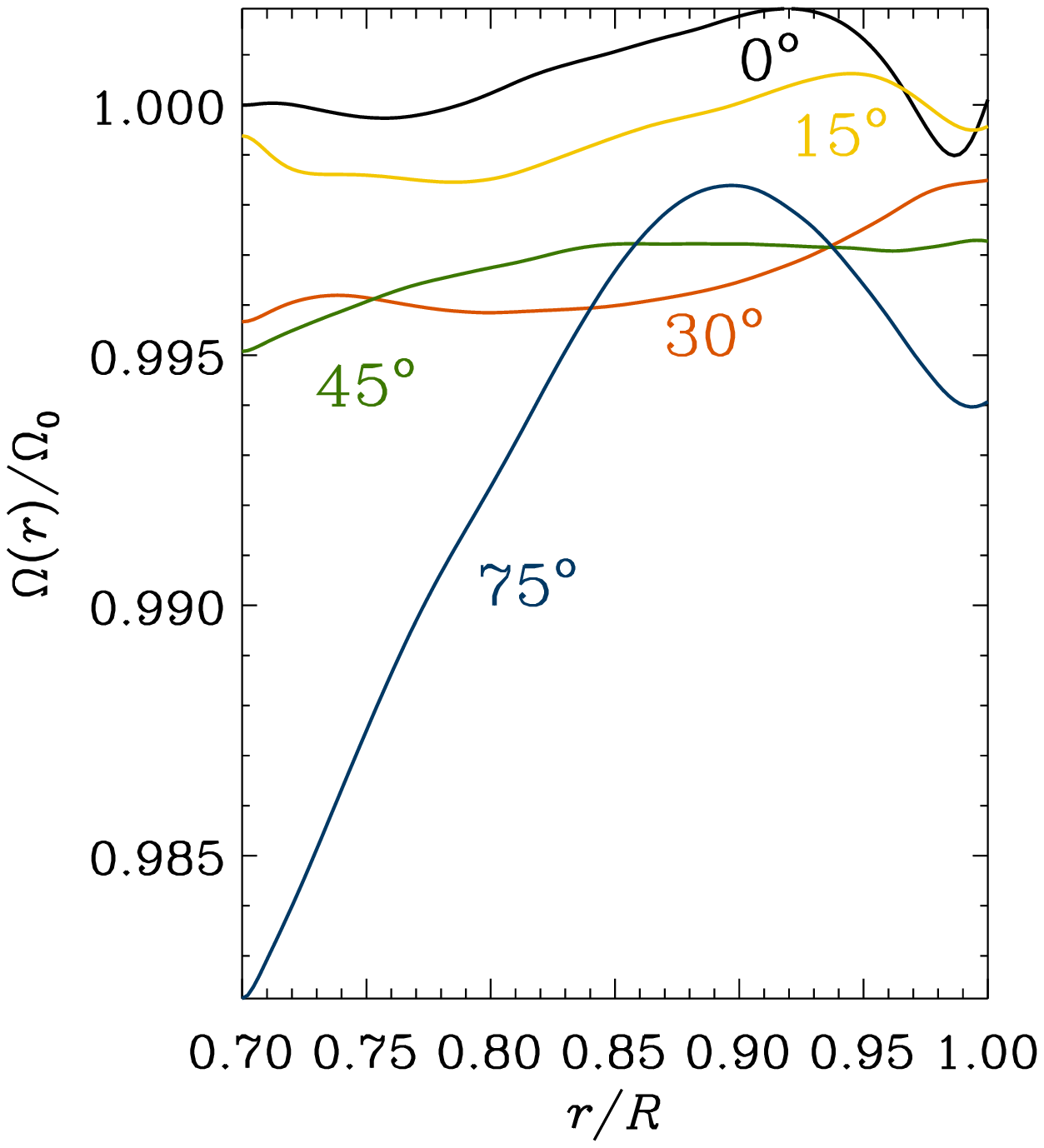}
\includegraphics[width=0.5\columnwidth]{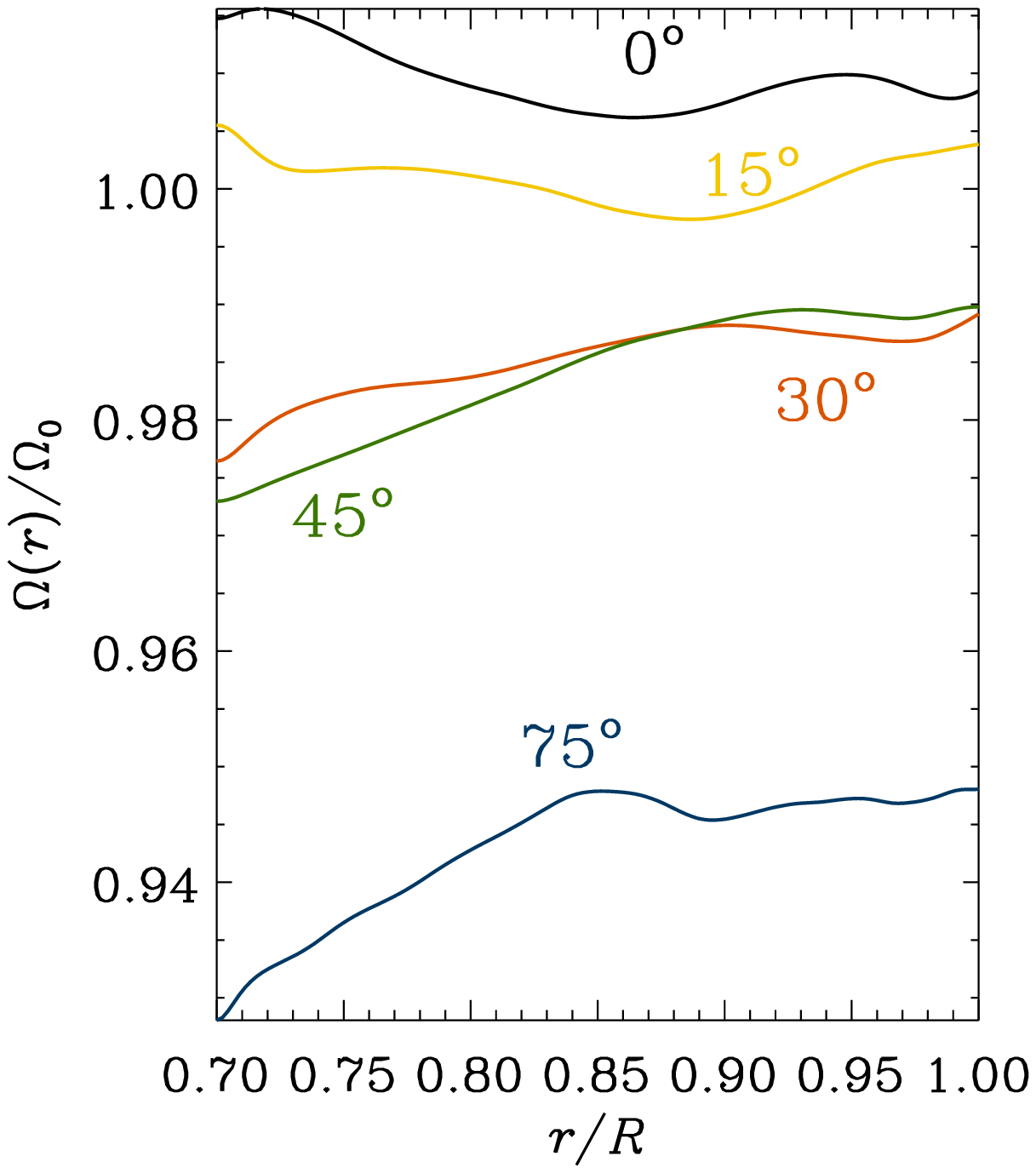}
%\includegraphics[width=0.49\columnwidth]{diffrot_conv_A.eps}
%\includegraphics[width=0.49\columnwidth]{diffrot_conv_B.eps}
%\includegraphics[width=0.24\columnwidth]{diffrotII_A.eps}
%\includegraphics[width=0.24\columnwidth]{diffrotII_Ab.eps}
%\includegraphics[width=0.24\columnwidth]{diffrotII_Ac.eps}
%\includegraphics[width=0.24\columnwidth]{diffrotII_B.eps}
%ApJ
\end{center}\caption[]{
%ApJ
%\small
%
Zoomed-in differential rotation profiles in the
northern hemisphere in the convection zone.
Top row: mean rotation profiles
$\mean{\Omega}(r,\theta)/\Omega_0$
for Runs~A and B.
The black dashed lines indicate the surface ($r=R$).
Bottom row: mean rotation profiles at four different latitudes
for the four Runs~A, Ab, Ac, and B:
$90^\circ-\theta=0^\circ$ (solid black), $90^\circ-\theta=15^\circ$
(yellow), $90^\circ-\theta=30^\circ$ (red),
$90^\circ-\theta_3=45^\circ$ (green), and $90^\circ-\theta_3=75^\circ$
(blue).
}
\label{diff}
\end{figure*}

In this work we focus on four simulations that are summarized
in Table~\ref{tab:runs}.
The main differences between these runs are their rotation rates
and the magnetic Reynolds numbers
Runs~Ab and Ac are a continuation of Run~A after $t/\tau=1350$ and
$t/\tau=1150$, respectively, but with smaller and higher diffusivities
$\eta$ in the convection zone.
Runs~A, Ab, and Ac have a higher Coriolis number $\Co$ and lower
values of $\Rey$ than Run~B.
The Coriolis number of Run~A is more than twice that of Run~B.
However, the nominal rotation rate determined by $\Omega_0$ is only 1.8
times larger.
We show the time evolution of the total rms velocity and magnetic
field, averaged over the whole domain,
$\Urms=\brac{u_r^2+u_{\theta}^2+u_{\phi}^2}_{r\theta\phi}^{1/2}$ and
$\Brms=\brac{B_r^2+B_{\theta}^2+B_{\phi}^2}_{r\theta\phi}^{1/2}$,
in all the four runs in Figure~\ref{purms}.  
Here, the subscripts on angle brackets denote averaging over
$r$, $\theta$, $\phi$.
Convection is sufficiently super-critical to develop during the first
few tens of turnover times. After 50--200 turnover times, the dynamo
starts to operate and a magnetic field grows at a rate that is higher
for faster rotation (compare Runs~A and B).
Due to the high rotation rate and the lower density in the corona, the
velocities there grow to higher values than in the convection zone.
As described in Section~\ref{model}, we use a higher viscosity to
suppress these velocities and associated numerical difficulties.
After the magnetic field in the convection zone has reached sufficient
strength and expanded throughout the whole domain, it quenches the
high velocities in the corona significantly,
as is evident from Figure~\ref{purms}. 
When the magnetic field reaches $\Brms/\Beq\approx0.3$, the rms velocity
decreases from $\Urms/\urms\approx6$ to $\approx1$, i.e., the contribution
from the corona is now sub-dominant.
This is caused by the Lorentz force, which becomes much stronger and comparable
to the Coriolis force in the corona.
In the saturated state we have $\Urms\approx\urms$,
which is reached after around $t/\tau=1000$ turnover times for Run~A.
Runs~Ab and Ac are restarted from Run~A after this saturation point.

For Run~B, at first it seems that the saturated state has been reached at
$t/\tau=1000$, but it turns out that both $\Urms$ and the magnetic field
start to grow again to reach another saturation level at
$t/\tau\approx1700$.
While the differential rotation profile remains roughly unchanged
despite of the growth of the energies (see Section~\ref{sec:diff}), the
magnetic field seems to undergo a mode change from an oscillatory to a
stationary solution or an oscillatory solution with a much longer
period in Run~B (see Section~\ref{sec:mgf}).
We note that an increase of $\Brms$ in Figure~\ref{purms}, where we
show the rms values of the magnetic field computed over the whole
domain, does not necessarily imply an increase of the magnetic field
inside the convection zone. The increase of the rms magnetic field
can also be attributed to the development of magnetic structures
ejected from the convection zone into the coronal region.

In the upper panel of Figure~\ref{pflux}, we show the balance of
various radial energy fluxes,
contributing to the total luminosity for Run~A.
The radial components of radiative, convective, kinetic, viscous, and
Poynting fluxes, as well as the flux due to the turbulent heat conductivity,
are defined as
\begin{eqnarray}
{\meanF_{\rm rad}} &=& \left\langle F_r^{\rm rad}\right\rangle,\\
{\meanF_{\rm conv}} &=& c_{\rm P}\left\langle(\rho u_r)^{\prime}T^{\prime} \right\rangle,\\
{\meanF_{\rm kin}} &=& \onehalf \left\langle\rho u_r{\bm
    u}^2\right\rangle,\\
{\meanF_{\rm visc}} &=& -2\nu \left\langle \rho u_i S_{ir} \right\rangle,\\
{\meanF_{\rm SGS}} &=& \left\langle F_r^{\rm SGS}\right\rangle,\\
{\meanF_{\rm Poy}} &=& \langle E_\theta B_\phi - E_\phi B_\theta
\rangle/\mu_0,
\end{eqnarray}
where ${\bm E}=\eta \mu_0 {\bm J}-{\bm u}\times{\bm B}$
is the electric field, the primes
denote fluctuations, and angle brackets imply averaging over $\theta$,
$\phi$, and a time interval over which the turbulence
is statistically stationary.
The resolved convective flux dominates inside the convection zone and
reaches much higher values here than in our earlier model \citep{WKMB12}.
In the corona the cooling keeps the total flux constant.
Note the convective overshoot into the exterior and the negative
radiative flux just above the surface ($r=R$), caused by the higher
temperature in the corona.
The kinetic energy flux has small negative values in the convection zone.
The luminosity due to viscosity and Poynting flux are too small to be visible.
In the lower panel of Figure~\ref{pflux}, we show the corresponding
latitudinal contributions $F_i$ normalized by the rms value of the expected
turbulent contribution from the latitudinal entropy gradient,
$-\chi_{t0} \mean{\rho}\mean{T} {\bm \nabla}_r\mean{s}$, where
$\chi_{t0}=\urms/3\kef$.
They are generally just a few per cent of the rms value of the
turbulent latitudinal heat flux and oriented mostly equatorward
(positive in the north and negative in the south).  These values are
small, indicating that the system is thermally relaxed also in the
$\theta$ direction.

\subsection{Differential Rotation}
\label{sec:diff}

\begin{figure*}[t!]
\begin{center}
\includegraphics[width=0.6\columnwidth]{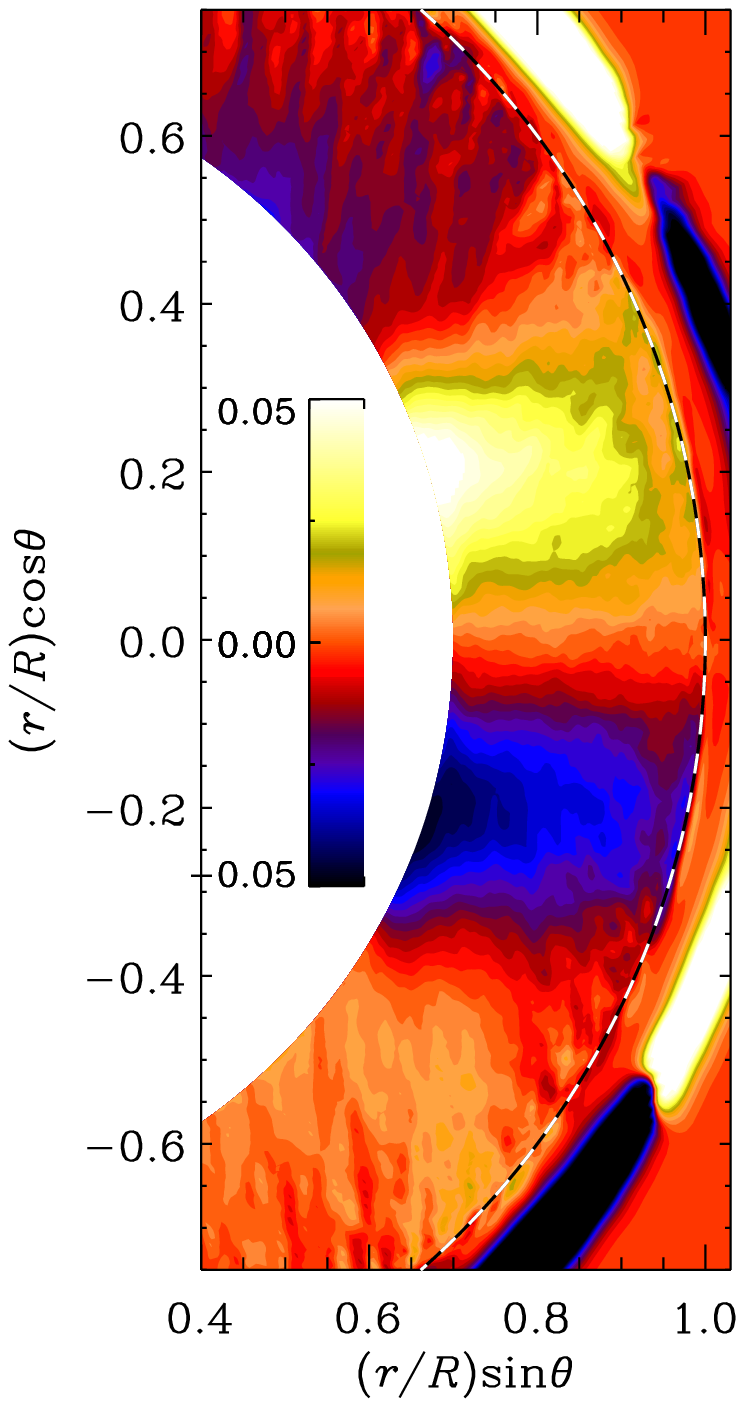}
\includegraphics[width=0.6\columnwidth]{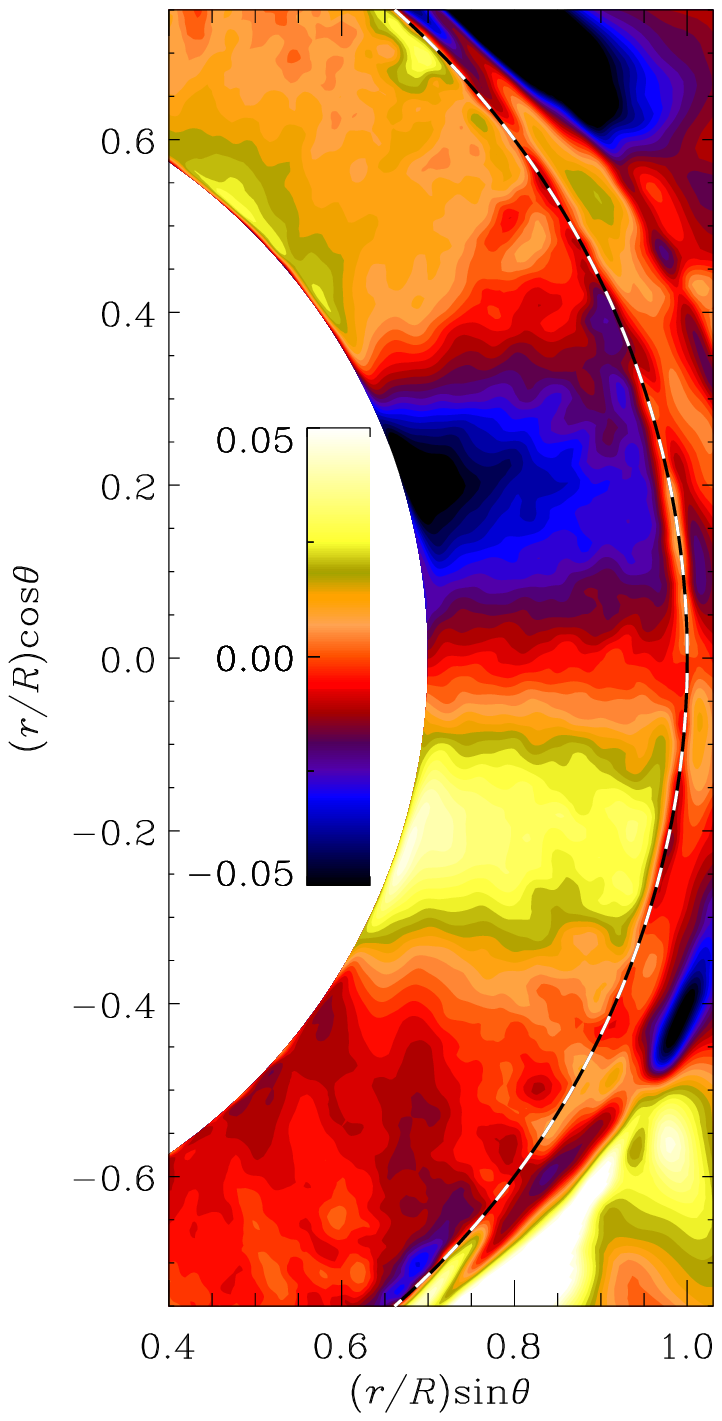}
\includegraphics[width=0.6\columnwidth]{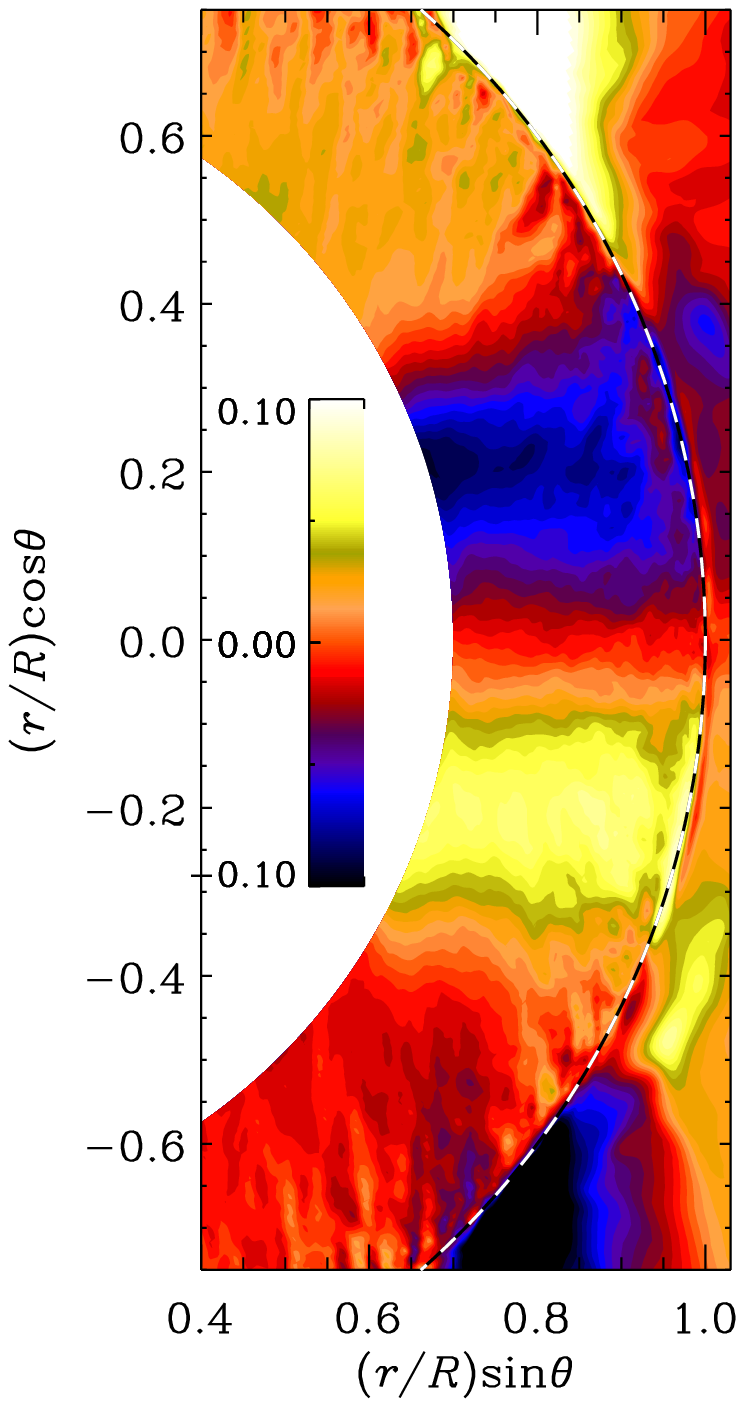}
%\includegraphics[width=0.3\columnwidth]{barocII_A.eps}
%\includegraphics[width=0.3\columnwidth]{angtr_A.eps}
%\includegraphics[width=0.3\columnwidth]{barocIII_A.eps}
%ApJ
\end{center}\caption[]{
%\small
%ApJ
Representations of the two dominant terms in the
evolution equation of mean azimuthal vorticity (see
Equation~(\ref{eq:baroc})) for Run~A:
$(\overline{\ve{\nabla}T\times\ve{\nabla} s})_\phi$ (left panel) and
$r\sin\theta\,\partial\mean{\Omega}^2/\partial z$ (middle panel), both normalized
by $\Omega_0^2$. The right-most panel shows the mean latitudinal
entropy gradient $R\nabla_\theta\overline{s}/c_{\rm P}$. The
dashed lines indicate the surface ($r=R$).
}
\label{baroc}
\end{figure*}
\begin{figure}[t!]
\begin{center}
\includegraphics[width=0.9\columnwidth]{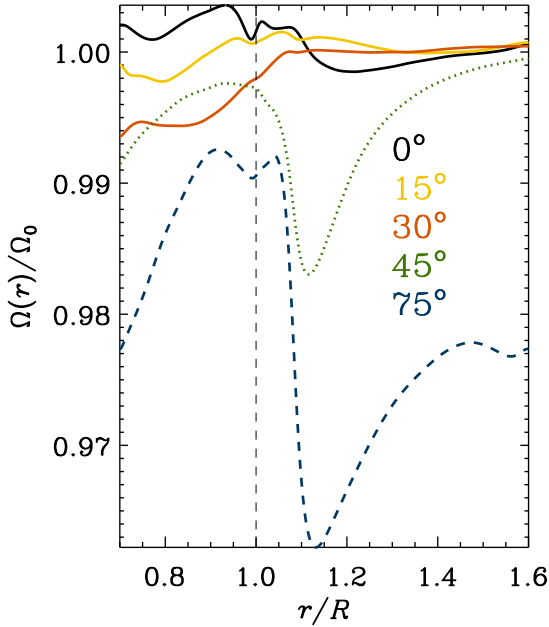}
%\includegraphics[width=0.6\columnwidth]{diffrot_cor_A.eps}
%ApJ
\end{center}\caption[]{
%\small
%ApJ
Differential rotation in the
northern hemisphere including the coronal layer.
Mean rotation profiles, $\mean{\Omega}(r,\theta)/\Omega_0$,
at five different latitudes for Run~A,
$90^\circ-\theta=0^\circ$ (solid black), $90^\circ-\theta=15^\circ$
(yellow), $90^\circ-\theta=30^\circ$ (red),
$90^\circ-\theta_3=45^\circ$ (dotted green), and $90^\circ-\theta_3=75^\circ$
(dashed blue).
The black dashed line indicates the surface ($r=R$).
}
\label{diffcor}
\end{figure}

In Figure~\ref{diff}, we show the mean rotation profiles
$\mean{\Omega}(r,\theta)=\Omega_0+\mean{u}_\phi/r\sin\theta$ for
Runs~A and B in the meridional plane and for Runs~A, Ab, Ac, and B at
four different latitudes.
The contours of constant rotation are clearly not cylindrical for any
of the four runs.
They show a ``spoke-like'' structure, i.e., the contours are
more radial than cylindrical, which is similar to the solar rotation
profile obtained by helioseismology \citep{Schouea98}
The equator is rotating faster than the poles, which has been seen in
many earlier simulations \citep{G83,BMT04,MBT06,KKBMT10,KMGBC11}
and resembles the observed rotation of the Sun
for our slower rotation case (Run~B).

The source of differential rotation is the
anisotropy of convection and is described by the
$r\phi$ and $\theta\phi$ components of the Reynolds stress.
Using a suitable parameterization of the Reynolds stress
in terms of the $\Lambda$ effect, one obtains differential rotation
where the equator rotates faster than the poles.
However, nonlinear mean-field hydrodynamic simulations have shown that
for rotation rates comparable with those of the Sun, the contours of
constant angular velocity become cylindrical \citep{BMT92,KR95}.
To produce spoke-like rotation contours, the Taylor--Proudman balance has
to be overcome by an important contribution in the evolution equation
for the mean azimuthal vorticity $\mean{\omega_\phi}$, which is given by:
\begin{eqnarray}
{\partial\mean{\omega_\phi}\over\partial
  t}&=&r\sin{\theta}{\partial\mean{\Omega}^2\over\partial z}
  +\left(\overline{\ve{\nabla} T\times \bm\nabla s}\right)_\phi + ...
 \label{eq:baroc}
\end{eqnarray}
where $\partial/\partial z=\cos\theta\,\partial/\partial r
-r^{-1}\!\sin\theta\,\partial/\partial\theta$ is the derivative
along the rotation axis.
The first term in Equation~(\ref{eq:baroc}) is related to the curl of the
Coriolis force and vanishes for cylindrical $\mean{\Omega}$ contours.
The second term is the mean baroclinic term,
which is caused mainly by latitudinal entropy variations. 
We ignore here additional contributions such as meridional
Reynolds and Maxwell stresses, which turn out to be small.
In Figure~\ref{baroc}, we plot the first and second terms
of Equation~(\ref{eq:baroc}) for Run~A.
These two contributions balance each other nearly perfectly.
This leads us to conclude that these two terms provides the dominant
contribution to the production of mean azimuthal vorticity and
that the Taylor--Proudman balance is broken by the baroclinic term.
It is remarkable that there is such a large and spatially coherent
latitudinal entropy gradient,
which is crucial to having a significant azimuthal baroclinic term
and is self-consistently produced in the simulations.

Similar results have been obtained in mean-field simulations
by including an anisotropic convective heat conductivity \citep{BMT92,KR95}
or by including a subadiabatic part of the tachocline \citep{Rempel05},
and in convection simulations by prescribing a latitudinal entropy
gradient at the lower radial boundary of the convection zone
\citep{MBT06}.
More recently, \cite{BMT11} showed that spoke-like contours can also
be obtained by including a lower stably stratified overshoot layer
in a purely hydrodynamical simulation.

The right-most panel of Figure~\ref{baroc} shows the mean latitudinal entropy
gradient $\nabla_{\theta}\overline{s}$ for Run~A.
The spatial distribution of the gradient agrees with the baroclinic term
as well, because $\nabla_r\overline{T}\approx\rm const$ in the
convection zone, so we can
conclude that the dominant contribution in the baroclinic term is due
to the product of the latitudinal entropy gradient and the radial
temperature gradient, which is more important than the radial entropy
gradient multiplying the latitudinal temperature gradient.
Compared with Run~A, the other three runs, not shown here,
have similar (Run~B) or even identical
(Runs~Ab and Ac) distributions of the two terms on the right hand side of
Equation~(\ref{eq:baroc}), as well as the latitudinal entropy gradient.
The location of the spoke-like differential rotation profile coincides
with a similarly shaped mean latitudinal entropy gradient.
The entropy gradient in the northern (southern) hemisphere is negative
(positive) below $\pm30^\circ$ latitude.
In Run~B, this region reaches to higher latitudes than in Runs~A, Ab and,
Ac, which leads to radial contours of angular velocity at higher
latitudes.

We also note that, owing to the coronal envelope, differential rotation is
able to develop a near-surface-shear layer.
This is manifested by the concentration of contours of
$\mean{\Omega}$ near the surface at lower latitudes for Runs~A and B, and is
also visible in the other runs as a negative gradient of $\mean{\Omega}$
in the same locations; see Figure~\ref{diff}.
In Run~B, there also exists a
concentration around $r=1.1\, R$.
In all the simulations, the shear layer is radially more extended
than in the Sun and penetrates deeper into the convection zone.
Further studies using higher stratification should prove if this is
just an artifact of weak stratification.
However, the spoke-like rotation profile with strong shear near the
surface
occurs mostly at lower latitudes ($90^\circ-\theta\le15^\circ$), i.e.,
close to the equator.
At latitudes above $\pm30^\circ$, the contours of constant rotation are more
complex, but show some indication of strong shear close to the surface and
only in Run~B do the contours become cylindrical beyond $\pm60^\circ$ latitude.
At higher latitudes ($90^\circ-\theta\le75^\circ$), the near-surface
shear layer is again visible.

We were not able to see spoke-like rotation profiles in our previous
work \citep{WKMB12}.
Therefore, the applied changes might play an important
role in the formation of spoke-like profiles.
There are three main differences between the two setups.
First, the fractional convective flux in \cite{WKMB12} is much lower than
in the present setup; compare Figure~2 of \cite{WKMB12} with Figure~\ref{pflux}.
A stronger convective flux can give rise to more vigorous heat fluxes
and thus a more efficient thermal
redistribution, causing a more pronounced latitudinal entropy gradient.
Second, the temperature in the current setup increases sharply above
the surface, generating a hot corona instead of being constant, as in
\cite{WKMB12}.
It is possible that the resulting steep temperature gradient is
important in providing enough thermal insulation between the
convection zone and the corona.
Third, the hot corona leads to a higher density stratification in both the
convection zone and the corona.

As seen in Figure~\ref{diffcor}, the corona rotates nearly uniformly with
$\mean{\Omega}(r,\theta)/\Omega_0=1$ at lower
latitudes ($90^\circ-\theta\le30^\circ$).
We suggest that the magnetic field, which connects the surface with the
corona, is responsible for this.
Near the poles ($90^\circ-\theta\ge45^\circ$), the rotation rate drops
sharply above the surface ($r\approx1.1\,R$).
This drop coincides with the steep temperature and density gradients
above the surface, but also with the increase in the applied
viscosity profile.
Further outside and away from this drop, the rotation profile is
cylindrical at high latitudes.
This behavior is similar in all four runs.

\subsection{Connection with Anisotropic Turbulent Diffusivity Tensor}
\label{sec:turb}
\begin{figure}[t!]
\begin{center}
\includegraphics[width=0.49\columnwidth]{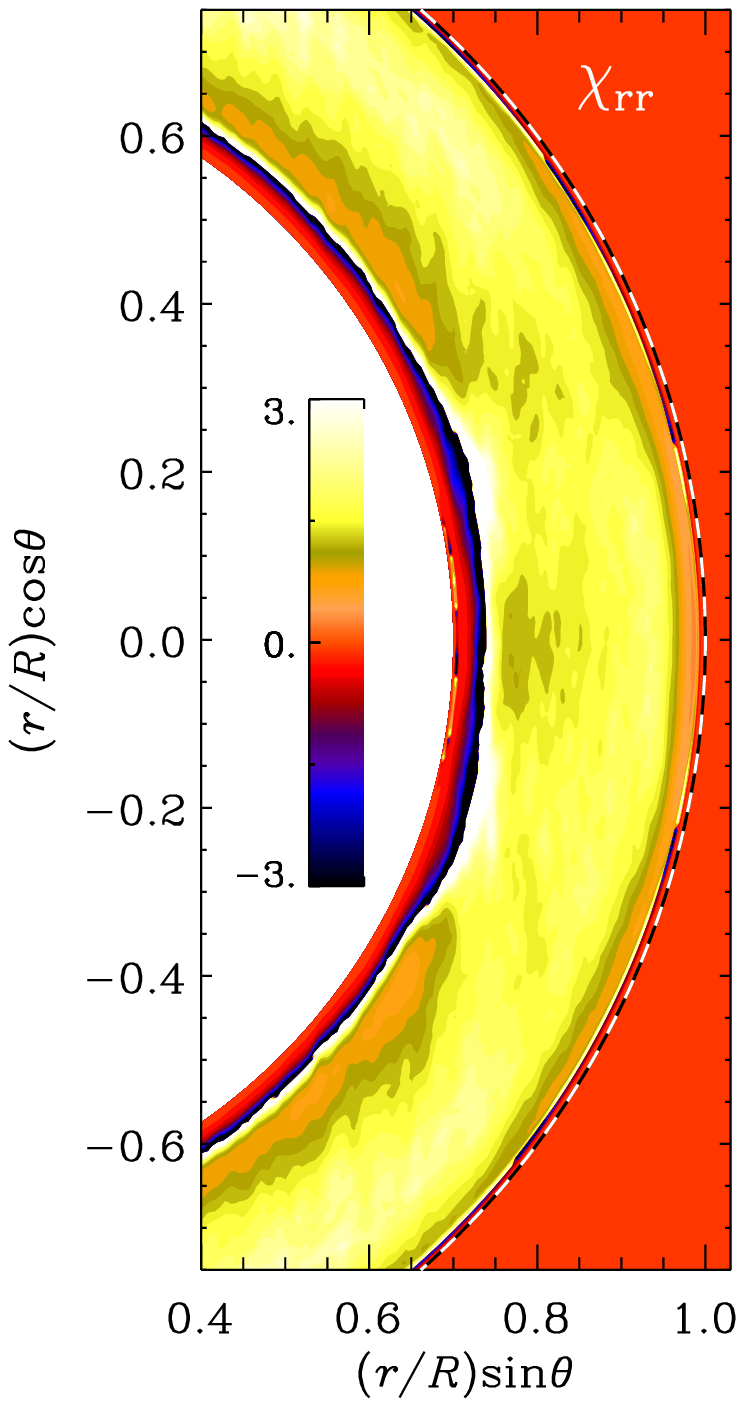}
\includegraphics[width=0.49\columnwidth]{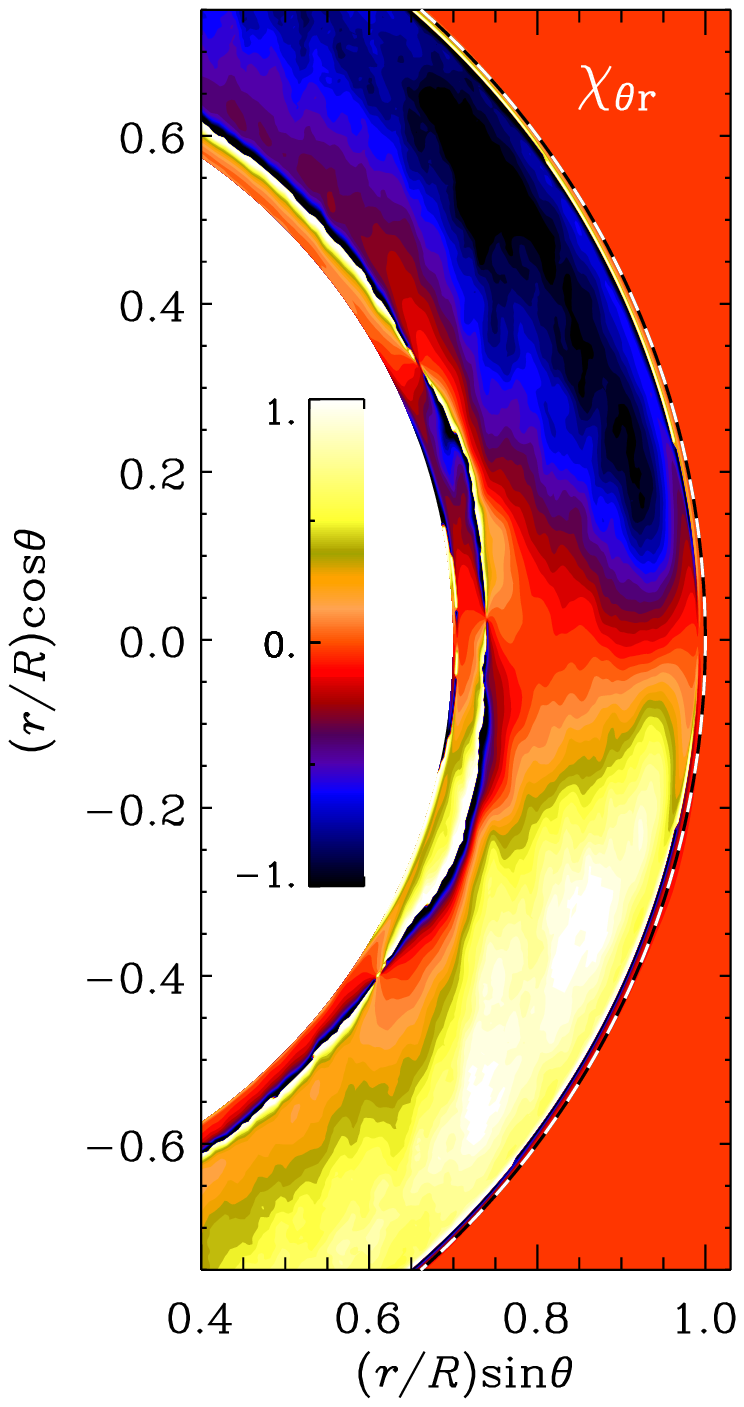}
%ApJ
%\includegraphics[width=0.3\columnwidth]{chitr_A.eps}
%\includegraphics[width=0.3\columnwidth]{chirr_A.eps}
%ApJ
\end{center}\caption[]{
%\small
%ApJ
Radial component $\chi_{r r}$ (left panel) and off-diagonal component
$\chi_{\theta r}$ (right panel) of the turbulent heat conductivity tensor
normalized by $\chi_{t0}=\urms/3\kef$ and calculated from
Equations~(\ref{eq:chitr}) and~(\ref{eq:chirr}) for Run~A.
Note the high values at the bottom of the convection zone, which
are due to the vanishing radial entropy gradient.
}
\label{chitr}
\end{figure}
\begin{figure*}[t!]
\begin{center}
\includegraphics[width=1.04\columnwidth]{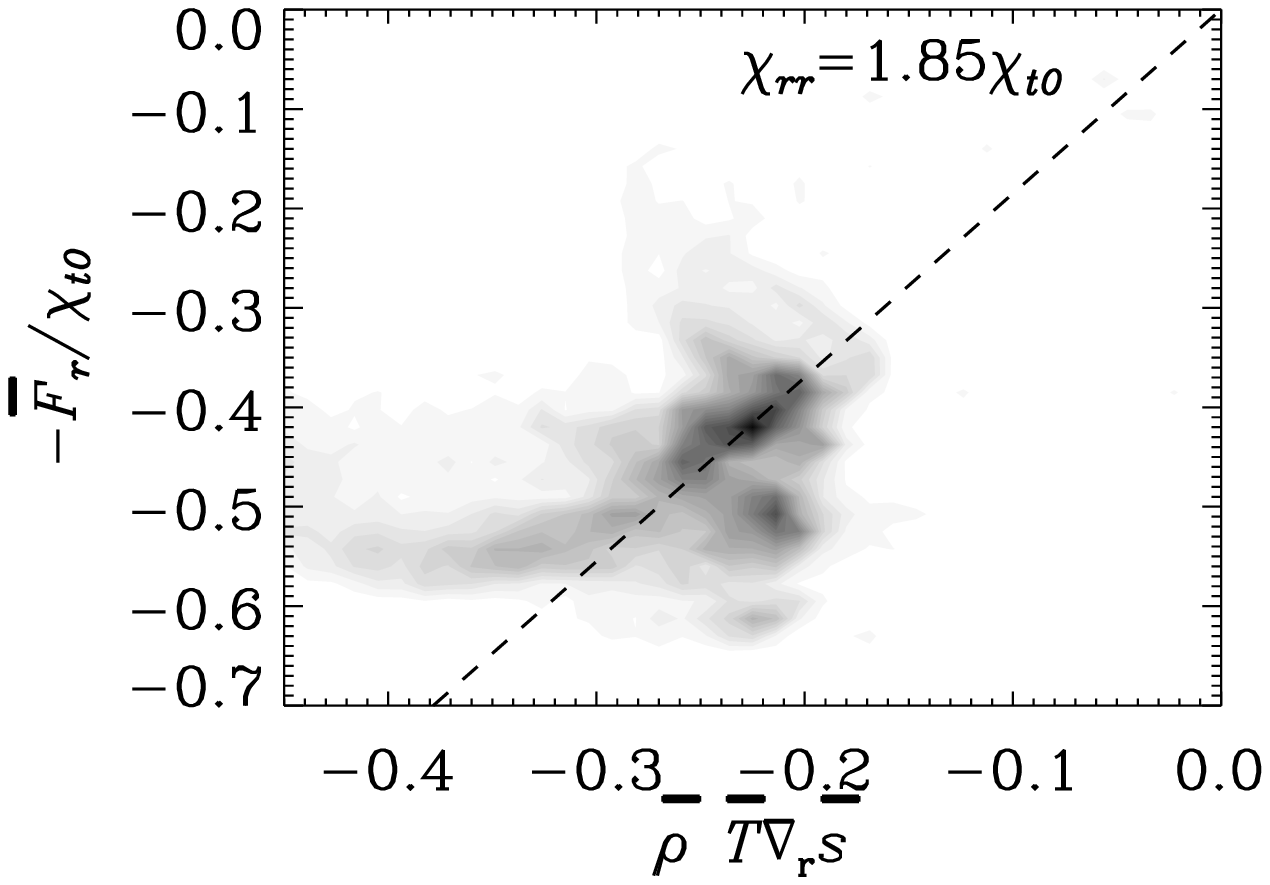}
\includegraphics[width=1.04\columnwidth]{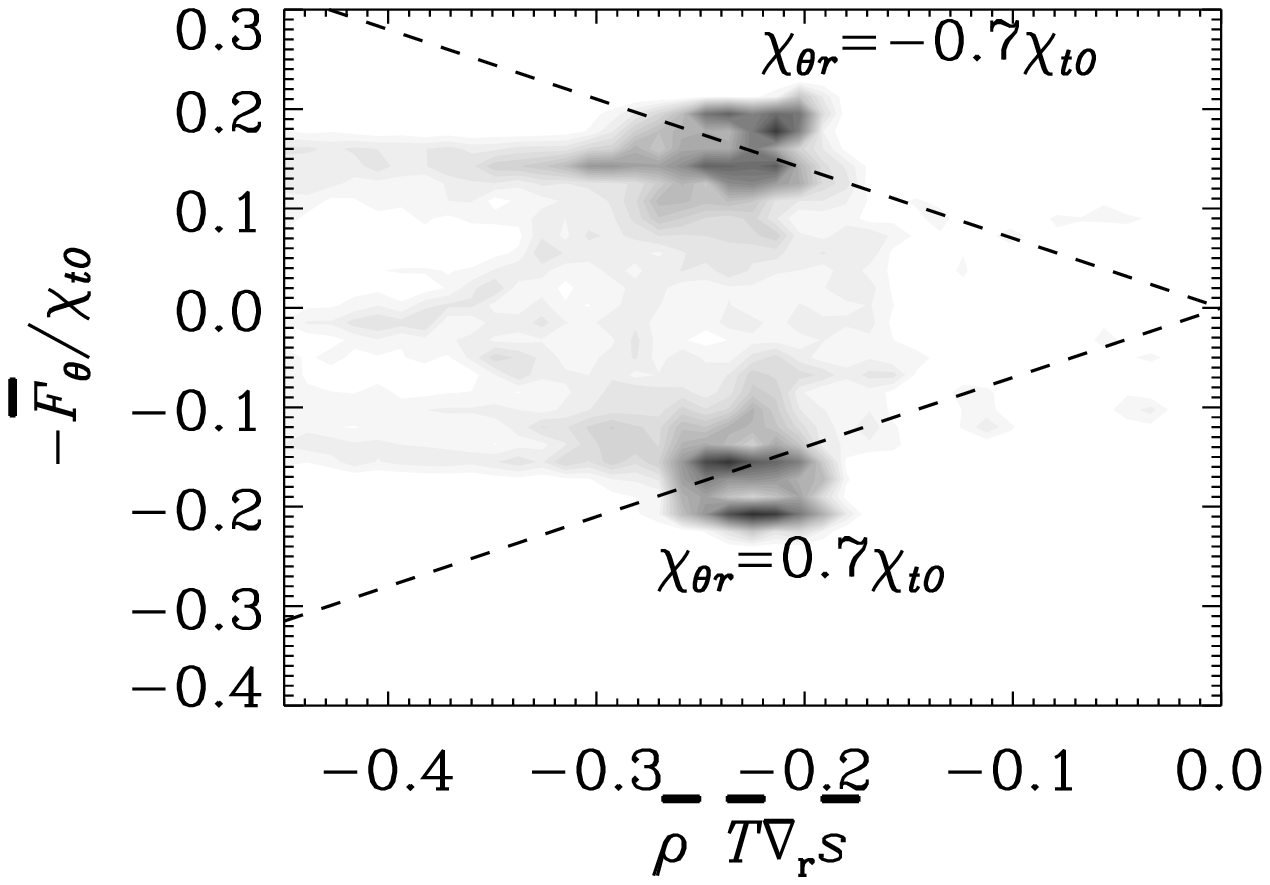}
\includegraphics[width=1.04\columnwidth]{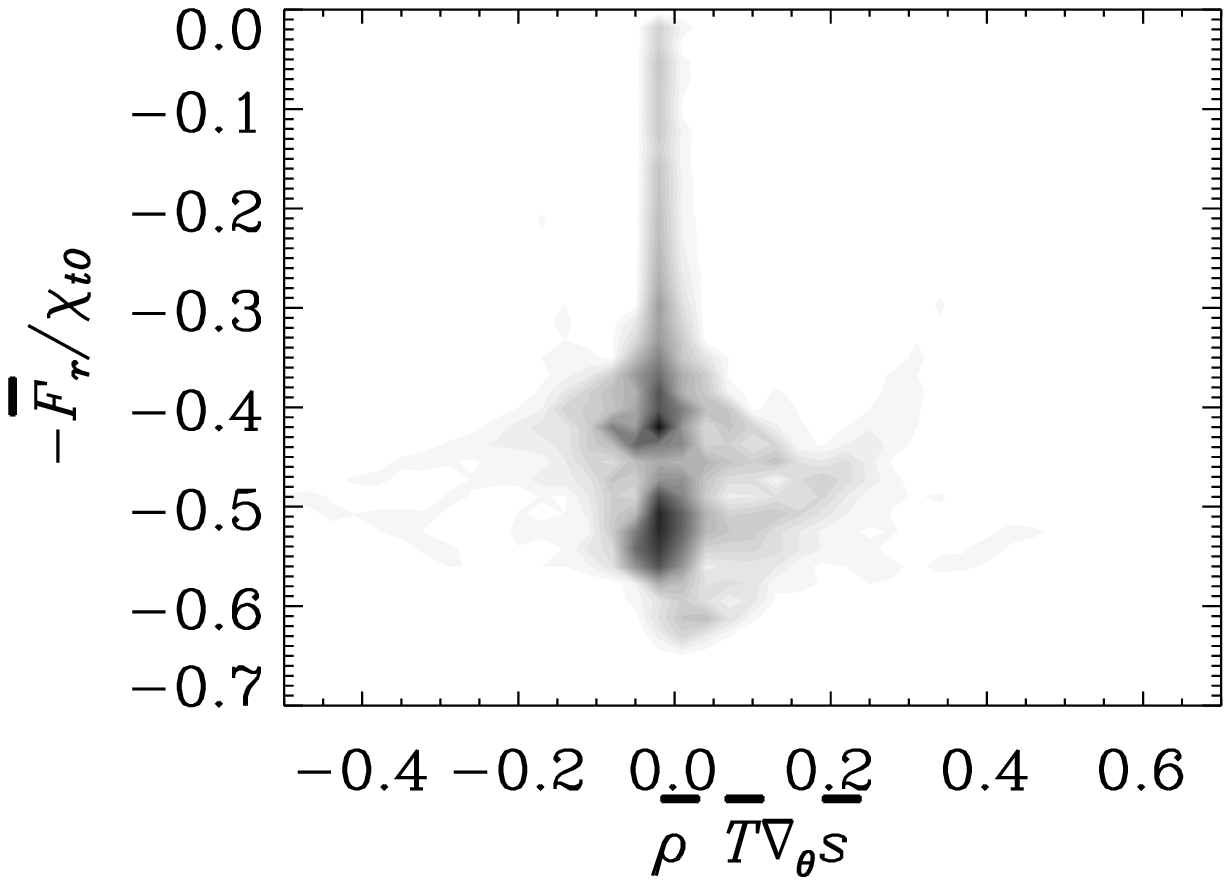}
\includegraphics[width=1.04\columnwidth]{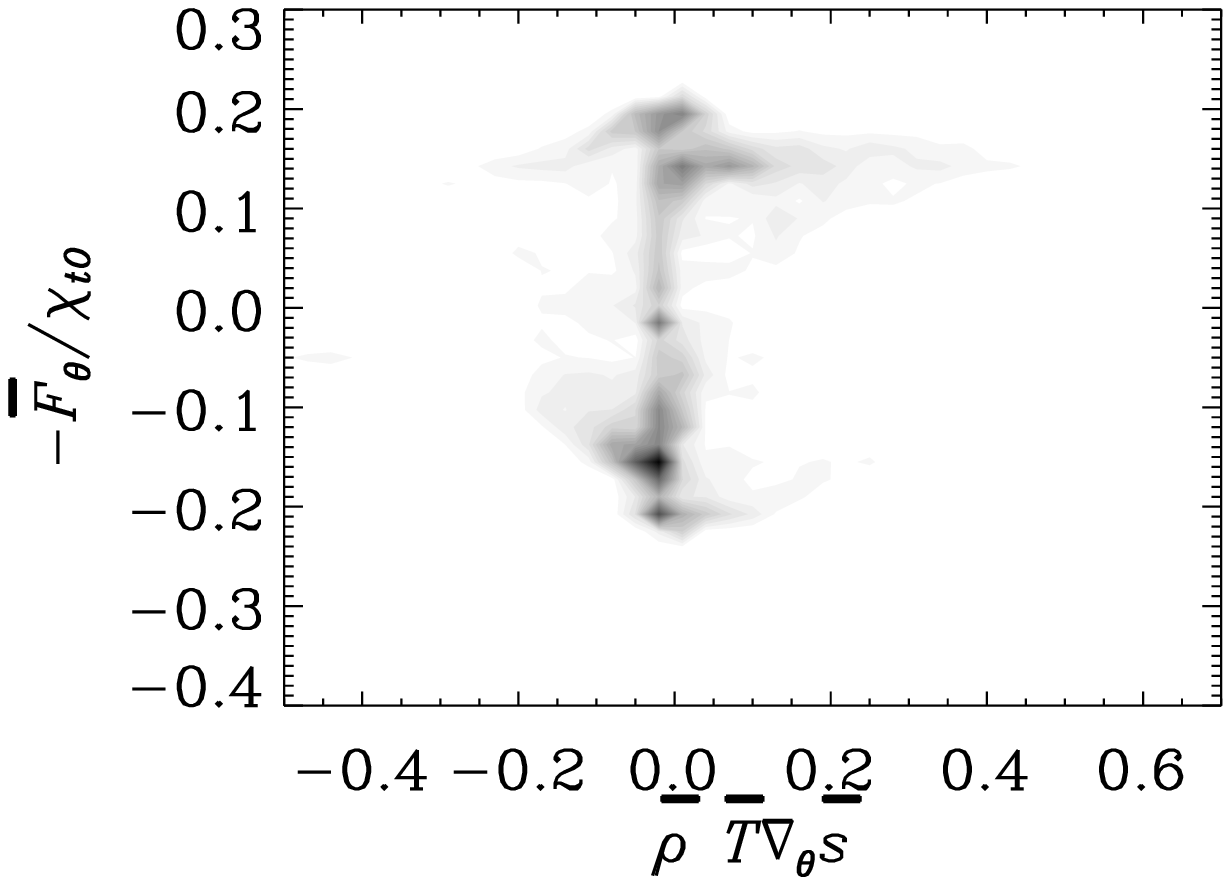}
%\includegraphics[width=0.49\columnwidth]{histo_rr_A.eps}
%\includegraphics[width=0.49\columnwidth]{histo_tr_A.eps}
%\includegraphics[width=0.49\columnwidth]{histo_rt_A.eps}
%\includegraphics[width=0.49\columnwidth]{histo_tt_A.eps}
%ApJ
\end{center}\caption[]{
%\small
%ApJ
Two-dimensional histograms of the radial and latitudinal heat flux vs.
the radial and latitudinal entropy gradient for Run~A.
The four panels show, from top left to bottom right,
$\meanF_r$ vs. $\mean{\rho}\mean{T} {\bm \nabla}_r\mean{s}$, 
$\meanF_\theta$ vs. $\mean{\rho}\mean{T} {\bm \nabla}_r\mean{s}$, 
$\meanF_r$ vs. $\mean{\rho}\mean{T} {\bm \nabla}_\theta\mean{s}$, and
$\meanF_\theta$ vs. $\mean{\rho}\mean{T} {\bm \nabla}_\theta\mean{s}$.
In the two top panels, we overplot the corresponding values of the
turbulent heat diffusivities, determined from Figure~\ref{chitr}
(dashed lines).
}
\label{histo}
\end{figure*}

\begin{figure*}[t!]
\begin{center}
\includegraphics[width=0.6\columnwidth]{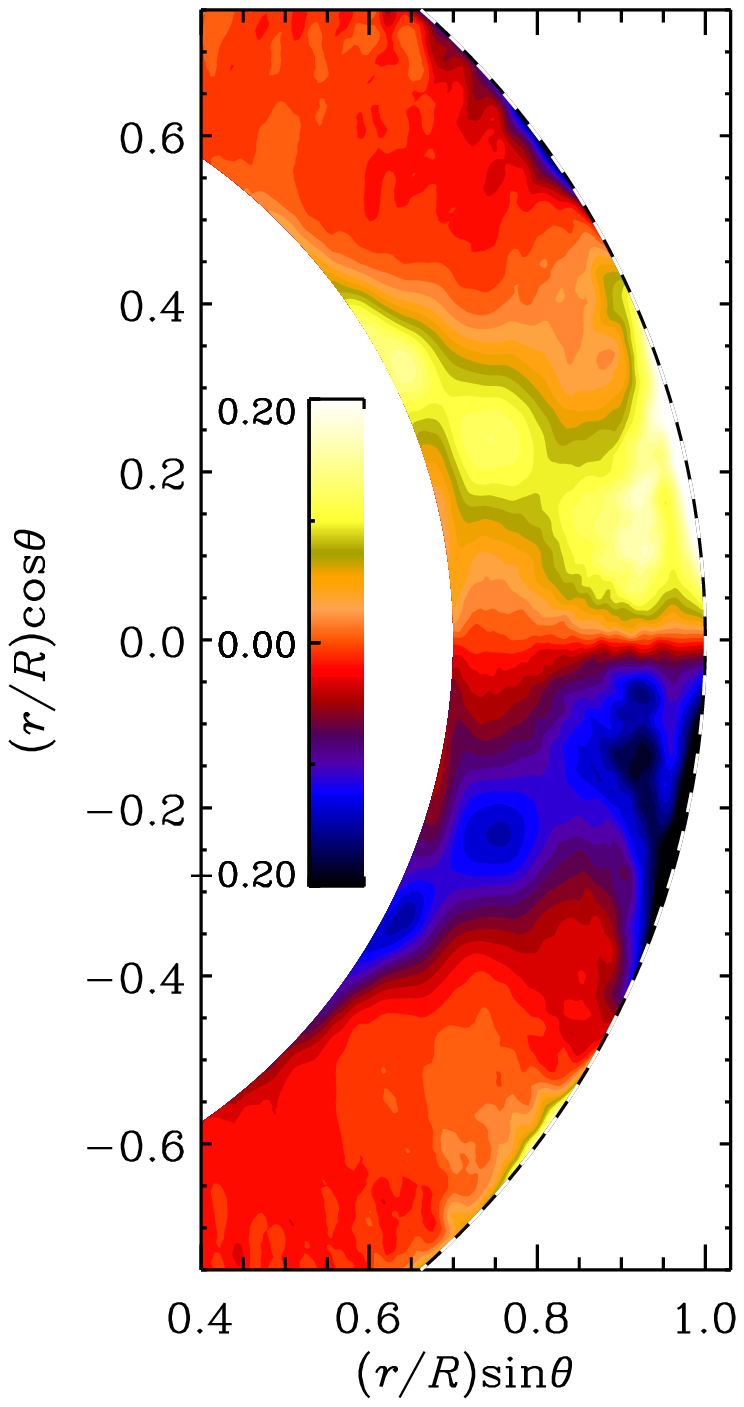}
\includegraphics[width=0.6\columnwidth]{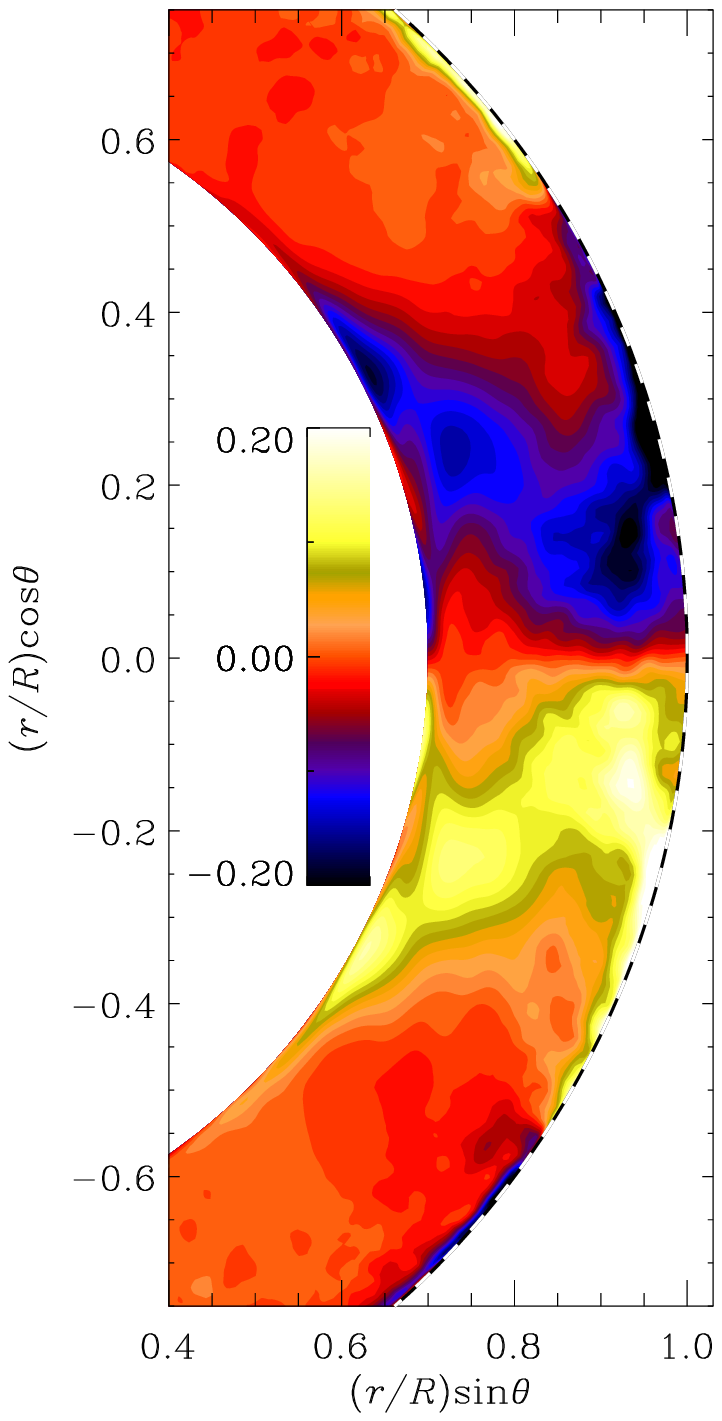}
\includegraphics[width=0.6\columnwidth]{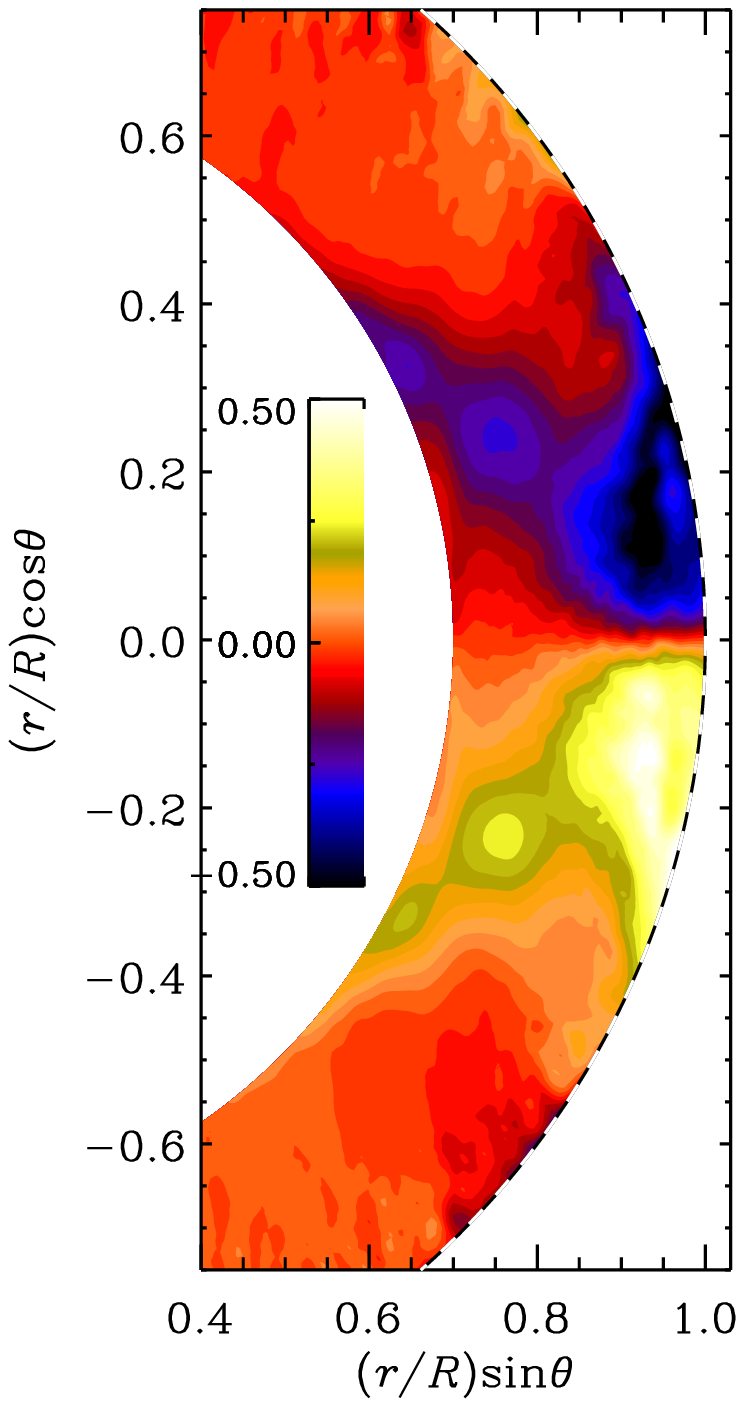}
%\includegraphics[width=0.3\columnwidth]{barocII_P.eps}
%\includegraphics[width=0.3\columnwidth]{angtr_P.eps}
%\includegraphics[width=0.3\columnwidth]{barocIII_P.eps}
%ApJ
\end{center}\caption[]{
%\small
%ApJ
Same as Figure~\ref{baroc}, but for Run~C1 of \cite{KMCWB13},
which is the same as Run~B4m of \cite{KMB12}.
}
\label{baroc_P}
\end{figure*}
\begin{figure}[t!]
\begin{center}
\includegraphics[width=0.49\columnwidth]{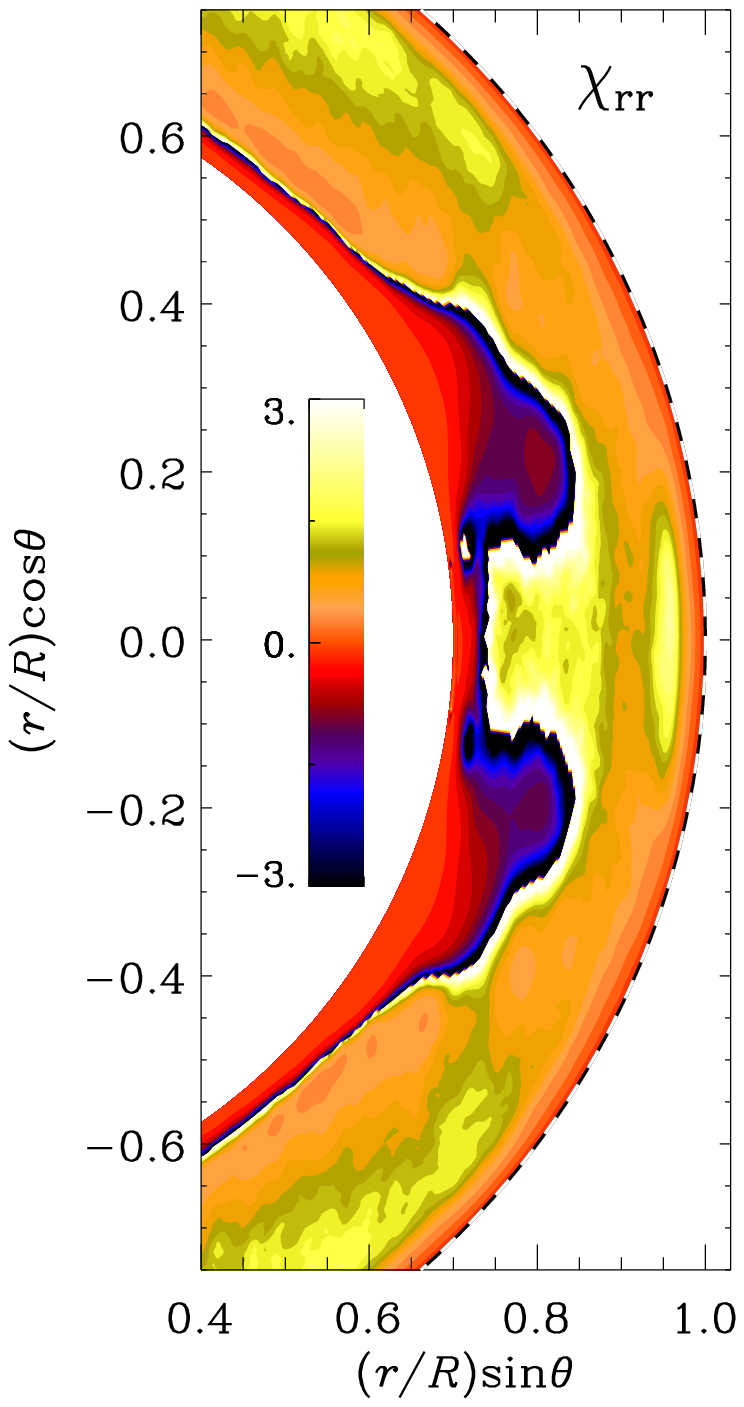}
\includegraphics[width=0.49\columnwidth]{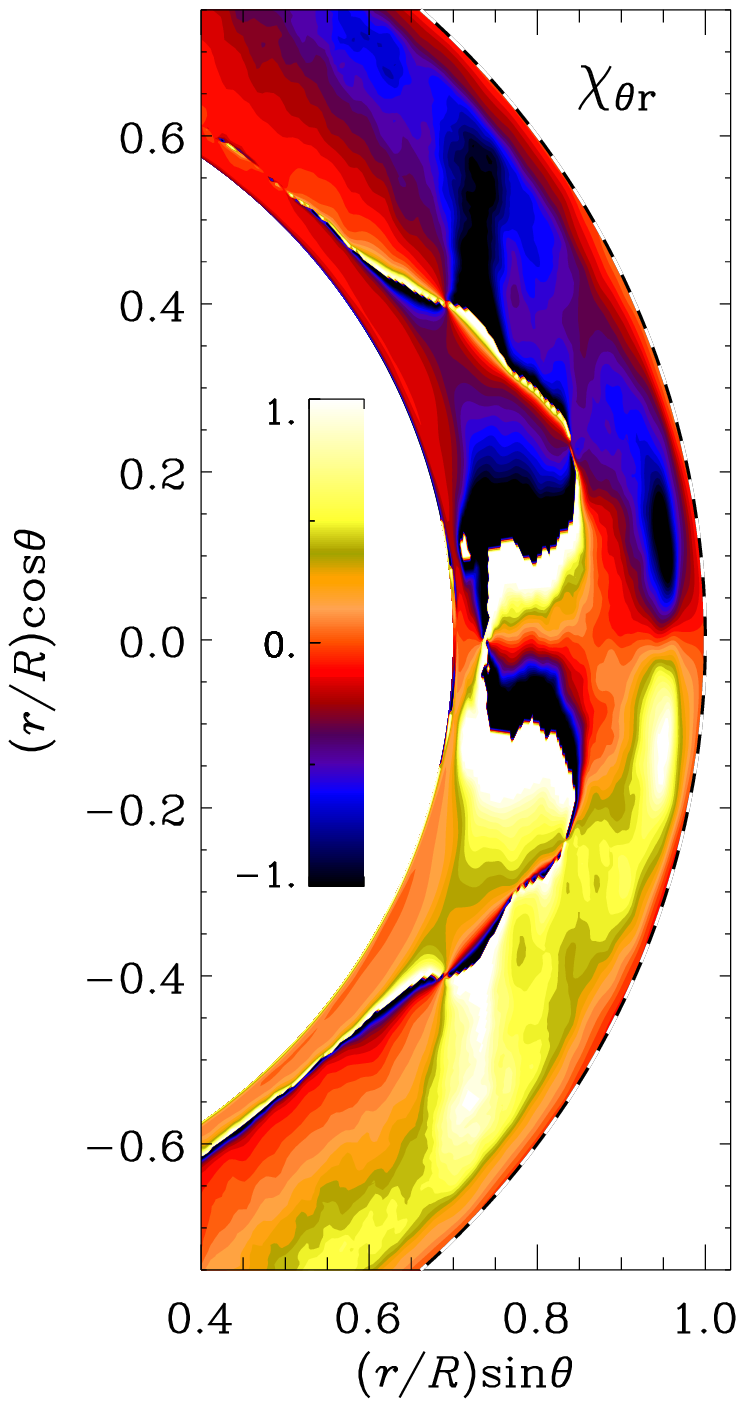}
%\includegraphics[width=0.3\columnwidth]{chirr_P.eps}
%\includegraphics[width=0.3\columnwidth]{chitr_P.eps}
%ApJ
\end{center}\caption[]{
%\small
%ApJ
Same as Figure~\ref{chitr}, but for Run~C1 of \cite{KMCWB13},
which is the same as Run~B4m of \cite{KMB12}.
This run is similar to those of the present work without a coronal
envelope.
}
\label{chitr_P}
\end{figure}

As discussed above, we expect the latitudinal entropy gradient
to be a consequence of an anisotropic convective (turbulent) diffusivity tensor.
Such anisotropies are caused by the rotational influence on the
turbulence \citep[see, e.g.,,][]{W65,R89}.
In particular, there is a term proportional to $\Omega_{0i}\Omega_{0j}$,
where, $\Omega_{0i}$ is the $i$th component of
${\bm\Omega}_0 =(\cos\theta,-\sin\theta,0)\Omega_0$,
which gives a symmetric contribution $\chi_{r\theta}=\chi_{\theta r}$
proportional to $\cos\theta\sin\theta$, so it vanishes at the poles and
at the equator.
In the presence of a latitudinal entropy gradient, it leads to
an additional contribution to the radial convective flux,
\begin{equation}
\meanF_r=-\chi_{rr} \mean{\rho}\mean{T} {\bm \nabla}_r\mean{s}
-\chi_{r\theta} \mean{\rho}\mean{T} {\bm \nabla}_\theta\mean{s}.
\end{equation}
Since $\chi_{r\theta}=\chi_{\theta r}$, and since there is a radial entropy
gradient, it also leads to a contribution in the latitudinal flux,
\begin{equation}
\meanF_\theta=-\chi_{\theta r} \mean{\rho}\mean{T} {\bm \nabla}_r\mean{s}
-\chi_{\theta \theta}\mean{\rho}\mean{T} {\bm \nabla}_\theta\mean{s}.
\end{equation}
If we were to ignore the second term proportional to
${\bm\nabla}_\theta\mean{s}$,
we could estimate $\chi_{\theta r}$ by measuring
\begin{equation}
\meanF_\theta=c_{\rm P} \mean{\rho} \mean{u_\theta^{\prime}T^{\prime}},
\end{equation}
so
\begin{equation}
\chi_{\theta r}\approx\left.
-c_{\rm P}
\mean{u_\theta^{\prime}T^{\prime}}\right/\mean{T}{\bm \nabla}_r\mean{s}.
\label{eq:chitr}
\end{equation}
The result is shown in Figure~\ref{chitr}, where we also plot a similar
estimate of the radial component,
\begin{equation}
\chi_{rr}\approx\left.
-c_{\rm P}
\mean{u_r^{\prime}T^{\prime}}\right/\mean{T}{\bm \nabla}_r\mean{s}.
\label{eq:chirr}
\end{equation}
We normalize $\chi_{ij}$ by $\chi_{t0}=\urms/3\kef$ and find
$\chi_{\theta r}/\chi_{t0}\approx1$ and $\chi_{rr}/\chi_{t0}\approx2$,
corresponding to $\chi_{\theta r}/\chitm\approx50$ and
$\chi_{rr}/\chitm\approx25$.
So, the SGS energy flux is small compared with the
resolved turbulent energy flux, as expected.

In reality, we cannot neglect the second term
proportional to ${\bm\nabla}_\theta\mean{s}$, even though this
gradient is about 10 times smaller than $|{\bm\nabla}_r\mean{s}|$ in our
simulations.
To test the accuracy of Equations~(\ref{eq:chitr})
and (\ref{eq:chirr}), we compute two-dimensional histograms of
latitudinal and radial heat fluxes versus latitudinal and radial
entropy gradients; see Figure~\ref{histo}.
The determined values of the turbulent heat diffusivities of
Figure~\ref{chitr} are consistent with those results.
However, a clear linear trend is not visible, except for a narrow
range in the case of $\chi_{rr}$.
In the first panel, the line with $\chi_{rr}=1.85\chi_{t0}$ fits the
maximum of the correlation well.
Indeed, looking at the left panel of Figure~\ref{chitr},
$\chi_{rr}\approx2\chi_{t0}$ is compatible with this.
A similar behavior can be seen in the top-right panel of
Figure~\ref{histo}.
We see that the lines of $\chi_{\theta r}=\pm0.7\chi_{t0}$ fit well through the
maxima of the data,
but we cannot find any indication of a linear correlation, as suggested
by Equations~(\ref{eq:chitr}) and (\ref{eq:chirr}).
The last two panels support our assumption that the latitudinal
entropy gradient can be neglected when calculating the turbulent heat diffusivity.
The main conclusion of the two-dimensional histogram is that the
correlation suggested by Equation~(\ref{eq:chitr}) and (\ref{eq:chirr})
is at best only true for the radial gradient of $\mean{s}$,
but not for the latitudinal one.
The ratio of the convective flux and the entropy gradient
is dominated by the ratio of two points rather than a
correlation.
Although the latitudinal entropy gradient is only 10 times smaller than the
radial one, we cannot find a linear correlation.
This is surprising, given that these mean field relations have been used
successfully to model the differential rotation profile as well as the
turbulent heat transport of the Sun---in good agreement with
observations \citep[see, e.g.,][]{KR95}.
In fact, as shown in the mean-field calculations of \cite{BMT92},
the $\chi_{r\theta}$ term tends to balance the first term so as to reduce the latitudinal
heat flux and thus produce a latitudinal entropy gradient
and a baroclinic term as we see it.

To investigate the baroclinic term and the turbulent
heat diffusivities as well as their influence on the differential
rotation in more detail,
we compare the present runs, where we include a coronal envelope, with
runs without a coronal envelope.
The runs without a coronal envelope are taken from \cite{KMB12} and
\cite{KMCWB13}.
Thus, we compare Figures~\ref{baroc} and \ref{chitr} for Run~A
with the corresponding ones for Run~C1
of \cite{KMCWB13}, which is the same as Run~B4m of \cite{KMB12}; see
Figures~\ref{baroc_P} and \ref{chitr_P}.
As in Run~A above, the baroclinic term of Run~C1 balances
the advection term.
However, the values are four times larger and the shape shows a clear
radial variation.
The baroclinic term is largest near the surface, whereas
in Figure~\ref{baroc} of Run~A, the term is stronger near the bottom
of the convection zone.
In Run~C1, on the other hand, the terms are small
near the bottom and close to the equator.
The component of the turbulent heat diffusivity tensor
(Figure~\ref{chitr_P}) looks quite different from Run~A; compare with
Figure~\ref{chitr}.  The radial heat conductivity is about two times
smaller than in the runs with a corona.  The mean radial entropy
gradient has a positive sign at $\pm15^\circ$ latitude,
and extends from the bottom to the middle of the
convection zone. How these two behaviors can change the solar-like
rotation to a more cylindrical rotation is unclear.
Stratification may be important, because in Run~C1 without a corona
the density contrast is $\rho_0/\rho_s=22$,
which is slightly higher than in the four runs of this work.

\subsection{Meridional Circulation}

\begin{figure*}[t!]
\begin{center}
\includegraphics[width=0.9\columnwidth]{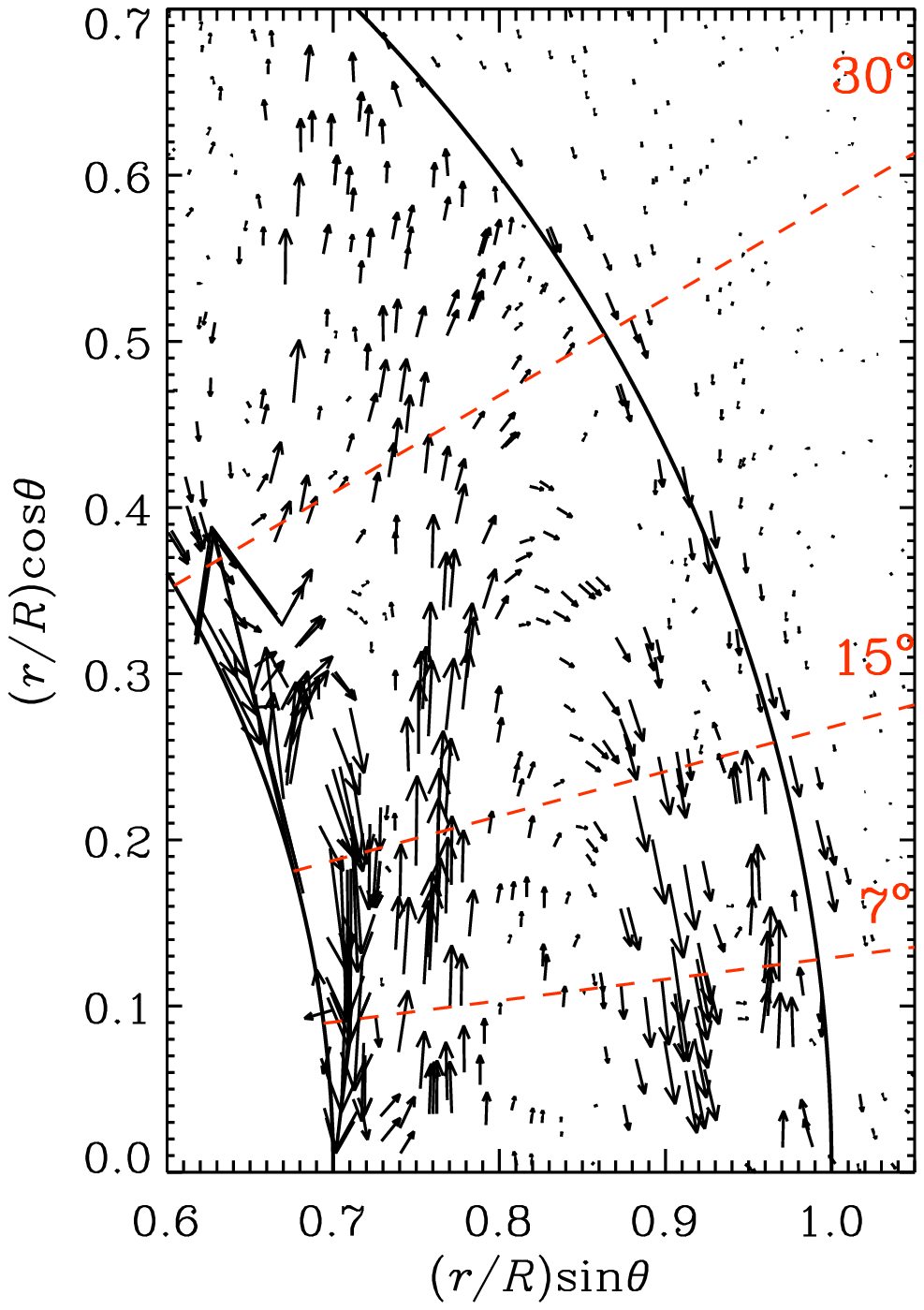}
\includegraphics[width=0.9\columnwidth]{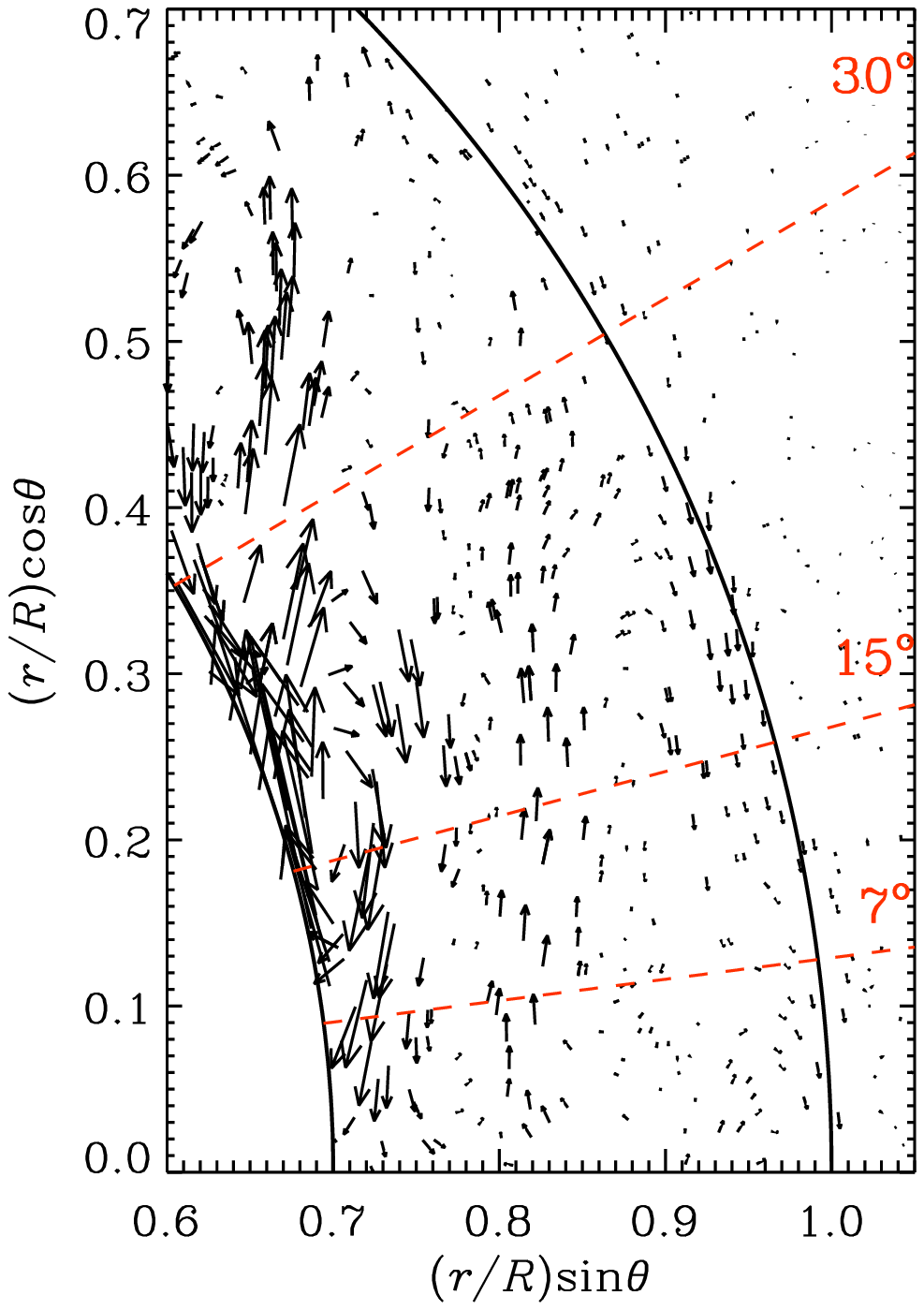}
\includegraphics[width=0.9\columnwidth]{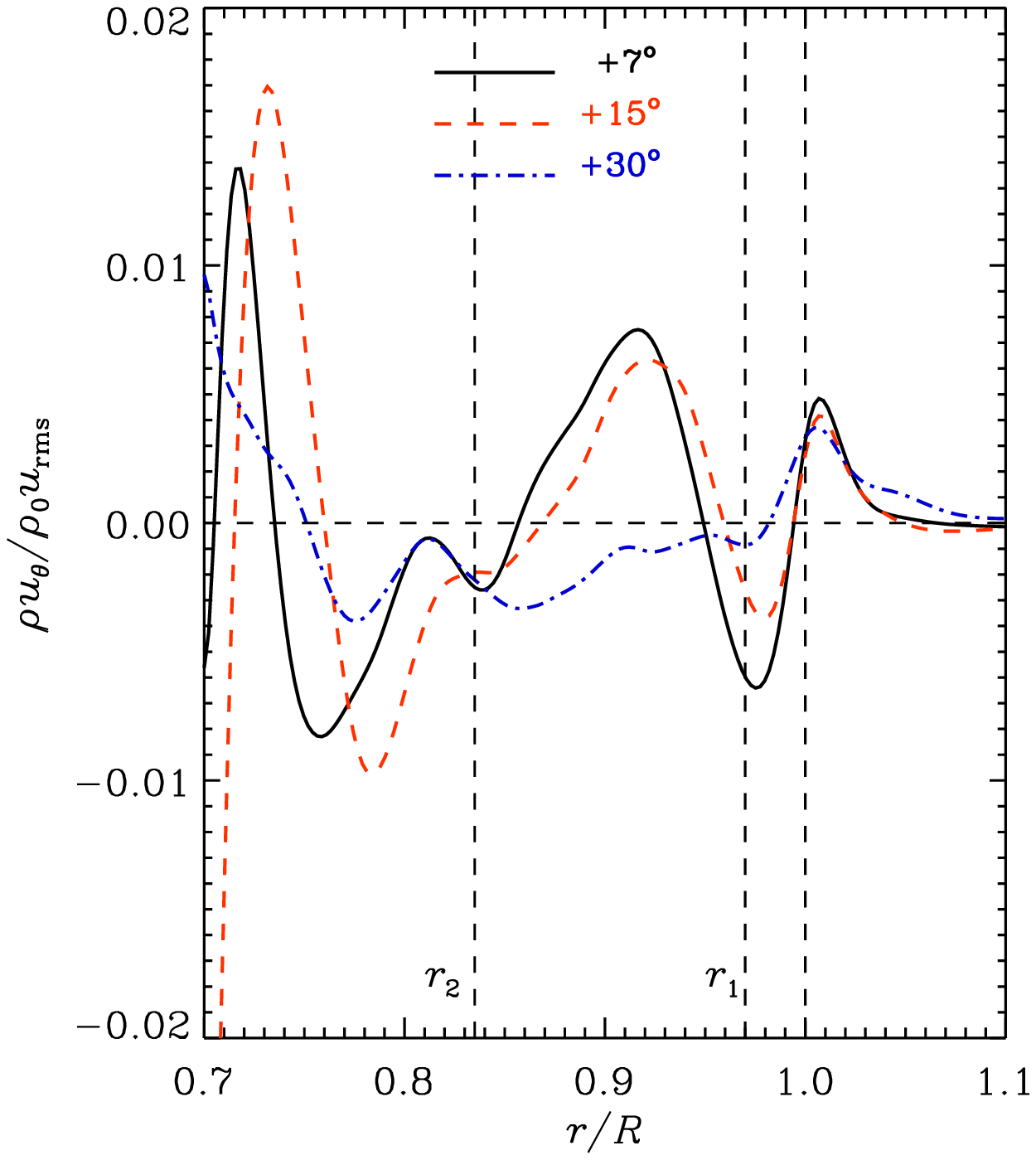}
\includegraphics[width=0.9\columnwidth]{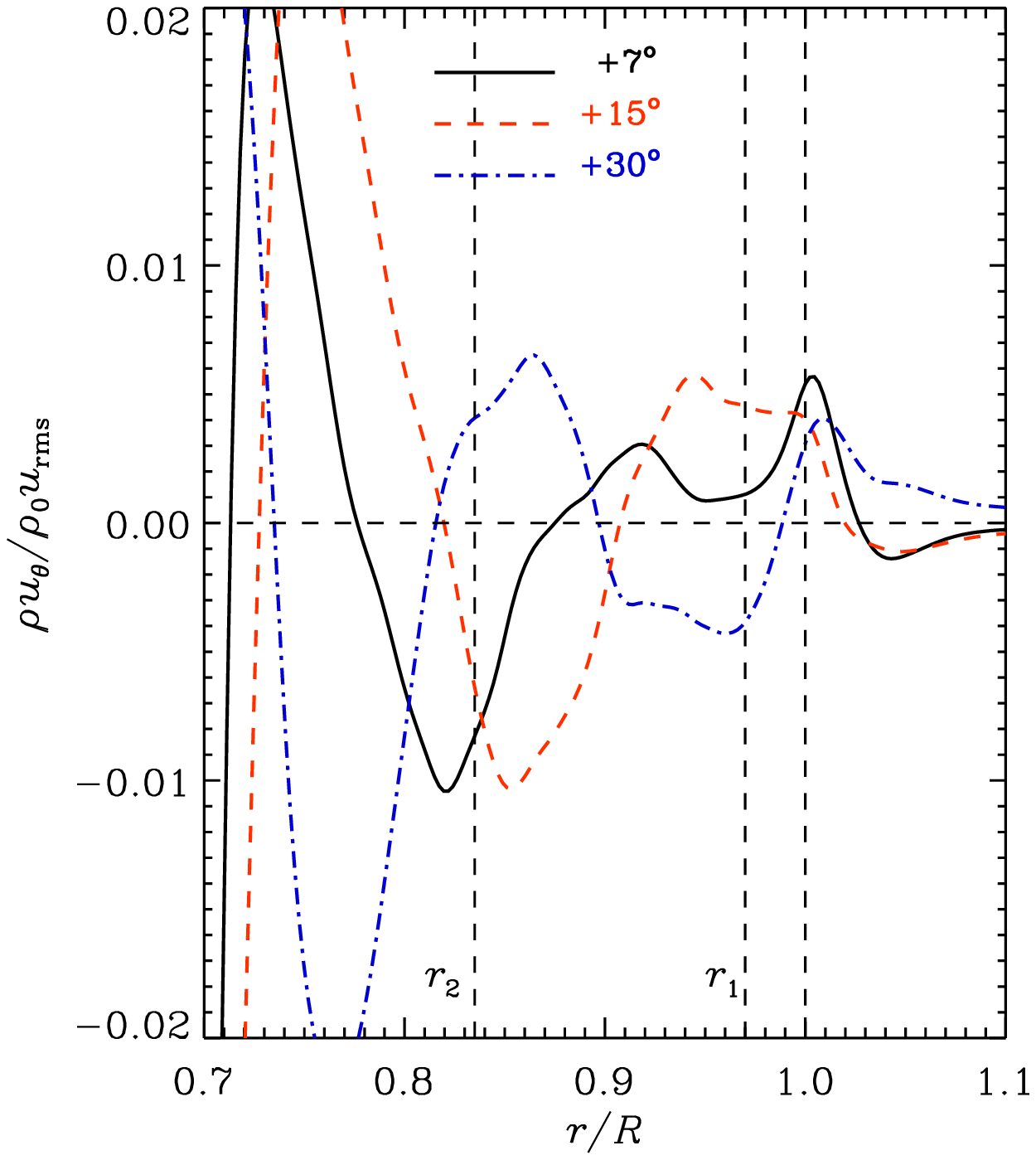}
%\includegraphics[width=0.43\columnwidth]{meri_conv_A.eps}
%\includegraphics[width=0.43\columnwidth]{meri_conv_B.eps}
%\includegraphics[width=0.41\columnwidth]{meri_cut_A.eps}
%\includegraphics[width=0.41\columnwidth]{meri_cut_B.eps}
%ApJ
\end{center}\caption[]{
%ApJ
%\small
%
Meridional circulation in the
northern hemisphere in the convection zone.
Top row: meridional circulation as vectors 
in terms of the mass flux $\mean{\rho}(\mean{u}_r,\mean{u}_\theta,0)$
for Runs~A (left) and B (right),
where for Run~B the arrows are reduced in size by a factor of three.
The three red dashed lines represent the
latitudes $90^\circ-\theta_1=30^\circ$, $90^\circ-\theta_2=15^\circ$,
and $90^\circ-\theta_3=7^\circ$, which are used in the bottom row.
The black solid lines indicate the surface ($r=R$) and the bottom of
the convection zone ($r=0.7\,R$).
Bottom row: latitudinal mass flux
$\mean{\rho}\mean{u_\theta}/\rho_0\urms$ plotted over radius $r/R$ for
three different latitudes $\theta_1$ (blue dot-dashed line), $\theta_2$
(red dashed line) and $\theta_3$ (black solid line) in the northern
hemisphere for Runs~A (left) and B (right).
The black dashed lines indicate the surface ($r=R$) and
the radii $r_1=0.97\, R$ and $r_2=0.84\, R$, which are also used in
Figures~\ref{butterfly} and \ref{butterfly2}.
}
\label{meri}
\end{figure*}

Another important result is the multi-cellular meridional circulation in the
convection zone with a poleward (solar-like) flow near the surface.
In Figure~\ref{meri}, we plot the
meridional circulation in terms of the mass flux as vectors of
$\mean{\rho}(\mean{u}_r,\mean{u}_\theta,0)$ and as radial cuts of
$\mean{\rho}\mean{u}_\theta$ through 
colatitudes $\theta_1$ = 60$^\circ$, $\theta_2$ = 75$^\circ$, and
$\theta_3$ = 83$^\circ$, corresponding to latitudes of $30^\circ$,
$15^\circ$, and $7^\circ$.
Note here that in the northern (southern) hemisphere a negative
$\mean{\rho}\mean{u_\theta}$ means a poleward (equatorward) flow, and a positive
one equatorward (poleward) flow.
Runs~A, Ab, and Ac show significant solar-like surface profiles of
meridional circulation,
while Run~B shows a different pattern.
Looking at Run~A, in the northern hemisphere at lower latitudes
($\leq 20^\circ$) just below the surface ($r_1=0.97\, R$),
the meridional circulation is poleward with
$\mean{\rho}\,\mean{u_\theta}=-0.007\rho_0 \urms$.
Above the surface there is a return flow in the equatorward
direction.
This return flow peaks above the surface with a similar flux.
The turning point $\mean{\rho}\,\mean{u_\theta}=0$ is just below the surface,
at around $r=0.985\, R$.
The location of this turning point is consistent with the location
where the entropy gradient changes from negative to positive,
i.e., from convectively unstable to convectively stable; see
Figure~\ref{strat}.
If we were to redefine the surface of the simulated star to this radius, we
would obtain a solar-like meridional circulation, where the circulation is
poleward at the surface.
The velocity near the surface is $u_\theta\approx0.07\urms$.
To compare this with the meridional circulation at the surface of the
Sun, which is $10-30\,$m\,s$^{-1}$ \citep{ZK04}, we calculate the
corresponding value of $\urms$ from the convective flux,
${\meanF_{\rm conv}}\approx\rho\urms^3$, with a typical density of
$\rho_{\rm conv}=2.5\,{\rm kg}\,{\rm m}^{-3}$ near the surface of the convection
zone at $r=0.996\,R$ \citep{Stix:02}.
Our estimate of corresponding meridional circulation gives
$u_{\rm m}=0.07(\meanF_{\rm conv}/\rho_{\rm conv})^{1/3}\approx
20\,$m\,s$^{-1}$,
which is consistent with the solar value.
Similar estimates apply to the southern hemisphere, but the
meridional circulation is a bit weaker here.
This behavior can also be found in Runs~Ab and Ac, where the flows are weaker
and the turning points lie slightly deeper.
Note that the strong return flow above the surface is a consequence of our
particular setup, which has much weaker stratification than the Sun;
see Figure~\ref{strat}. Higher stratification should lead to a much
weaker return flow in this location. 
In Run~B, a poleward flow
develops in the northern hemisphere only close to poles
($\theta=\theta_1$).  The meridional circulation has a latitudinal
dependence.  In Run~A the return flow reaches higher velocities at
lower latitudes ($\leq \pm20^\circ$).  The same is true for the
poleward circulation below the solar surface.  In Run~B, we find the
opposite and both the return flow and the meridional circulation
increase with latitude.

From the bottom row of Figure~\ref{meri}, we can estimate
the number of meridional circulation cells at low latitudes.
We find that there are at least two cells in the convection zone.
In Run~A, there is one cell with poleward flow maxima around $r=r_1$ and
minima around $r=0.91\, R$, where the mass flux closer to the surface
is as large as that in the return flow deeper in the convection zone.
A second cell is deeper down in the convection zone and has
similar extent and flux.
A similar two-cell meridional flow pattern has recently been reported
by \citet{ZBKDH13} from helioseismic inversions of {\it Solar Dynamics
  Observatory}/Helioseismic and Magnetic Imager data.
In Runs~Ab and Ac, the pattern is different and the flux is weaker, but there
are indications of a third cell.
Note that the stratification leads to stronger mass fluxes over a
smaller cross-sectional area deeper in
the convection zone than near the surface, while the velocity is
similar.
In Run~B, there are two strong cells of meridional circulation.
The cells seem to be more cylindrical than latitudinal, which can also be
seen in the phase shift of the pattern for different latitudes.
Even deeper down in the convection zone, the meridional flow is
much stronger than in Runs~A, Ab, and Ac.
This is consistent with results from models with anisotropic viscosity
(or lowest order $\Lambda$ effect), which show a maximum of meridional
circulation for Taylor numbers around $10^7$ \citep{Ko70}.
Our Taylor number is above this value, so the circulation decreases
with faster rotation, which is also in agreement with
numerical simulations \citep{BBT07,BBBMT08,ABBMT12}.
We emphasize that this does not apply to the non-dimensional meridional
circulation, normalized by viscosity, which does not show a maximum.

\begin{figure*}[t!]
\begin{center}
\includegraphics[width=1.03\columnwidth]{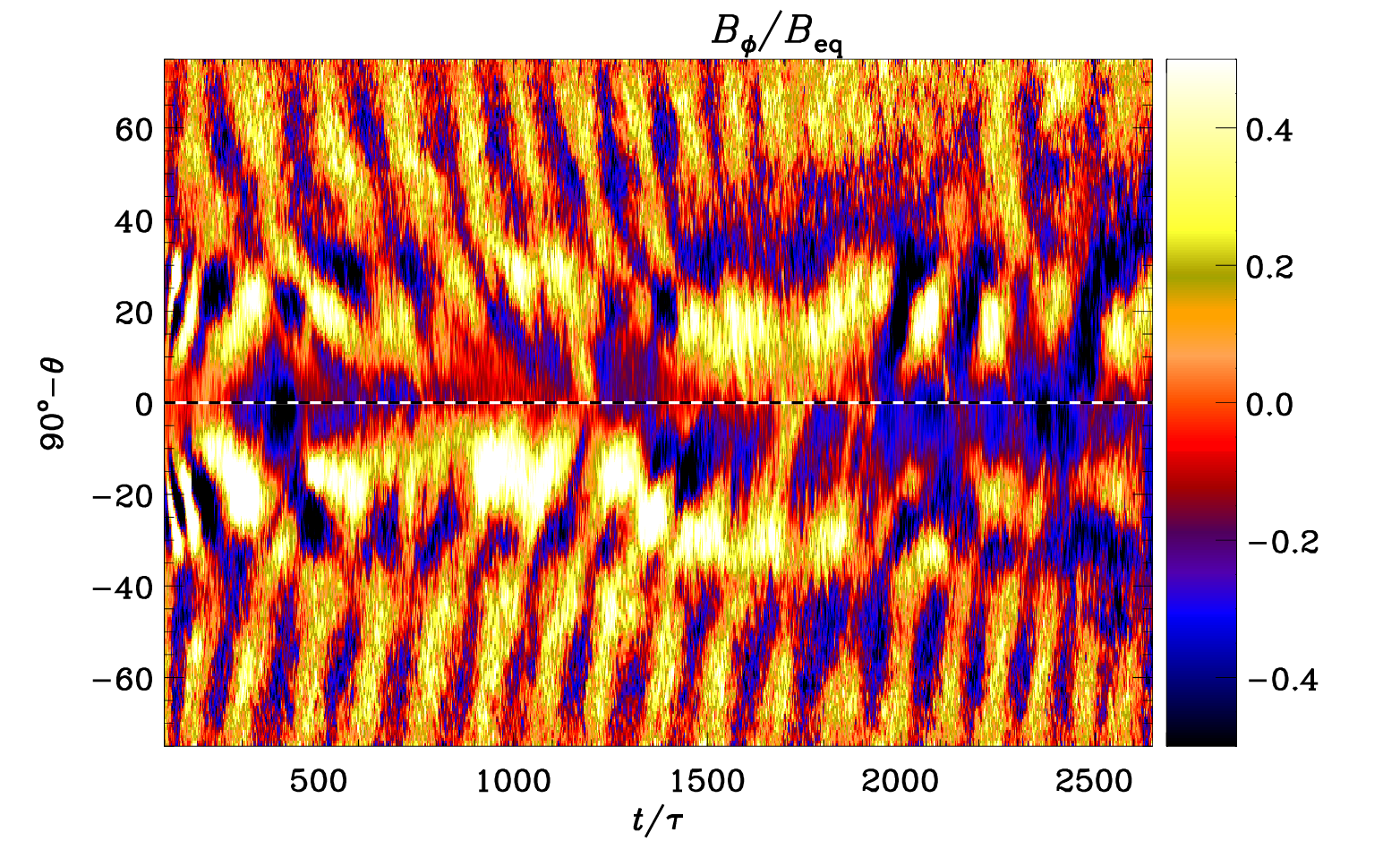}
\includegraphics[width=1.03\columnwidth]{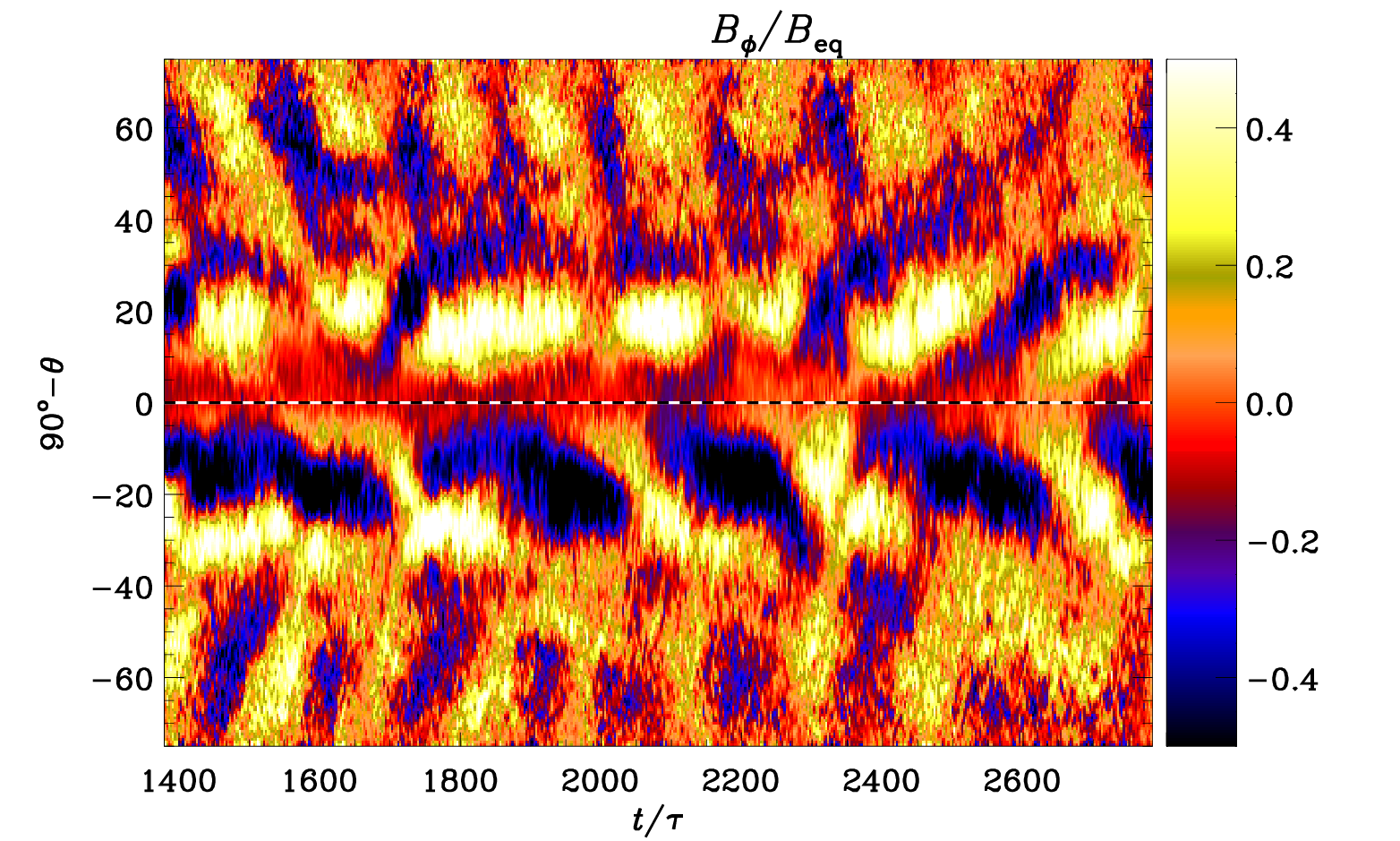}
\includegraphics[width=1.03\columnwidth]{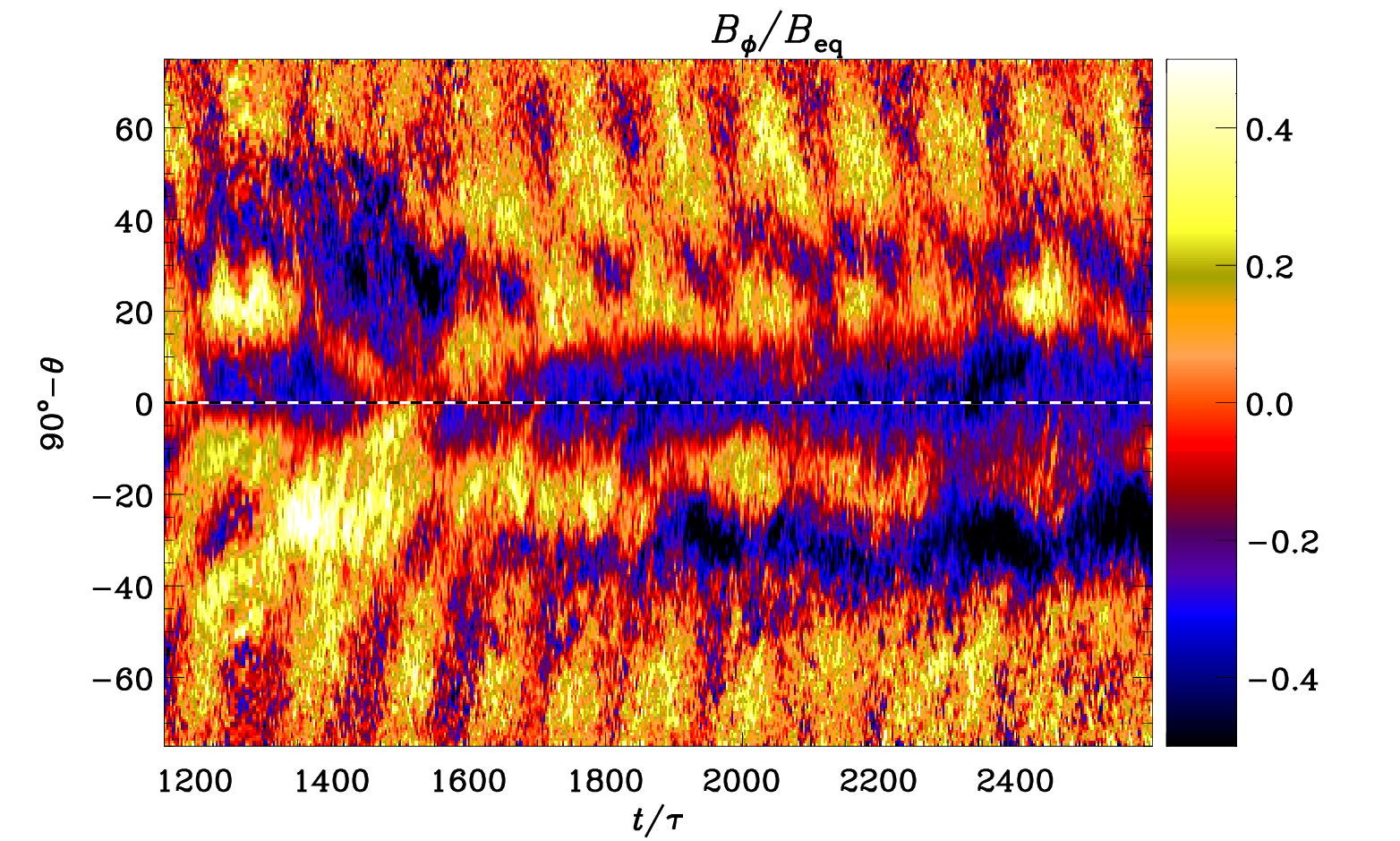}
\includegraphics[width=1.03\columnwidth]{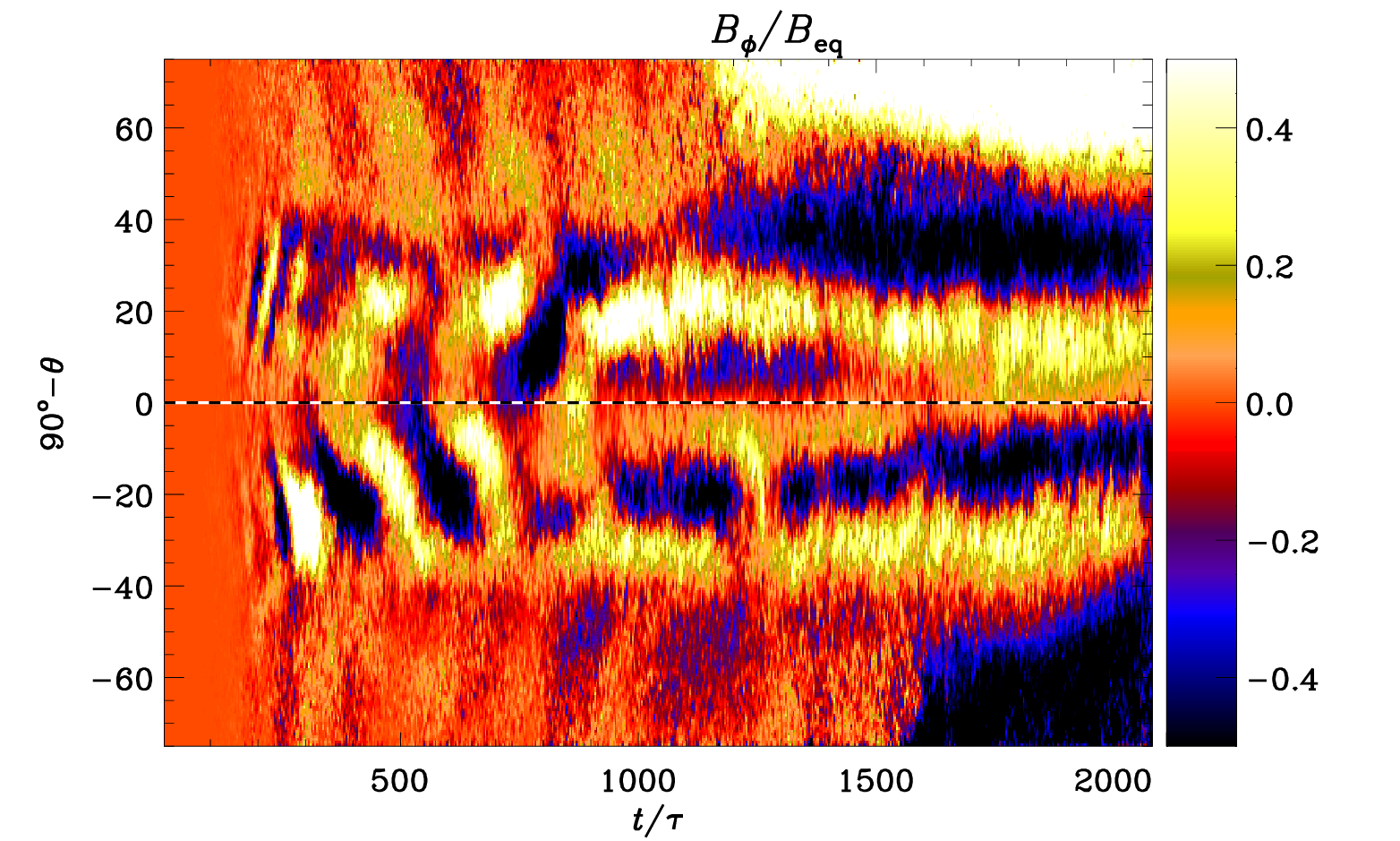}
%\includegraphics[width=0.49\columnwidth]{but1_A.eps}
%\includegraphics[width=0.49\columnwidth]{but1_Ab.eps}
%\includegraphics[width=0.49\columnwidth]{but1_Ac.eps}
%\includegraphics[width=0.49\columnwidth]{but1_B.eps}
%ApJ
\end{center}\caption[]{
%\small
%ApJ
Time evolution of the mean magnetic field in the convection zone.
From top left to bottom right, we show $\mean{B}_{\phi}$ for
Runs~A, Ab, Ac, and B at $r_1=0.97R$. 
Dark blue shades represent negative values and light yellow shades positive values.
The dashed horizontal lines show the location of the equator at
$\theta=\pi/2$.
The magnetic field is normalized by its equipartition value, $\Beq$.
}
\label{butterfly}
\end{figure*}

\subsection{Mean Magnetic Field Evolution}
\label{sec:mgf}

\begin{figure}[t!]
\begin{center}
\includegraphics[width=\columnwidth]{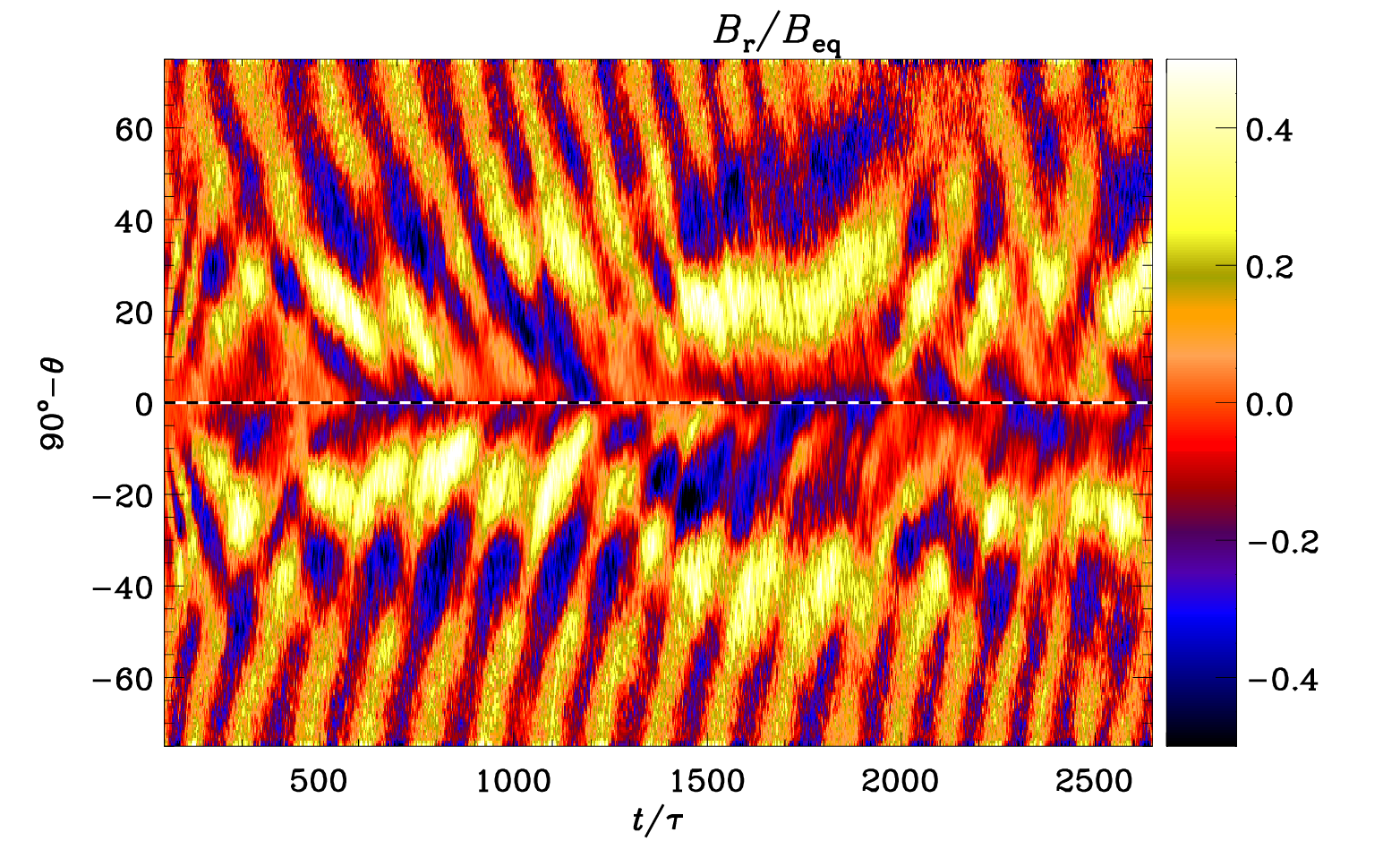}
\includegraphics[width=\columnwidth]{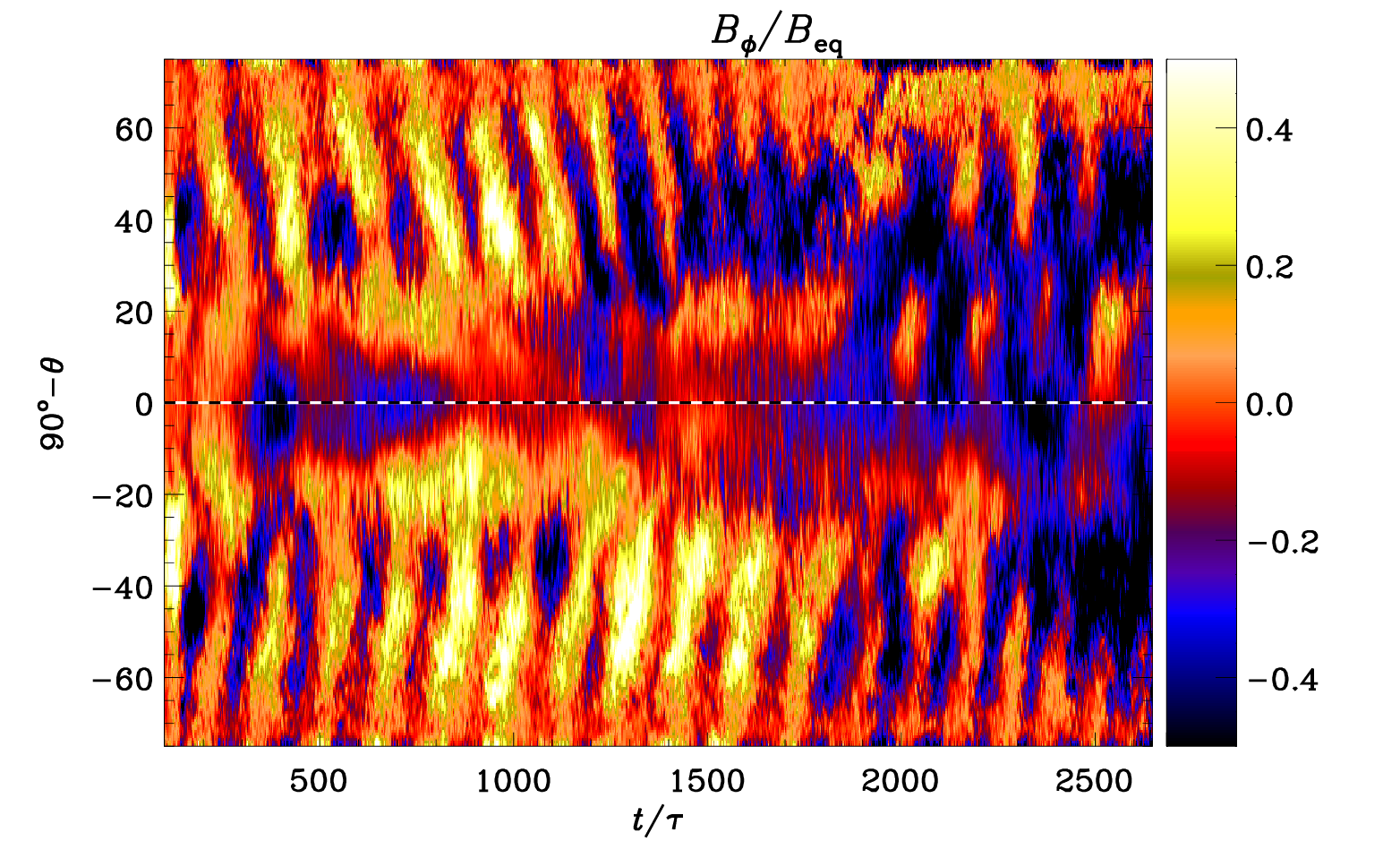}
\includegraphics[width=\columnwidth]{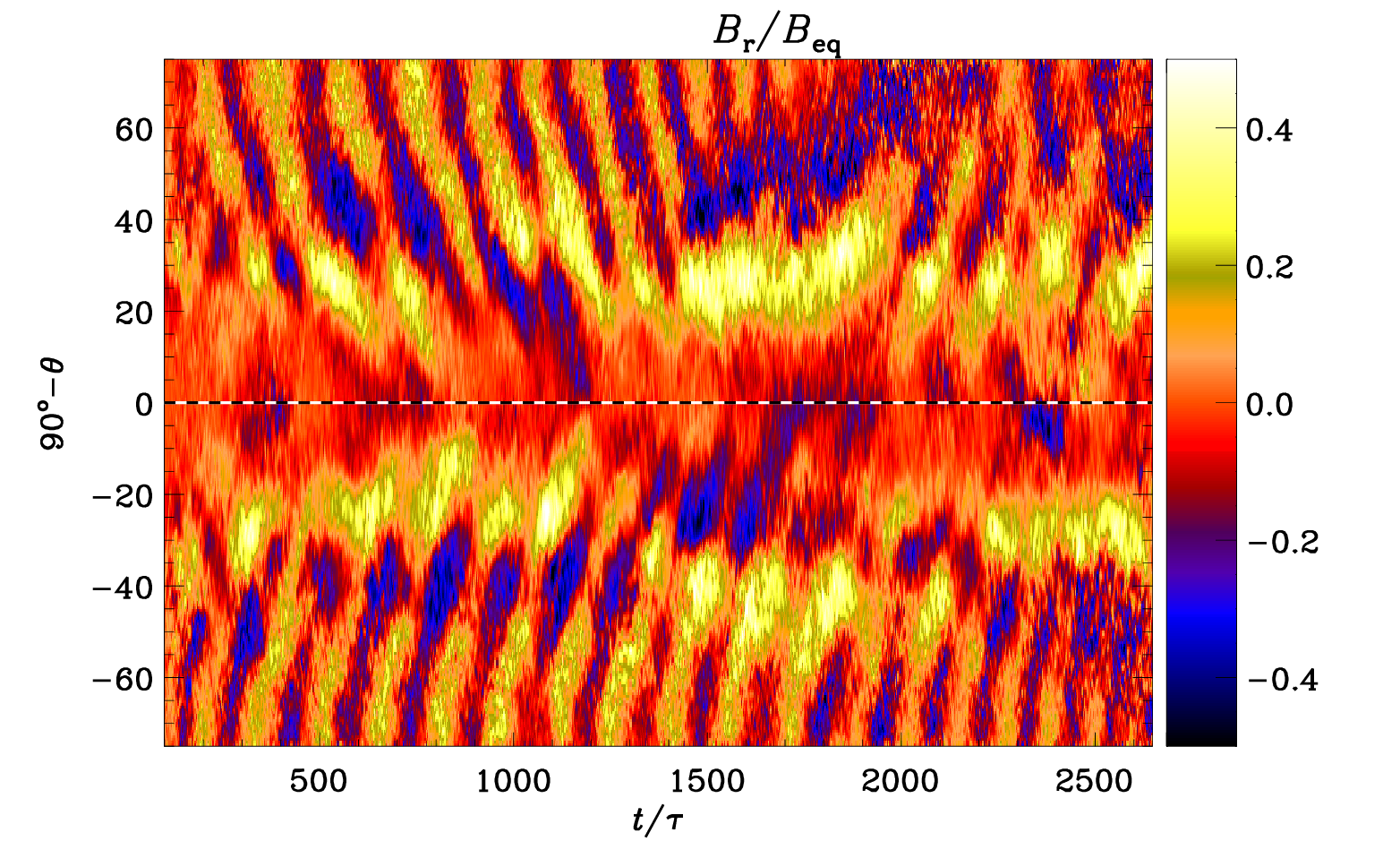}
\includegraphics[width=\columnwidth]{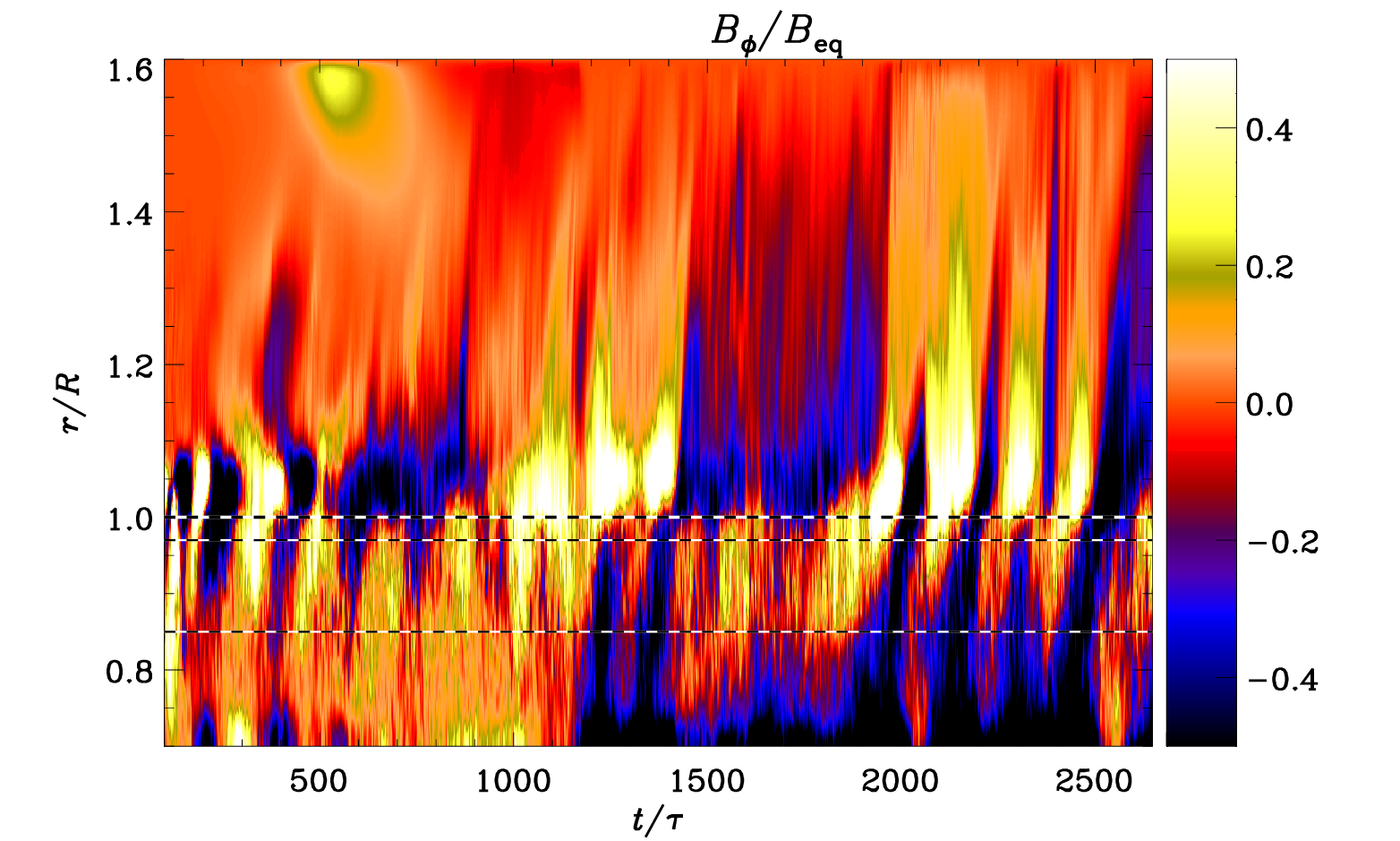}
%\includegraphics[width=0.51\columnwidth]{but2_A.eps}
%\includegraphics[width=0.51\columnwidth]{but3_A.eps}
%\includegraphics[width=0.51\columnwidth]{but4_A.eps}
%\includegraphics[width=0.51\columnwidth]{but5_A.eps}
%ApJ
\end{center}\caption[]{
%\small
%ApJ
Time evolution of the mean magnetic field in the convection zone for
Run~A.
From top to bottom:
mean radial field $\mean{B}_{r}$ at $r_1=0.97R$, mean azimuthal
field $\mean{B}_{\phi}$ at $r_2=0.84R$, mean radial field $\mean{B}_{r}$ at
$r_2=0.84R$ and mean azimuthal field $\mean{B}_{\phi}$ at $25^\circ$
latitude for the whole radial extent.
Otherwise the same as in Figure~\ref{butterfly}.
In the last panel the dashed lines indicate the surface ($r=R$)
and the radii $r_1=0.97R$ and $r_2=0.84R$.
}
\label{butterfly2}
\end{figure}

The turbulent helical motions generated by convective heat transport,
together with differential rotation, produce a large-scale magnetic
field inside the convection zone.
It grows exponentially and shows an initial saturation after around
$t/\tau=100$ for Run~A; see Figure~\ref{purms}.
Run~B shows a more peculiar behavior: the field seems to have saturated
at around $t/\tau=300$, but it starts growing again at around
$t/\tau=700$ and appears to saturate at $t/\tau=1700$.
The latter growth is possibly related to a change of the oscillatory
mode into a
stationary one; see Figure~\ref{butterfly} and the discussion below.
The magnetic and fluid Reynolds numbers of Run~B are higher than for the
other cases, which
should lead to a higher growth rate.
However, the rotation rate measured by $\Co$ is around half
that of Run~A, which leads to a slower amplification of the field.
At a later time, around $t/\tau=1000$, the field of Run~B becomes
comparable to or even stronger than that of Run~A.
The value of $\brms$ reaches around 0.5 of the equipartition field strength in
Runs~A, Ab, and Ac and 0.6 in Run~B; see Table~\ref{tab:runs}.
In comparison, $\Brms$ is around 20\% lower because the field is mainly
concentrated in the convection zone.
The equatorward migration pattern is visible in three of the
four runs at high latitudes.
In Run~A, the pattern seems to transform into a slow poleward migration
at lower latitudes, but the equatorward migration pattern re-appears at
$t/\tau=2300$.
We suggest that the equatorward migrating dynamo mode is dominant after
$t/\tau=500$, while being overcome by other modes between
$t/\tau=1500$ and $t/\tau=2300$.

Comparing our results with those of \cite{KMB12} without a corona
but an otherwise comparable setup, the magnetic field in the current
simulations is slightly weaker.
In Figure~2 of \cite{KMB12}, the mean toroidal magnetic 
field strength is close to
super-equipartition ($\mean{B}_{\phi}\approx\Beq$), whereas in
Figure~\ref{butterfly} the mean magnetic field strength is roughly
$\mean{B}_{\phi}=0.5\Beq$.
Additionally, the growth rate of the dynamo is greater than in the
models without corona, where it takes up to five times longer to
reach dynamically important field strengths.
This is not surprising because the dynamo in the two-layer model is
less restricted and has more freedom for different dynamo modes to be
excited.
There is no restriction due to the magnetic boundary at the surface,
which is open in our simulations, but restricted to vertical
fields in the convection zone simulations of \cite{KMB12}.
This could explain the fast growth in the beginning, but not
the decreased saturation level.
On the other hand, the runs in this work and the runs of
\cite{KMB12,KMCWB13} also show differences in other parameters, such
as stratification, rotation rate, and Reynolds numbers, so a direct
comparison might not be possible.

Recently, other authors also have reported magnetic cycles.
In the works by \cite{BMBBT11} and \cite{NBBMT13}, using anelastic LES, the authors
were able to produce an oscillatory field, but without a clear pattern
and no equatorward migration.
In the simulations by \cite{GCS10} and \cite{Racine11}, who
used an implicit method, the mean magnetic field shows a clear
oscillatory behavior, but only a weak tendency for equatorward
migration; see Figure~4 of \cite{GCS10} and
Figure~8 of \cite{Racine11}.
There is evidence that oscillatory solutions are favored when the density
stratification is strong \citep{Gastine_etal12}, but such dynamos might
well be of the $\alpha^2$ type \citep{Mitra_etal10,Schrinner_etal11}, while
strong shear favors poleward migration \citep{Schrinner_etal12}.
At the moment, only the work by \cite{KMB12,KMCWB13} and the present work
show clear evidence of equatorward migration.
\begin{figure*}[t!]
\begin{center}
\includegraphics[width=0.48\columnwidth]{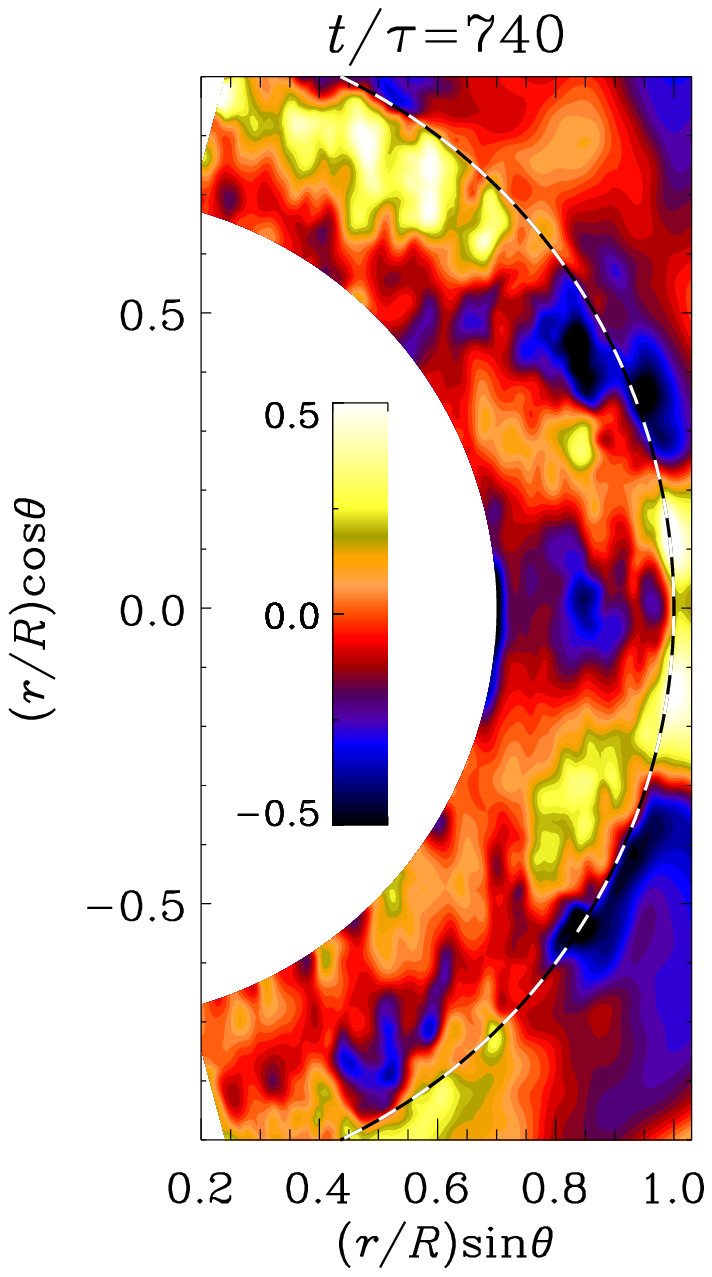}
\includegraphics[width=0.48\columnwidth]{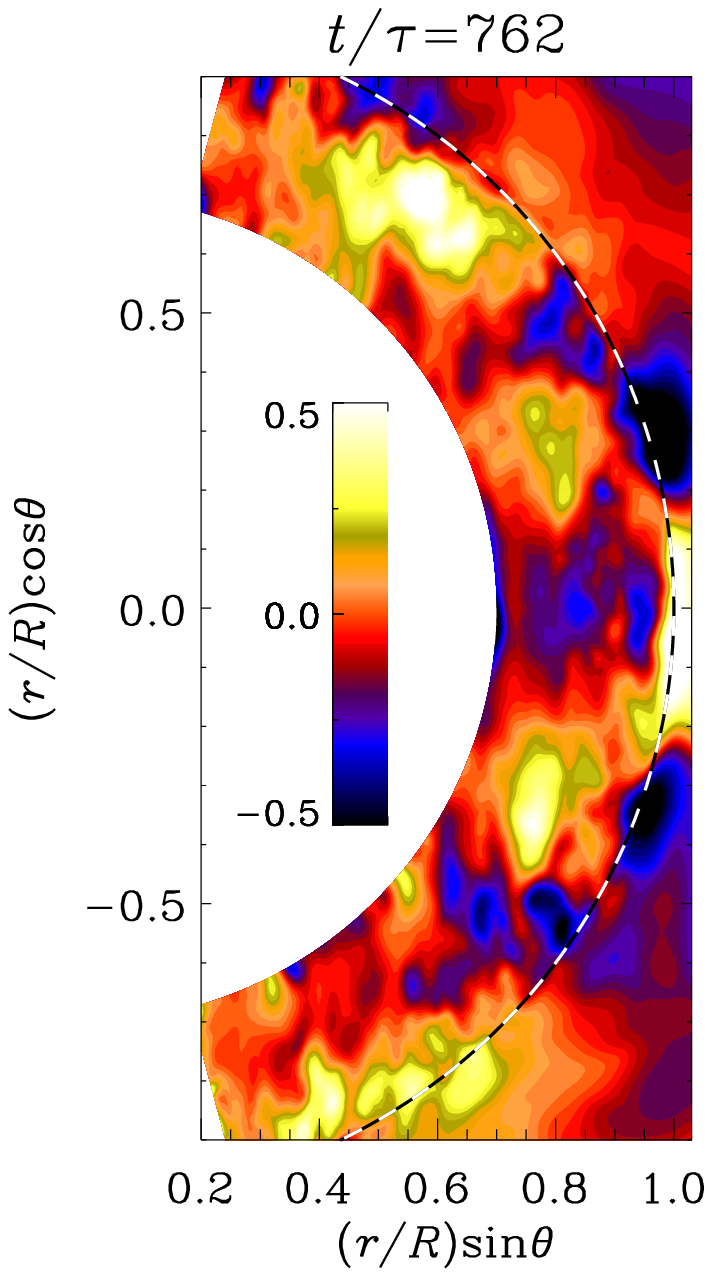}
\includegraphics[width=0.48\columnwidth]{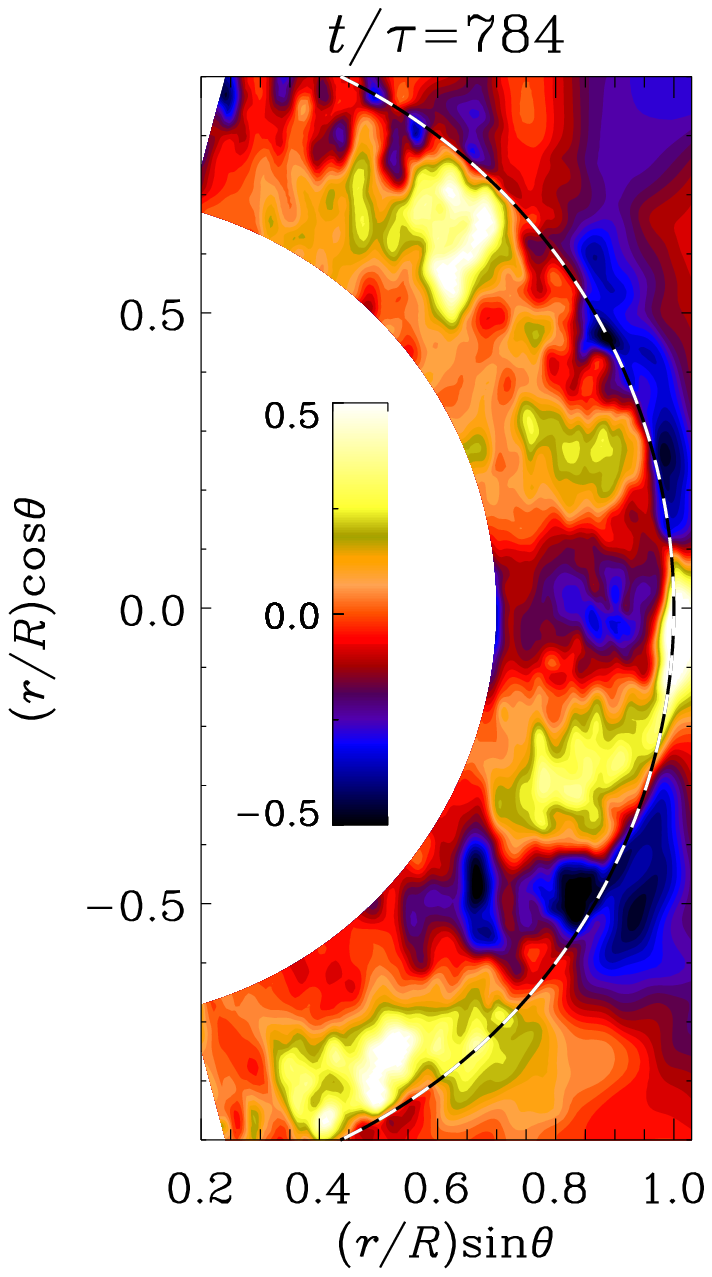}
\includegraphics[width=0.48\columnwidth]{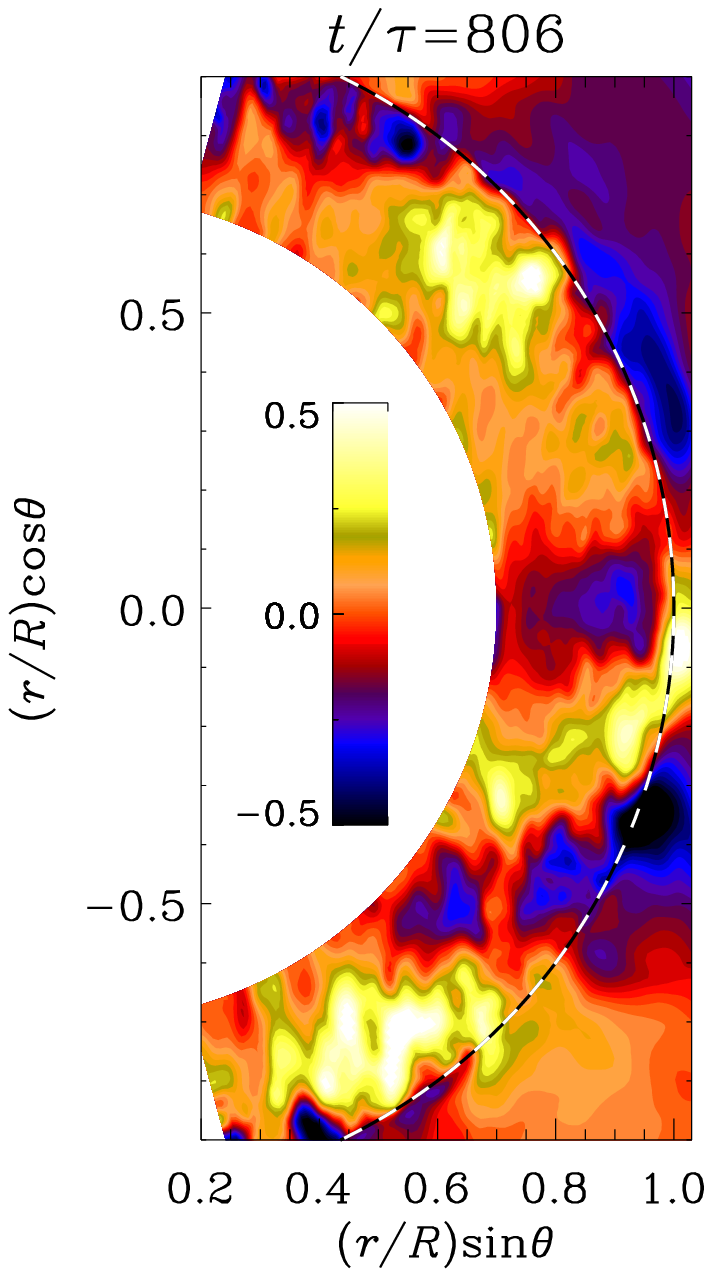}
\includegraphics[width=0.48\columnwidth]{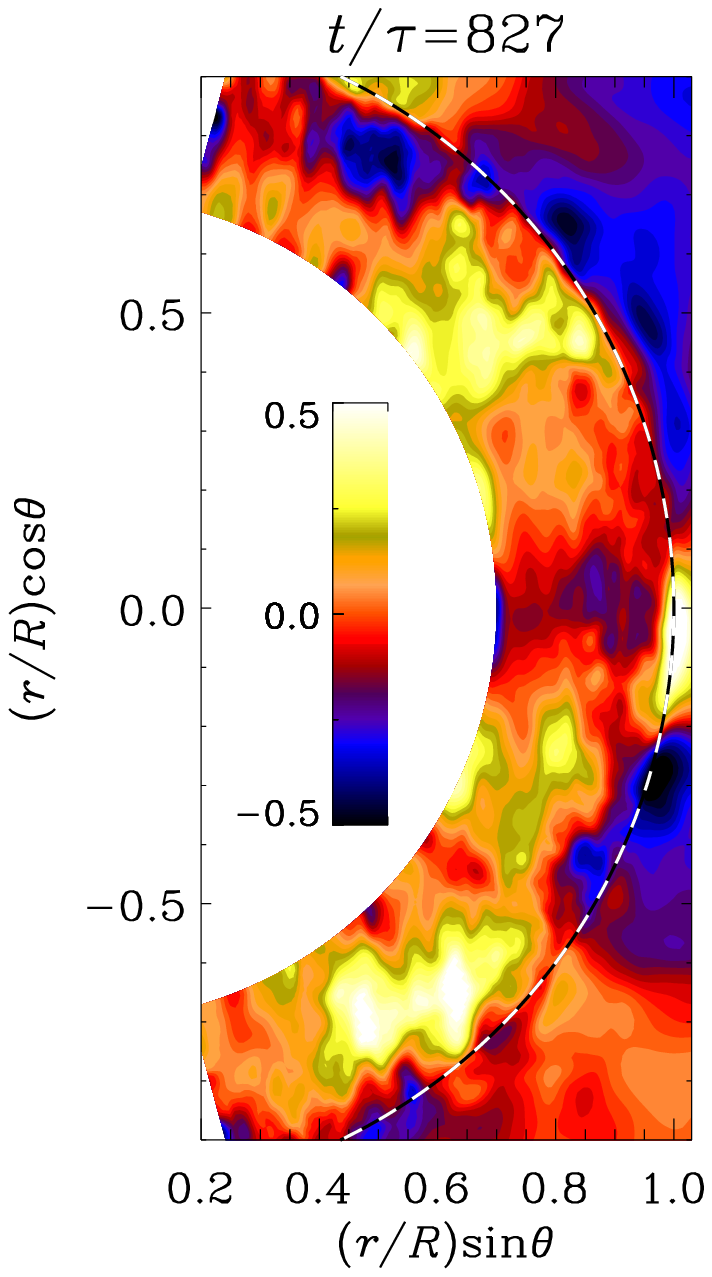}
\includegraphics[width=0.48\columnwidth]{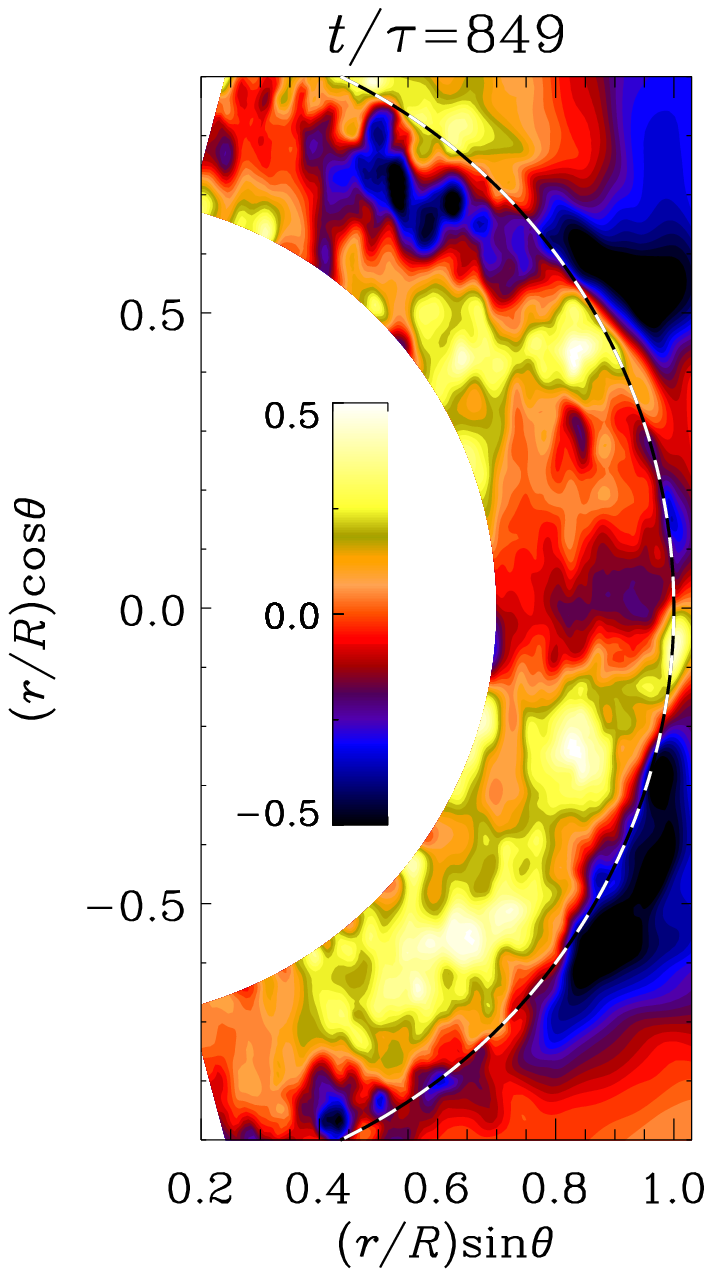}
\includegraphics[width=0.48\columnwidth]{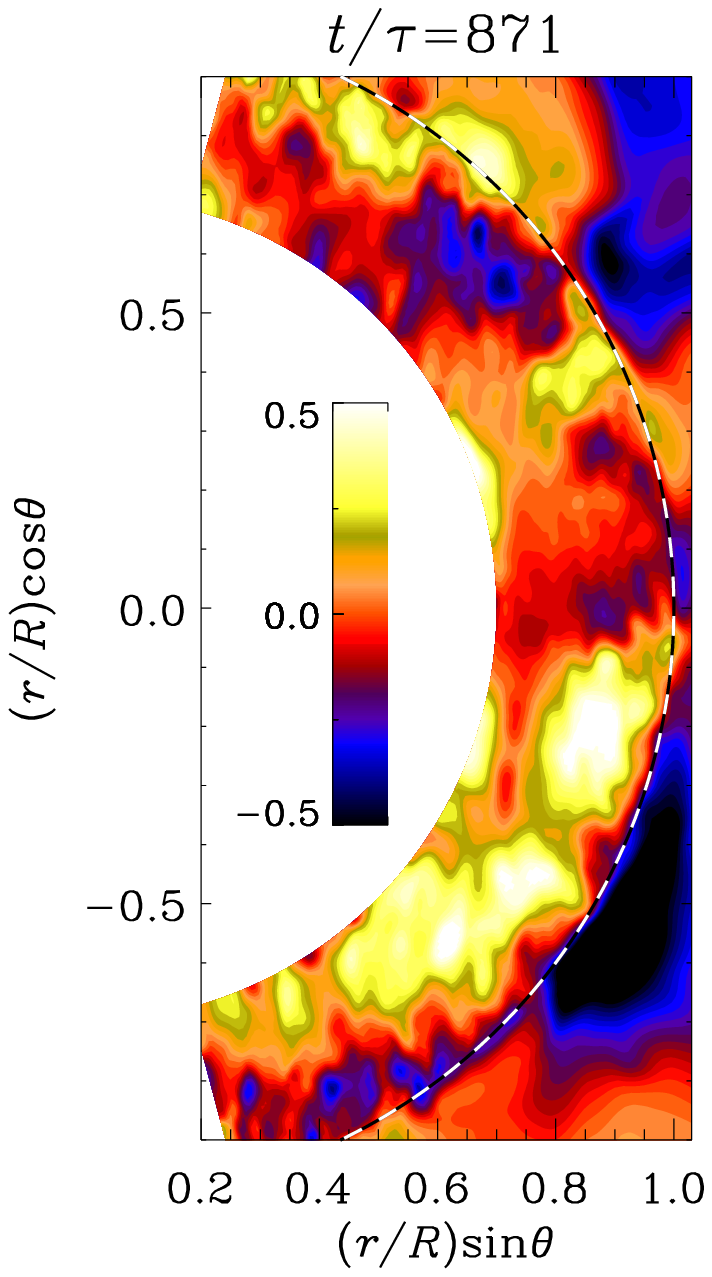}
\includegraphics[width=0.48\columnwidth]{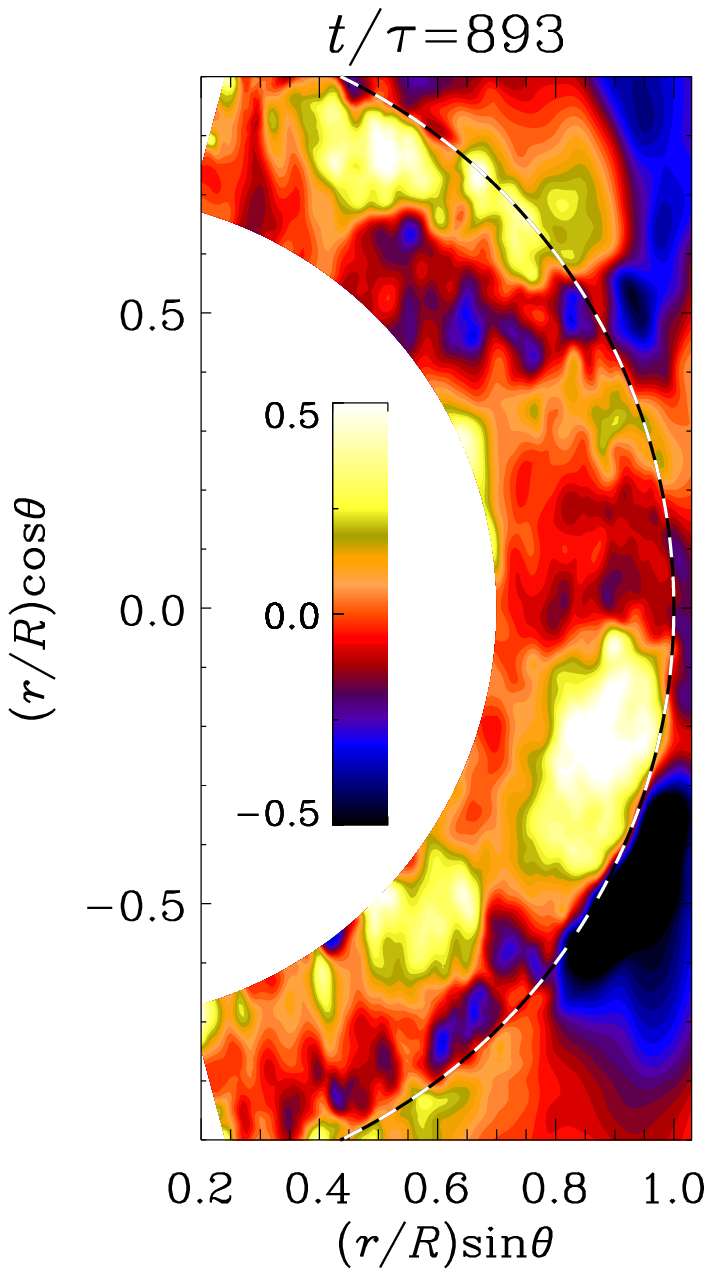}
%ApJ
%\includegraphics[width=0.22\columnwidth]{bphi1_A.eps}
%\includegraphics[width=0.22\columnwidth]{bphi2_A.eps}
%\includegraphics[width=0.22\columnwidth]{bphi3_A.eps}
%\includegraphics[width=0.22\columnwidth]{bphi4_A.eps}
%\includegraphics[width=0.22\columnwidth]{bphi5_A.eps}
%\includegraphics[width=0.22\columnwidth]{bphi6_A.eps}
%\includegraphics[width=0.22\columnwidth]{bphi7_A.eps}
%\includegraphics[width=0.22\columnwidth]{bphi8_A.eps}
\end{center}\caption[]{
%\small
%ApJ
Time series of eight snapshots of the mean azimuthal magnetic field 
$\mean{B}_\phi$ separated by 22 turnover times and covering one full 
magnetic cycle from Run~A.
Dark blue shades represent negative values and light yellow shades
represent positive values.
The dashed line indicates the surface ($r=R$).  The field is
normalized by the equipartition field strength $\Beq$.  }
\label{bphi}
\end{figure*}

Looking just at $\brms$ in the convection zone or at $\Brms$ in
the whole domain in Figure~\ref{purms}, we find evidence of
cyclical behavior of the field for Runs~A, Ab, and Ac.
The cycle period is $\approx100\tau$.
In Run~B, there is no clear evidence of cyclical behavior.
Investigating the different
components of the mean magnetic field we find signs of
oscillatory behavior for all runs, except that Run~B
shows oscillations only at early times.
In Figure~\ref{butterfly}, we plot the
azimuthal mean magnetic field $\mean{B}_{\phi}$ over time and latitude
at $r_1$ for Runs~A, Ab, Ac, and B, while in Figure~\ref{butterfly2}
we show $\mean{B}_{\phi}$ at $r_2$ and the radial mean field
$\mean{B}_{r}$ at $r_1$ and $r_2$ for Run~A.
The structure of the magnetic
field changes as the dynamo evolves from the kinematic regime, where
the magnetic field is weak and does not significantly influence the flow.
In Run~A, the azimuthal and radial mean fields migrate poleward
close to the equator in the kinematic regime.
The cycle period is short, just around $20\tau$.
In Run~B, we find a similar behavior.
The fast poleward migration happens at
low latitudes ($\pm 40^\circ$) for both runs.
We recall that in Run~A, after a short time ($t/\tau\sim 100$), the field is 
strong enough to backreact on the flow.
At that time, two things happen simultaneously: 
an oscillating mean magnetic field starts to migrate equatorward
at higher latitudes and the fast poleward migration becomes slower.
The period of the equatorward oscillation is longer and is between
$100\tau$ and $150\tau$ for the rest of the run.
This period is consistent with those obtained from the $\Brms$ time
series.  The poleward migration near the equator slows down until it
finally turns into an equatorward migration aligned with the migration
at higher latitudes ($t/\tau=500$).  Thus, we have equatorward
migration of the mean radial and azimuthal fields at all latitudes
until around $t/\tau=1500$ and again after $t/\tau=2300$; see
Figure~\ref{butterfly2}.  During this interval, the dynamo mode
changes and, consequently, its latitudinal migration pattern changes.
The equatorward migrating and oscillating field near the poles show a
stable pattern during the whole simulation, but near the equator the
field changes with time.  In the northern hemisphere there is a
transient poleward migration, which is in phase with the equatorward
migration near the poles.  In the southern hemisphere the equatorward
migration is still dominant, but a stationary mode is superimposed near
the equator.  After $t/\tau=2300$, the equatorward migration returns
and penetrates again to lower latitudes.

The migration patterns are
not just features appearing close to the surface, but they penetrate
the entire convection zone until the bottom, as seen in
Figure~\ref{butterfly2}.  This makes it implausible that meridional
circulation is the main driver of this migration.  As
discussed in Section~\ref{sec:diff}, the meridional circulation shows
strong variability in radius and has at least two cells.

Runs~Ab and Ac have been restarted from a snapshot of Run~A after
$t/\tau\approx1350$ and $1150$, respectively.
In Run~Ab we decrease the magnetic Prandtl number, while in Run~Ac we
increase it; see Table~\ref{tab:runs}.
This was done to investigate the influence of the magnetic Reynolds
number on the equatorward migration.
As seen from Figure~\ref{butterfly} the pattern of the mean magnetic
field is not strongly affected by this change.
There is clear equatorward migration near the poles and some
indication of poleward migration at low latitudes.
It seems that Run~Ac, with a higher magnetic Reynolds number,
shows a clearer equatorward migration pattern.
However, some stationary fields are superimposed on the field near the
equator.
In Run~Ab, the field tends to migrate poleward near the equator and
equatorward near the poles.
In any case, a higher magnetic Reynolds number (keeping the other
parameters the same) seems to support equatorward migration.
This is promising because it goes in the right direction toward explaining the
Sun, although the differences in Reynolds numbers are not large enough to
draw strong conclusions.

In Run~B, where the fluid and magnetic Reynolds numbers are higher and
the rotation is slower than in Run~A, the structure of the mean field
evolution shows some differences.
In the kinematic regime, the field is similar to that of Run~A in which it
migrates poleward at lower latitudes.
Also, as the field gets stronger, it begins to migrate from higher latitudes
toward the equator and the low-latitude fast poleward branch becomes slower.
The main difference from Run~A is that the poleward migration does
not turn into equatorward migration near the equator.
In Runs~A, Ab, and Ac, the field strengths have no clear latitudinal
dependence.
In contrast, in Run~B the field strength near the poles is around half the
strength near the equator.
Only during late times does the high-latitude branch increase in
strength.
In Runs~A and Ac, the radial and azimuthal components have approximately
the same strength, whereas in Run~B, the radial mean magnetic field
seems to be weaker by a factor of two.
Also, Run~B shows no clear radial dependence in the structure of the mean field.
The period of the equatorward migrating field is $\approx200\tau$,
which is a bit longer than in Run~A.
The poleward migration near the equator has an irregular oscillation
and is usually not in phase with the equatorward migration near the
poles.
At $t/\tau\approx1000$, the dynamo mode changes significantly.
Not only does the magnetic field start to grow (see Figure~\ref{purms}),
but the magnetic field also changes from an oscillatory pattern to a
stationary one, or at least an oscillatory one with a much longer period; see
Figure~\ref{butterfly}.
In particular, in the northern hemisphere, the mean azimuthal field shows
a strong increase in strength.
The field pattern now consists of a strong time-independent component
with a latitudinal dependence.
Near the surface (at $r_1$; see Section~\ref{model})
the field seems to migrate slowly toward the equator,
but it is not possible to identify a migrating pattern in the present
run.

If one translates the cycle period of the equatorward migration to solar values
using a turnover time $\tau$ of 1\,month, we obtain a
cycle period of 12 and 16 yr for Runs~A and B, respectively.
This would be a typical value in the middle of the convection zone.
However, if one uses the Coriolis numbers of our runs (see
Table~\ref{tab:runs}),
then $\tau=P_{\rm sun}\Co/4\pi$ would be 0.7 months,
which would lead to 9 yr and 12 yr for Runs~A and B, respectively.
The regular magnetic cycles in the work of \cite{GCS10} and \cite{Racine11}
have a somewhat longer period of 60 years.
\cite{NBBMT13} and \cite{BMBBT11} were only able to generate highly irregular
cycles, with no clear reversal in both hemispheres.
They therefore found a large range of cycle periods, which
span from 1 yr to around 60 yr.

To investigate the equatorward migration, we plot the mean azimuthal
magnetic
field $\mean{B}_\phi$ for eight different times for Run~A, resolving one cycle;
see Figure~\ref{bphi}.
The field penetrates the entire convection zone and has up to four
regions with different polarities in one hemisphere.
These polarities are migrating toward the equator.
In the northern hemisphere at around $45^\circ$ latitude, there is a
magnetic field concentration with positive polarity.
After $\Delta t/\tau=827$--$740=87$ (panel 5), we find a negative
magnetic field concentration at the same location, and again after $\Delta t/\tau=65$
(panel 8) the same polarity as in the beginning of the cycle appears.
One can see a clear cyclical equatorward migration of the field, but
it is irregular.
The two hemispheres do not show the same magnetic field strength and
it seems that, from time to time, there is only one dominant polarity
in one hemisphere, while in the other there are three.
Note also the strong negative magnetic fields near and above the
surface, which also seem to show cyclical behavior.

It is still unclear why the equatorward migration
takes place, so we can only speculate about it.
There are several candidate explanations.
One is the meridional circulation, which shows a solar-like pattern in
Runs~A, Ab, and Ac.
But, as shown in Figure~\ref{butterfly2}, the equatorward migration
is present throughout the bulk of the convective zone, while the
meridional circulation becomes more incoherent with depth.
The incoherence is a manifestation of the multi-cellular structure of the
meridional circulation.
The shape and number of cells are similar to those obtained in
recent simulations by
\cite{KMB12} and \cite{NBBMT13} and observations by \cite{ZBKDH13}.
This is quite different from the single cell circulation postulated in
flux transport dynamo models to drive equatorward migration
\citep[e.g.,][]{CSD95,DC99,KO11}.
We can therefore conclude that we find no evidence for magnetic field
generation similar to the mechanism proposed by the flux transport
dynamo models.
A second candidate is the contribution of current helicity of small-scales
in the magnetic $\alpha$ tensor.
However, preliminary studies suggest that the isotropic part does not
seem to play an important role here \citep{WKMB13}.
This point is not fully conclusive, because we have not yet determined
the full anisotropic contributions to the magnetic quenching term
\citep[see][for a detailed description and discussion]{BS07}.
Finally, \cite{KMCWB13} used the phase difference of the poloidal and toroidal
magnetic fields to argue that an $\alpha^2$ dynamo is responsible for
the equatorward migration in their model.
This might also be the case here, which is also indicated by the fact
that the amplitude ratio of poloidal to toroidal field is near unity;
see Figure~\ref{butterfly2}.

The radial dependence of the mean azimuthal magnetic field
(see the last panel of Figure~\ref{butterfly2}) suggests that most
of the contribution to the cyclical behavior comes from the surface layers.
First of all, the field is strong near and above the surface where
the density stratification is large, but also at the bottom of the
convection zone, at least after saturation.
The oscillation pattern seems to be predominantly a surface phenomenon
with extension to the bulk rather than one rooted deep in the
convection zone.
The field at the bottom of the convection zone has constant
polarity, while in the bulk and at the surface of the convection the
oscillation is quite pronounced.
For example, at $t/\tau\approx2000$, a negative magnetic field rises from
the bottom and gets concentrated near the surface while, at the same time,
a positive field seems to be formed close to the
surface and emerges above the surface where it gets concentrated. 
This suggests that the strong density stratification, which is present
only very close to the surface of the Sun, might be responsible for the
oscillation and the equatorward migration of the solar magnetic field.

Further investigations measuring the turbulent transport
coefficients in their full tensorial form are necessary to determine the
reason for the equatorward migration.
Measuring the components of $\alpha$ by neglecting the contributions
of the turbulent magnetic diffusivity $\etat$ as by \cite{Racine11},
can be misleading.
A more sophisticated approach is to use the so-called test-field method
\citep{SRSRC07,BCDHKR10} adapted for spherical coordinates.

\section{Conclusions}
\label{sec:conclusions}

We have used a model that combines the turbulent convective dynamo
with a coronal layer to reproduce properties of the Sun.
We found a solar-like differential rotation with roughly radial contours of
angular velocity at low latitudes.
This is accompanied by a multi-cellular meridional circulation, 
which is manifested as a solar-like poleward flow near the surface.
Additionally, the differential rotation
profiles show a near-surface shear layer in all of the four
simulations we perform.  In one of the four simulations, there also
exists a similar layer above the surface. 
We identify the self-consistently generated, non-zero latitudinal
entropy gradient as the main cause of the spoke-like differential
rotation.

The mean magnetic field shows a pattern of equatorward migration
similar to \cite{KMB12,KMCWB13}.
This pattern is mostly visible at higher latitudes and only
in two of the simulation at lower latitudes.
However, at intermediate times of the simulation, the equatorward migration
is only visible at high latitudes, while at lower latitudes poleward
migration or stationary modes occur.
In one of the simulations, the dynamo mode changed to a
stationary one on all latitudes at later stages.
The dynamo has a shorter excitation time than in the earlier work of
\cite{KMB12}.

The present work leads to the conclusion that the inclusion of a coronal layer
in convective dynamo simulations has an influence on the fluid and magnetic
properties of the interior.
In recent simulations, we were able to produce recurrent coronal
ejections from the solar surface \citep{WKMB12} using a two layer
approach.
In earlier models of forced turbulence with a coronal layer
\citep{WB10,WBM11,WBM12}, we also found ejection of magnetic
helicity out of the dynamo region.
These ejections can support and amplify the magnetic
field due to significant magnetic helicity fluxes.

Here, we present evidence that even the fluid properties
in the bulk of the convection zone might be influenced by the coronal layer.
Spoke-like rotation profiles could not be obtained by earlier
DNS of convective dynamos \citep{KMB12} without
prescribing a latitudinal entropy gradient at the bottom of the
convection zone \citep{MBT06} or adding a stably stratified layer
below the convection zone \citep{BMT11} in purely
hydrodynamical LES.
However, to have more convincing evidence in support of this, we need to perform a
detailed parameter study using different coronal sizes and
compare them with simulations without a corona.

Another extension of our work is the measurement of magnetic helicity fluxes through the
surface and their dependence on the size of the corona.
To investigate the mechanism of the equatorward migration, which is
crucial for understanding the solar dynamo, one should measure the
turbulent transport coefficients through approaches like the test-field method
\citep{SRSRC07}.

In further work, we plan to investigate the possibility of producing
coronal ejections using the setup of these runs.
In comparison with \cite{WKMB12}, we use here a corona with a much
higher temperature and a lower plasma beta.
It will be interesting to see how the coronal ejections are influenced
by these changes.

\acknowledgements 
We thank the anonymous referee for many useful suggestions.
We acknowledge the allocation of computing resources provided by the
Swedish National Allocations Committee at the Center for
Parallel Computers at the Royal Institute of Technology in
Stockholm, the National Supercomputer Centers in Link\"oping and the
High Performance Computing Center North in Ume\aa.
Part of the computations have been carried out in 
the facilities hosted by the CSC---IT Center for Science in Espoo, Finland, 
which are financed by the Finnish ministry of education.
This work was supported in part by
the European Research Council under the AstroDyn Research Project No.\ 227952, 
the Swedish Research Council Grant No.\ 621-2007-4064, and the Academy 
of Finland grants 136189, 140970 (P.J.K) and 218159, 141017 (M.J.M),
the University of Helsinki ``Active Suns'' research project, as well
as the HPC-Europa2 project, funded by the European Commission--DG
Research in the Seventh Framework Programme under grant agreement No.\ 228398.
The authors thank NORDITA for hospitality during their visits.
\bibliography{paper}

\begin{thebibliography}{53}
\expandafter\ifx\csname natexlab\endcsname\relax\def\natexlab#1{#1}\fi

\bibitem[{{Augustson} {et~al.}(2012){Augustson}, {Brown}, {Brun}, {Miesch}, \&
  {Toomre}}]{ABBMT12}
{Augustson}, K.~C., {Brown}, B.~P., {Brun}, A.~S., {Miesch}, M.~S., \&
  {Toomre}, J. 2012, \apj, 756, 169

\bibitem[{{Ballot} {et~al.}(2007){Ballot}, {Brun}, \&
  {Turck-Chi{\`e}ze}}]{BBT07}
{Ballot}, J., {Brun}, A.~S., \& {Turck-Chi{\`e}ze}, S. 2007, \apj, 669, 1190

\bibitem[{{Blackman} \& {Brandenburg}(2003)}]{BB03}
{Blackman}, E.~G., \& {Brandenburg}, A. 2003, \apjl, 584, L99

\bibitem[{{Brandenburg}(2005)}]{B05}
{Brandenburg}, A. 2005, \apj, 625, 539

\bibitem[{{Brandenburg} {et~al.}(2010){Brandenburg}, {Chatterjee}, {Del Sordo},
  {Hubbard}, {K{\"a}pyl{\"a}}, \& {Rheinhardt}}]{BCDHKR10}
{Brandenburg}, A., {Chatterjee}, P., {Del Sordo}, F., {Hubbard}, A.,
  {K{\"a}pyl{\"a}}, P.~J., \& {Rheinhardt}, M. 2010, PhST, 142, 014028

\bibitem[{{Brandenburg} {et~al.}(1992){Brandenburg}, {Moss}, \&
  {Tuominen}}]{BMT92}
{Brandenburg}, A., {Moss}, D., \& {Tuominen}, I. 1992, \aap, 265, 328

\bibitem[{{Brandenburg} \& {Sandin}(2004)}]{BS04}
{Brandenburg}, A., \& {Sandin}, C. 2004, \aap, 427, 13

\bibitem[{{Brandenburg} \& {Subramanian}(2005)}]{BS05}
{Brandenburg}, A., \& {Subramanian}, K. 2005, \physrep, 417, 1

\bibitem[{{Brandenburg} \& {Subramanian}(2007)}]{BS07}
{Brandenburg}, A., \& {Subramanian}, K. 2007, AN, 328, 507

\bibitem[{{Brown} {et~al.}(2008){Brown}, {Browning}, {Brun}, {Miesch}, \&
  {Toomre}}]{BBBMT08}
{Brown}, B.~P., {Browning}, M.~K., {Brun}, A.~S., {Miesch}, M.~S., \& {Toomre},
  J. 2008, \apj, 689, 1354

\bibitem[{{Brown} {et~al.}(2011){Brown}, {Miesch}, {Browning}, {Brun}, \&
  {Toomre}}]{BMBBT11}
{Brown}, B.~P., {Miesch}, M.~S., {Browning}, M.~K., {Brun}, A.~S., \& {Toomre},
  J. 2011, \apj, 731, 69

\bibitem[{{Brun} {et~al.}(2004){Brun}, {Miesch}, \& {Toomre}}]{BMT04}
{Brun}, A.~S., {Miesch}, M.~S., \& {Toomre}, J. 2004, \apj, 614, 1073

\bibitem[{{Brun} {et~al.}(2011){Brun}, {Miesch}, \& {Toomre}}]{BMT11}
{Brun}, A.~S., {Miesch}, M.~S., \& {Toomre}, J. 2011, \apj, 742, 79

\bibitem[{{Choudhuri} {et~al.}(1995){Choudhuri}, {Sch\"ussler}, \&
  {Dikpati}}]{CSD95}
{Choudhuri}, A.~R., {Sch\"ussler}, M., \& {Dikpati}, M. 1995, \aap, 303, L29

\bibitem[{{Dikpati} \& {Charbonneau}(1999)}]{DC99}
{Dikpati}, M., \& {Charbonneau}, P. 1999, \apj, 518, 508

\bibitem[{{Gastine} {et~al.}(2012){Gastine}, {Duarte}, \&
  {Wicht}}]{Gastine_etal12}
{Gastine}, T., {Duarte}, L., \& {Wicht}, J. 2012, \aap, 546, A19

\bibitem[{{Ghizaru} {et~al.}(2010){Ghizaru}, {Charbonneau}, \&
  {Smolarkiewicz}}]{GCS10}
{Ghizaru}, M., {Charbonneau}, P., \& {Smolarkiewicz}, P.~K. 2010, \apjl, 715,
  L133

\bibitem[{{Gilman}(1983)}]{G83}
{Gilman}, P.~A. 1983, \apjs, 53, 243

\bibitem[{{Hubbard} \& {Brandenburg}(2012)}]{HB12}
{Hubbard}, A., \& {Brandenburg}, A. 2012, \apj, 748, 51

\bibitem[{{K{\"a}pyl{\"a}} {et~al.}(2010){K{\"a}pyl{\"a}}, {Korpi},
  {Brandenburg}, {Mitra}, \& {Tavakol}}]{KKBMT10}
{K{\"a}pyl{\"a}}, P.~J., {Korpi}, M.~J., {Brandenburg}, A., {Mitra}, D., \&
  {Tavakol}, R. 2010, AN, 331, 73

\bibitem[{{K{\"a}pyl{\"a}} {et~al.}(2006){K{\"a}pyl{\"a}}, {Korpi}, \&
  {Tuominen}}]{KKT06}
{K{\"a}pyl{\"a}}, P.~J., {Korpi}, M.~J., \& {Tuominen}, I. 2006, AN, 327, 884

\bibitem[{{K{\"a}pyl{\"a}} {et~al.}(2011{\natexlab{a}}){K{\"a}pyl{\"a}},
  {Mantere}, \& {Brandenburg}}]{KMB11}
{K{\"a}pyl{\"a}}, P.~J., {Mantere}, M.~J., \& {Brandenburg}, A.
  2011{\natexlab{a}}, AN, 332, 883

\bibitem[{{K{\"a}pyl{\"a}} {et~al.}(2012){K{\"a}pyl{\"a}}, {Mantere}, \&
  {Brandenburg}}]{KMB12}
{K{\"a}pyl{\"a}}, P.~J., {Mantere}, M.~J., \& {Brandenburg}, A. 2012, \apjl,
  755, L22

\bibitem[{{K{\"a}pyl{\"a}} {et~al.}(2013){K{\"a}pyl{\"a}}, {Mantere}, {Cole},
  {Warnecke}, \& {Brandenburg}}]{KMCWB13}
{K{\"a}pyl{\"a}}, P.~J., {Mantere}, M.~J., {Cole}, E., {Warnecke}, J., \&
  {Brandenburg}, A. 2013, \apj, 778, 41

\bibitem[{{K{\"a}pyl{\"a}} {et~al.}(2011{\natexlab{b}}){K{\"a}pyl{\"a}},
  {Mantere}, {Guerrero}, {Brandenburg}, \& {Chatterjee}}]{KMGBC11}
{K{\"a}pyl{\"a}}, P.~J., {Mantere}, M.~J., {Guerrero}, G., {Brandenburg}, A.,
  \& {Chatterjee}, P. 2011{\natexlab{b}}, \aap, 531, A162

\bibitem[{{Karak}(2010)}]{K10}
{Karak}, B.~B. 2010, \apj, 724, 1021

\bibitem[{{Kitchatinov} \& {Olemskoy}(2012)}]{KO11}
{Kitchatinov}, L.~L., \& {Olemskoy}, S.~V. 2012, \solphys, 276, 3

\bibitem[{{Kitchatinov} \& {R\"udiger}(1995)}]{KR95}
{Kitchatinov}, L.~L., \& {R\"udiger}, G. 1995, \aap, 299, 446

\bibitem[{{K{\"o}hler}(1970)}]{Ko70}
{K{\"o}hler}, H. 1970, \solphys, 13, 3

\bibitem[{Krause \& R{\"a}dler(1980)}]{KR80}
Krause, F., \& R{\"a}dler, K.-H. 1980, {Mean-field Magnetohydrodynamics and
  Dynamo Theory} (Oxford: Pergamon)

\bibitem[{{Miesch} {et~al.}(2006){Miesch}, {Brun}, \& {Toomre}}]{MBT06}
{Miesch}, M.~S., {Brun}, A.~S., \& {Toomre}, J. 2006, \apj, 641, 618

\bibitem[{{Mitra} {et~al.}(2009){Mitra}, {Tavakol}, {Brandenburg}, \&
  {Moss}}]{MTBM09}
{Mitra}, D., {Tavakol}, R., {Brandenburg}, A., \& {Moss}, D. 2009, \apj, 697,
  923

\bibitem[{{Mitra} {et~al.}(2010){Mitra}, {Tavakol}, {K{\"a}pyl{\"a}}, \&
  {Brandenburg}}]{Mitra_etal10}
{Mitra}, D., {Tavakol}, R., {K{\"a}pyl{\"a}}, P.~J., \& {Brandenburg}, A. 2010,
  \apjl, 719, L1

\bibitem[{{Nelson} {et~al.}(2013){Nelson}, {Brown}, {Brun}, {Miesch}, \&
  {Toomre}}]{NBBMT13}
{Nelson}, N.~J., {Brown}, B.~P., {Brun}, A.~S., {Miesch}, M.~S., \& {Toomre},
  J. 2013, \apj, 762, 73

\bibitem[{{Racine} {et~al.}(2011){Racine}, {Charbonneau}, {Ghizaru}, {Bouchat},
  \& {Smolarkiewicz}}]{Racine11}
{Racine}, {\'E}., {Charbonneau}, P., {Ghizaru}, M., {Bouchat}, A., \&
  {Smolarkiewicz}, P.~K. 2011, \apj, 735, 46

\bibitem[{{Rempel}(2005)}]{Rempel05}
{Rempel}, M. 2005, \apj, 622, 1320

\bibitem[{{R\"udiger}(1980)}]{R80}
{R\"udiger}, G. 1980, GApFD, 16, 239

\bibitem[{{R\"udiger}(1989)}]{R89}
{R\"udiger}, G. 1989, {Differential Rotation and Stellar Convection. Sun and
  Solar-type Stars} (Berlin: Akademie)

\bibitem[{{Schou} {et~al.}(1998){Schou}, {Antia}, {Basu}, {Bogart}, {Bush},
  {Chitre}, {Christensen-Dalsgaard}, {di Mauro}, {Dziembowski}, {Eff-Darwich},
  {Gough}, {Haber}, {Hoeksema}, {Howe}, {Korzennik}, {Kosovichev}, {Larsen},
  {Pijpers}, {Scherrer}, {Sekii}, {Tarbell}, {Title}, {Thompson}, \&
  {Toomre}}]{Schouea98}
{Schou}, J., {et~al.} 1998, \apj, 505, 390

\bibitem[{{Schrinner} {et~al.}(2011){Schrinner}, {Petitdemange}, \&
  {Dormy}}]{Schrinner_etal11}
{Schrinner}, M., {Petitdemange}, L., \& {Dormy}, E. 2011, \aap, 530, A140

\bibitem[{{Schrinner} {et~al.}(2012){Schrinner}, {Petitdemange}, \&
  {Dormy}}]{Schrinner_etal12}
{Schrinner}, M., {Petitdemange}, L., \& {Dormy}, E. 2012, \apj, 752, 121

\bibitem[{{Schrinner} {et~al.}(2007){Schrinner}, {R{\"a}dler}, {Schmitt},
  {Rheinhardt}, \& {Christensen}}]{SRSRC07}
{Schrinner}, M., {R{\"a}dler}, K.-H., {Schmitt}, D., {Rheinhardt}, M., \&
  {Christensen}, U.~R. 2007, GApFD, 101, 81

\bibitem[{{Spiegel} \& {Weiss}(1980)}]{SW80}
{Spiegel}, E.~A., \& {Weiss}, N.~O. 1980, \nat, 287, 616

\bibitem[{{Stix}(2002)}]{Stix:02}
{Stix}, M. 2002, {The Sun: An Introduction} (Berlin: Springer)

\bibitem[{{Warnecke} \& {Brandenburg}(2010)}]{WB10}
{Warnecke}, J., \& {Brandenburg}, A. 2010, \aap, 523, A19

\bibitem[{{Warnecke} {et~al.}(2011){Warnecke}, {Brandenburg}, \&
  {Mitra}}]{WBM11}
{Warnecke}, J., {Brandenburg}, A., \& {Mitra}, D. 2011, \aap, 534, A11

\bibitem[{{Warnecke} {et~al.}(2012{\natexlab{a}}){Warnecke}, {Brandenburg}, \&
  {Mitra}}]{WBM12}
{Warnecke}, J., {Brandenburg}, A., \& {Mitra}, D. 2012{\natexlab{a}}, JSCSW, 2,
  A11

\bibitem[{{Warnecke} {et~al.}(2012{\natexlab{b}}){Warnecke}, {K{\"a}pyl{\"a}},
  {Mantere}, \& {Brandenburg}}]{WKMB12}
{Warnecke}, J., {K{\"a}pyl{\"a}}, P.~J., {Mantere}, M.~J., \& {Brandenburg}, A.
  2012{\natexlab{b}}, \solphys, 280, 299

\bibitem[{{Warnecke} {et~al.}(2013{\natexlab{a}}){Warnecke}, {K{\"a}pyl{\"a}},
  {Mantere}, \& {Brandenburg}}]{WKMB13}
{Warnecke}, J., {K{\"a}pyl{\"a}}, P.~J., {Mantere}, M.~J., \& {Brandenburg}, A.
  2013{\natexlab{a}}, in IAU Symposium, Vol. 294, IAU Symposium, ed. A.~G.
  {Kosovichev}, E.~{de Gouveia Dal Pino}, \& Y.~{Yan}, 307--312

\bibitem[{{Warnecke} {et~al.}(2013{\natexlab{b}}){Warnecke}, {Losada},
  {Brandenburg}, {Kleeorin}, \& {Rogachevskii}}]{WLBKR13}
{Warnecke}, J., {Losada}, I.~R., {Brandenburg}, A., {Kleeorin}, N., \&
  {Rogachevskii}, I. 2013{\natexlab{b}}, \apjl, 777, L37

\bibitem[{{Weiss}(1965)}]{W65}
{Weiss}, N.~O. 1965, Obs, 85, 37

\bibitem[{{Zhao} {et~al.}(2013){Zhao}, {Bogart}, {Kosovichev}, {Duvall}, \&
  {Hartlep}}]{ZBKDH13}
{Zhao}, J., {Bogart}, R.~S., {Kosovichev}, A.~G., {Duvall}, T.~L., \&
  {Hartlep}, T. 2013, \apjl, 774, L29

\bibitem[{{Zhao} \& {Kosovichev}(2004)}]{ZK04}
{Zhao}, J., \& {Kosovichev}, A.~G. 2004, \apj, 603, 776

\end{thebibliography}
\end{document}